\documentclass[12pt]{article}
\usepackage{newtxtext,newtxmath}
\usepackage{microtype} % improved typesetting
\usepackage{graphicx}
\usepackage[letterpaper,margin=1in]{geometry}
\usepackage[
    backend=biber,
    style=apa,
  ]{biblatex}
\usepackage{rotating} 
\usepackage{amsmath}
\usepackage{titling}
\usepackage{bbm}
\usepackage{url}
\usepackage{makecell}
\usepackage{booktabs}
\usepackage{multirow}
\usepackage{float}
\usepackage{longtable}
\usepackage{tikz}
\usetikzlibrary{arrows.meta, positioning, shapes.geometric, fit}
\usepackage{animate}
\usepackage{algorithm}
\usepackage{algpseudocode}
\usepackage{color}
\usepackage{caption}
\usepackage{subcaption}
\usepackage{fontawesome5}
\usepackage{titlesec}
\usepackage{mathtools} % dcases environment
\usepackage[hidelinks]{hyperref} % clickable links in pdf
\usepackage{siunitx}
\usepackage{csquotes}
\usepackage{placeins}
\MakeOuterQuote{"} % opening and closing quotes

\titleformat{\paragraph}[block]{\normalfont\normalsize\bfseries}{\theparagraph}{1em}{}
\titlespacing*{\paragraph}{0pt}{1.5ex plus 0.5ex minus .2ex}{1ex}
\newcommand{\svec}{\mathbf{s}} % bold s for location
\newcommand{\ind}{\mathbbm{1}} % indicator function

\newcommand{\prob}{\operatorname{\mathsf{P}}} % P for proba
\newcommand{\E}{\operatorname{\mathsf{E}}} % E for expected value
\newcommand{\Cov}{\operatorname{\mathsf{cov}}} % covariance

\newcommand{\Normal}{\operatorname{\mathcal{N}}} % Normal dist
\newcommand{\HalfNormal}{\mathcal{HN}}
\newcommand{\bu}{\boldsymbol{u}}

\newcommand{\taildep}[3]{\chi_{#1,#2}(#3)}
\newcommand{\taildepest}[3]{\hat{\chi}_{#1,#2}(#3)}

\newcommand{\abs}[1]{\left\lvert #1 \right\rvert}
\newcommand{\hvec}{\boldsymbol{h}} % euclidian distance h
\newcommand{\Rregion}{\mathcal{R}} % Region symbole
\newcommand{\period}{\ensuremath{N}} % return period
\newcommand{\zlevel}{z_N} % Return level 
\newcommand{\prop}{\mathcal{P}}      % property of interest
\newcommand{\propval}{X^{(\prop)}}  % value od property 
\newcommand{\Ngh}{\mathcal{A}}

\newcommand{\covariates}{\mathbf{c}}
\newcommand{\extremalI}{\theta}
\newcommand{\Tthres}{\vartheta_{\svec}^{\text{thres}}}
\newcommand{\TthresEmp}{\hat{\vartheta}_{\svec}^{\text{thres}}}
\newcommand{\paramsDEP}{\Theta^\text{DEP}_{\Rregion}}
\newcommand{\paramsDEPboot}{\Theta^\text{DEP,(b)}_{\Rregion}} 

\newcommand{\Anisomat}{\boldsymbol{T}}

\newcommand{\paramsGEV}{\Theta^{\mathrm{GEV}}_{\Rregion}}
\newcommand{\paramsGEVboot}{\Theta^{\mathrm{GEV,(b)}}_{\Rregion}}

\newcommand{\paramsBSQR}{\Theta^{\mathrm{QR}}_{\Rregion}}
\newcommand{\paramsBSQRboot}{\Theta^{\mathrm{QR,(b)}}_{\Rregion}}

\newcommand{\paramsQR}{\boldsymbol{\beta}^{\text{QR}}}
\newcommand{\paramsQRest}{\hat{\boldsymbol{\beta}}^{\text{QR}}}

\renewenvironment{abstract}
{
\vspace{-1 em}
  \begin{center}
    \bfseries Abstract
  \end{center}
  \quotation
}
{
  \endquotation
}

\date{}

\makeatletter
\renewcommand{\fnum@figure}{\textbf{Figure \thefigure}}
\renewcommand{\fnum@table}{\textbf{Table \thetable}}
\makeatother

\usepackage{url}

\def\scititle{
	A spatio-temporal statistical framework for heatwave attribution under climate change
}

\title{\bfseries \boldmath \scititle}

\author{
	Kamal~Gasser$^{1\ast}$,
	Johan~Segers$^{1,2}$,
	Francesco~Ragone$^{3}$\and
	\small$^{1}$ LIDAM/ISBA, UCLouvain, Belgium.\and
	\small$^{2}$Dept.\ of Mathematics, KU Leuven, Belgium.\and
	\small$^{3}$School of Computing and Mathematical Sciences, University of Leicester, UK.\and
	\small$^\ast$Lead and corresponding author. Email: kamal.gasser@uclouvain.be
}
\definecolor{darkteal}{rgb}{0,0.35,0.35}

\begin{document} 

% Insert the title and author list
\maketitle

\begingroup
\renewcommand\thefootnote{}
\footnotetext{%
\begin{tabular}{@{}l p{0.78\textwidth}@{}}
Funding: & This work was supported by the EXALT project funded by the Projet d’Actions de Recherche Concertées No.\ 24/29-146 of the Communauté française de Belgique.\\[0.3em]
Code availability: & \url{https://github.com/kamalgasser/heatwave-attribution-paper}
\end{tabular}%
}
\endgroup

% Abstract
\begin{abstract}
We develop a unified statistical framework for attributing heatwaves as spatio-temporal phenomena under climate change. We quantify  the impact of anthropogenic forcing on the probability and persistence of heatwaves not captured by standard marginal extreme-value approaches. Our methodology constructs a generative model for daily temperature fields that separates marginal nonstationarity from spatio-temporal dependence. We combine three components: a Bayesian spatial quantile regression model for the bulk of the data; a nonstationary spatial generalized extreme value model for tail behavior; and a copula-based model capturing both asymptotic dependence and independence in the extremes. The framework is applied to the CMIP6 MRI-ESM2 climate model, contrasting factual and counterfactual scenarios for probabilistic attribution. Our results show that the approach captures key heatwave characteristics inaccessible to traditional methods, enabling direct estimation of event-level attribution metrics. Overall, it provides a flexible basis for analyzing and attributing complex climate extremes as space-time objects.
\end{abstract}

\newpage

\section{Introduction}\label{sec:Intro}

In recent decades, the global climate has undergone significant warming due primarily to anthropogenic activity, with global mean surface temperature (GMST) rising by approximately $1.1^{\circ}$C since the pre-industrial period \parencite{ipcc2021}. The IPCC Sixth Assessment Report emphasizes that even modest shifts in mean temperature can induce substantial increases in the frequency, intensity, and duration of extreme events \parencite{Seneviratne2021}. Among these, heatwaves pose severe risks to human health, agriculture, and ecosystems \parencite{perkins2015review, Risser2025}. Europe has been particularly affected: the 2003 heatwave led to more than \num{70000} excess deaths \parencite{Zaitchik2006}, and more recent events in 2019 and 2022 have broken temperature records by unprecedented margins \parencite{Vautard2020,ballester2023heat}.

From a statistical perspective, a key challenge is to quantify how anthropogenic forcing alters the probability of such events. However, heatwaves are inherently spatio-temporal phenomena, characterized jointly by high temperatures, temporal persistence over consecutive days, and spatial coherence across regions. Many quantities of interest such as the probability, duration, or spatial extent of a regional heatwave are therefore defined at the level of the entire space-time field, rather than at a single location or time point. Standard approaches based on marginal extreme-value theory are not designed to estimate such quantities, as they discard temporal persistence and spatial dependence.

The statistical literature on extreme event attribution is currently dominated by two main approaches. First, index-based methods model spatially aggregated quantities, such as regional mean temperature or predefined heatwave indices, using parametric extreme value theory (EVT) models with covariates such as GMST anomalies \parencite{vanOldenborgh2022, wake2025heatwave}. While computationally convenient, these approaches collapse the multi-dimensional structure of heatwaves into a one-dimensional summary and do not provide a probabilistic description of their spatial extent or temporal evolution. Second, methods based on pointwise or spatially smoothed extreme-value models estimate marginal distributions at individual locations \parencite{auld2023changes}, but treat spatial and temporal dependence only indirectly. As a result, they cannot directly quantify the probability of events defined by simultaneous exceedances across space and time.

In the spatial extremes literature, dependence is often modeled using max-stable processes, which arise as asymptotic limits of normalized maxima \parencite{davison2019spatial, huser2014space,genton2015multivariate}. While theoretically appealing, these models are restricted to asymptotic dependence and can be too rigid for environmental applications, where data frequently exhibit weakening dependence at extreme levels \parencite{huser2022advances}. This limitation has motivated the development of sub-asymptotic models capable of capturing both asymptotic dependence and independence \parencite{huser2017bridging,huser2019modeling,simpson2021conditional,dell2024flexible, wadsworth2018spatial}. In the context of heatwaves, accurately representing this dependence structure is crucial, as it directly affects the probability of spatially extended and temporally persistent events \parencite{li2025importance}.

In this paper, we propose a statistical framework for modeling heatwaves as coherent spatio-temporal events under nonstationary climate conditions. The key idea is to separate marginal nonstationarity from spatio-temporal dependence, allowing each component to be modeled flexibly while yielding a coherent joint representation of daily temperature fields. This enables the estimation of event-level quantities, such as the maximum intensity  and duration of regional heatwaves, which are not directly accessible using standard marginal approaches.

Our approach combines three components. First, a Bayesian spatial quantile regression (BSQR) model \parencite{reich2011bayesian} is used to capture the nonstationarity in the bulk of the temperature distribution and to transform observations to a uniform scale. Second, a nonstationary spatial generalized extreme value (GEV) model is employed to represent the marginal tail behavior of seasonal maxima \parencite{sang2009hierarchical,dyrrdal2015bayesian,cooley2010spatial}. Third, a flexible sub-asymptotic spatio-temporal dependence model \parencite{dell2024flexible,huser2019modeling} is used to capture the joint behavior of daily temperatures across space and time. Together, these components define a generative model for daily temperature fields that preserves both marginal nonstationarity and realistic dependence structures. 

The contribution of this work is the development of a coherent statistical framework that enables the analysis of heatwaves as spatio-temporal objects through the combination of several model components. By combining marginal modeling with flexible dependence structures, the proposed approach allows the estimation of quantities relevant for extreme event attribution that cannot be obtained from standard marginal EVT methods. In particular, it provides a probabilistic characterization of heatwave occurrence, persistence, and spatial extent under both factual and counterfactual climate scenarios.

The remainder of the paper is organized as follows. Section~\ref{sec:heat_def} describes the climate model data and introduces the spatio-temporal definition of a heatwave. Section~\ref{sec:methods} presents the statistical methodology. Section~\ref{sec:Data} applies the framework to CMIP6 simulations to quantify the influence of anthropogenic forcing on heatwave characteristics. Section~\ref{sec:Discussion} concludes with a discussion of the results and implications for extreme event attribution.

\section{Climate Model Data and Heatwave Definition}\label{sec:heat_def}
\subsection{Climate Model Data and Temperature Anomalies}
\label{sec:Climate model data and Temperature Anomalies}
For the attribution analysis, we use climate model simulations from the Coupled Model Intercomparison Project Phase~6 (CMIP6). Among the available CMIP6 models, we focus exclusively on the Meteorological Research Institute Earth System Model version~2 (MRI-ESM2) \parencite{yukimoto2019meteorological}, and which has been widely used in recent detection and attribution analyses \parencite{bone2023detection,engdaw2023attribution}. The MRI-ESM2 model provides daily maximum near-surface air temperature (TASMAX) under multiple forcing scenarios. In particular, we use:
\begin{itemize}
    \item the historical experiment (``hist''), which includes both anthropogenic and natural forcings and represents the \emph{factual} world;
    \item the natural-forcing-only experiment (``hist-nat''), which excludes anthropogenic forcings and represents the \emph{counterfactual} world.
\end{itemize}
Both experiments cover the period 1850--2014, with the hist-nat simulation extending to 2020. To extend the factual world beyond 2014, we append the Shared Socioeconomic Pathway~2-4.5 (SSP245) scenario to the historical simulation which is a common practice in attribution study \parencite{quilcaille2025systematic,engdaw2023attribution,philip2022rapid}. SSP245 corresponds to a ``middle-of-the-road'' socioeconomic pathway and a radiative forcing level of \SI{4.5}{\watt\per\meter\squared} by 2100, leading to an expected global mean temperature increase of approximately \SIrange{2}{3}{\degreeCelsius} above preindustrial levels by the end of the century \parencite{tebaldi2021climate}. This extension allows the factual simulations to cover the period up to 2020. In summary, the factual world is represented by the MRI-ESM2 historical and SSP245 simulations, while the counterfactual world is represented by the hist-nat simulation. Comparing these two ensembles enables the quantification of the influence of anthropogenic forcing on extreme temperature events.

Let $Y^{\mathrm{F}}(\svec,d,t)$ and $Y^{\mathrm{C}}(\svec,d,t)$ denote daily maximum summer (June--July--August, JJA) temperatures in the factual and counterfactual worlds, respectively. Here, $\svec=(s_x,s_y)$ denotes the longitude and latitude of a grid-cell center, with $\svec\in\{\svec_1,\dots,\svec_n\}$, while $d\in\{1,\dots,D\}$ indexes the day of the summer season with $D=92$ and $t\in\{1,\dots,T\}$ indexes the year with $T=171$. For each grid cells we have a total of $DT=\num{15732}$ recorded values. To remove the mean seasonal cycle and focus on extremes relative to a preindustrial baseline, temperature values are converted into daily anomalies by using the $N=100$ first years:
\begin{equation}
\label{eq:anomalies}
X(\svec,d,t)
=
Y(\svec,d,t)
-
\frac{1}{N}\sum_{t'=1}^{N} Y(\svec,d,t').
\end{equation}
The transformation is applied to both factual and counterfactual simulations. The baseline period in both cases is 1850--1949, which corresponds to an early-industrial climate with minimal anthropogenic influence.

\subsection{Definition of heatwaves}
% Extreme weather events are typically characterized by both intensity and persistence. While phenomena such as hurricanes have well-established operational definitions \parencite{taylor2010saffir} , no universal definition of a heatwave exists. Instead, definitions vary across institutions and studies, reflecting differences in climatic context and scientific objectives. Meteorological agencies emphasize sustained periods of abnormally high temperatures, sometimes in conjunction with humidity \parencite{NWSHeatWave,NOAAHeatWave,WMOHeatWave}. In the scientific literature, heatwaves are most often defined using percentile-based thresholds relative to local climatology combined with a minimum duration criterion. For example, \textcite{fischer2010consistent} and \textcite{vautard2013simulation} define heatwaves based on exceedances of high local percentiles over several consecutive days. A detailed review of existing definitions is provided by \textcite{perkins2015review}. Despite methodological differences, two core components are common: an extreme-temperature threshold and a spatio-temporal persistence.
Heatwaves do not have a universally agreed definition and are typically characterized by both extreme temperature and persistence. In the literature, they are commonly defined using high percentile thresholds relative to local climatology combined with a minimum duration criterion \parencite{fischer2010consistent,vautard2013simulation,perkins2015review}.
In this work, we adopt a percentile-based and duration-based definition aligned with the climate extremes literature \parencite{stefanon2012heatwave,perkins2015review}. The definition below is intended as an operational indicator for attribution and simulation purposes rather than as a new conceptual definition of heatwaves. A grid cell at location $\svec$ is defined as experiencing a heatwave on summer day $d$ in year $t$ if the following indicator equals one:
\begin{equation}
\label{eq:heatwave-def}
H(\svec,d,t)
=
\ind\bigl\{
X(\svec,d-2,t)\ge \Tthres,\,
X(\svec,d-1,t)\ge \Tthres,\,
X(\svec,d,t)\ge \Tthres
\bigr\},
\end{equation}
that is, if daily temperature anomalies exceed a high threshold for at least three consecutive days. The threshold $\Tthres$ is defined as the 95th percentile of the counterfactual anomaly distribution, which is assumed to be stationary both within and across years:
\begin{equation}
\label{eq:threshold}
\Tthres
=
\inf\left\{x\in\mathbb{R}:\prob\left(X^{\mathrm{C}}(\svec,d,t)\le x\right)\ge 0.95\right\}.
\end{equation}
In practice, we use the empirical counterpart of the threshold, defined as
\begin{equation}
\label{eq:threshold_empirical}
\TthresEmp
=
\inf\left\{
x \in \mathbb{R} :
\frac{1}{DT}\sum_{d=1}^D \sum_{t=1}^T
\ind\!\left\{ X^{\mathrm{C}}(\svec,d,t) \le x \right\}
\ge 0.95
\right\}.
\end{equation}
Defining the threshold with respect to the counterfactual climate ensures that heatwaves are identified relative to a stationary baseline unaffected by anthropogenic warming, facilitating a consistent comparison between the factual and counterfactual worlds.

Finally, let $\Rregion \subset \mathbb{R}^2$ denote a spatial region of interest, composed of a collection of grid cell centers $\{\svec_1,\dots,\svec_{n_{\Rregion}}\}$. Let $a(\svec)$ denote the surface area (in $\mathrm{km}^2$) of the cell of which $\svec$ is the center. We define the regional heatwave indicator as
\begin{equation}
\label{eq:regional-heatwave}
H_{\Rregion}(d,t)
=
\ind\!\left\{
\sum_{\svec \in \Rregion}
a(\svec)\,H(\svec,d,t)
\;\ge\;
\alpha \sum_{\svec \in \Rregion} a(\svec)
\right\},
\end{equation}
where $H(\svec,d,t)$ is the local heatwave indicator defined in \eqref{eq:heatwave-def}, and $\alpha \in (0,1)$ denotes the minimum fraction of the region's total surface area that must be affected to declare a regional heatwave event. In this work, we set $\alpha = 0.6$. This fractional area criterion is in accordance with \textcite{stefanon2012heatwave,rousi2022accelerated,tian2024characterizing}. Under this definition, a region $\Rregion$ is said to experience a heatwave on day $d$ of year $t$ if all grid cells experiencing a heatwave simultaneously cover at least $60\%$ of the region's total surface area.

\section{Methods}\label{sec:methods}

Our objective is to model and simulate nonstationary daily summer temperature fields under the factual climate scenario, and stationary temperature fields under the counterfactual scenario, in order to study the occurrence, persistence, and spatial extent of heatwaves. Heatwaves, as defined in Section~\ref{sec:heat_def}, Eqs.~\eqref{eq:heatwave-def} and~\eqref{eq:regional-heatwave}, depend jointly on high quantiles of the daily temperature distribution, persistence over consecutive days, and spatial coherence across a region. Capturing these features requires a framework that accommodates nonstationary marginal behavior driven by external covariates while allowing flexible spatio-temporal dependence at the daily scale. Recall that, as discussed in Section~\ref{sec:Climate model data and Temperature Anomalies}, seasonality has been approximately removed by restricting the analysis to summer months, under the assumption that seasonal effects primarily impact the mean of the temperature distribution.

We adopt the working assumption that, for a given year~$t$, daily temperature anomalies are marginally stationary within the summer season. Let $ F_{\svec,d,t}$ be the distribution function of $X(\svec,d,t)$. We assume that
\begin{equation}
\label{eq:daily-stationarity}
    \forall\, \svec \in \Rregion,\ \forall\, t \in \{1,\dots,T\},\ \exists\, F_{\svec,t}
    \quad \text{such that} \quad
    F_{\svec,1,t} = \cdots = F_{\svec,D,t} = F_{\svec,t},
\end{equation}
where $F_{\svec,t}$ denotes the distribution function of $X(\svec,d,t)$ in year $t$. This assumption implies that, conditional on the year, daily observations at a given location are identically distributed, while the marginal distribution is allowed to evolve across years. Temporal nonstationarity is introduced exclusively through year-specific covariates, while within-season variability is captured through the dependence structure.

To address these challenges, we decompose the modeling problem into three complementary components, each targeting a specific aspect of the heatwave process:
\begin{itemize}
\item A BSQR model is used to capture the nonstationary evolution of the bulk of the temperature distribution across space and time. This component allows us to define heatwave thresholds relative to a counterfactual baseline and to account for changes in the distribution under climate forcing. However, as we show in Section~\ref{sec:Data}, the BSQR model alone is insufficient to accurately represent the upper tail of the distribution.

\item A GEV model to address the limitation of the BSQR. We model the marginal tail behavior using a nonstationary spatial GEV model applied to seasonal maxima. This component provides a theoretically justified representation of extreme values and corrects the underestimation of tail risk observed when relying solely on the bulk model.

\item Finally, a copula-based spatio-temporal dependence model is introduced to capture the joint behavior of daily temperatures across space and time. This component is essential to represent the persistence and spatial coherence of heatwaves, which cannot be recovered from marginal models alone.
\end{itemize}
Together, these three components define a coherent generative model for daily temperature fields that preserves both marginal nonstationarity and realistic dependence structures.

\subsection{Bayesian Spatial Quantile Regression Model}
This section presents the Bayesian hierarchical model corresponding to the \emph{approximate method} of \textcite[Sec.~3]{reich2011bayesian}. We use the BSQR framework to model the conditional quantile function of temperature anomalies $X(\svec,d,t)$ given covariates $\covariates(t)$. Let $q(\tau\mid\covariates(t),\svec)$ denote the conditional $\tau$-th quantile of $X(\svec,d,t)$, modeled as
\begin{equation}
q(\tau\mid\covariates(t),\svec)
=
\covariates(t)^\top \paramsQR(\tau,\svec),
\end{equation}
where $\paramsQR(\tau,\svec)$ are spatially varying quantile regression coefficients. Following \textcite{reich2011bayesian}, inference is performed using a two-stage approximation. First, for each grid cell center $\svec$, classical quantile regression is applied at levels $\tau_1,\ldots,\tau_K$, yielding estimates $\paramsQRest(\svec)\in\mathbb{R}^{pK}$ and associated covariance matrices $\hat\Sigma(\svec)$. These estimators are consistent and asymptotically normal; details on their computation are provided in Supplement~\ref{sec:supp_BSQR}.

In the second stage, these estimates are treated as observations and linked to the latent coefficients through a Gaussian likelihood:
\begin{equation}
\label{eq:approx_likelihood_short}
\paramsQRest(\svec)
\sim
\mathcal N \bigl(\paramsQR(\svec),\,\hat\Sigma(\svec)\bigr),
\qquad \svec\in\Rregion.
\end{equation}

To model variation across quantile levels, each coefficient function is represented using a Bernstein polynomial basis,
\begin{equation}
\beta^{\mathrm{QR}}_j(\tau,\svec)
=
\sum_{m=1}^{M} B_m(\tau)\,\alpha_{jm}(\svec),
\end{equation}
with $M$ fixed. To enforce non-crossing quantiles, we adopt the increment parametrization of \textcite{reich2011bayesian}, defining $\alpha_{jm}(\svec)=\sum_{\ell=1}^{m}\delta_{j\ell}(\svec)$, where constraints on $\delta_{jm}(\svec)$ ensure monotonicity in $\tau$. Spatial dependence is introduced by assigning Gaussian process priors to latent variables $\delta^{*}_{jm}(\svec)$, which determine the increments and hence the quantile coefficient functions. Their prior mean is centered on a parametric base distribution indexed by $\Omega$, while $\tilde{\sigma}_j^2$ and $\rho_j$ control respectively the marginal variance and spatial range of the Gaussian process associated with covariate $j$. For notational convenience, let $\boldsymbol{\beta}^{\mathrm{QR}}(\svec)\in\mathbb{R}^{pK}$ denote the vector obtained by evaluating $\beta_j^{\mathrm{QR}}(\tau,\svec)$ at $\tau_1,\ldots,\tau_K$ and stacking over $j=1,\ldots,p$.

Let $\Theta^{\mathrm{QR}}_{\Rregion}$ denote the collection of all unknown parameters. The posterior distribution is given by
\begin{equation}
\label{eq:posterior_bsqr_short}
\begin{aligned}
\pi\left(\Theta^{\mathrm{QR}}_{\Rregion} \mid \{\paramsQRest(\svec)\}_{\svec \in \Rregion}\right)
\;\propto\;&
\prod_{\svec \in \Rregion}
\Normal\!\left(
\paramsQRest(\svec) \mid
\boldsymbol\beta^{\mathrm{QR}}(\svec),
\hat\Sigma(\svec)
\right)
\\
&\times
\pi\bigl(\{\delta^{*}_{jm}(\cdot)\}\mid\Omega,\{\tilde{\sigma}_j^2,\rho_j\}\bigr)
\,
\pi_\Omega(\Omega)
\prod_{j=1}^{p}
\pi(\tilde{\sigma}_j^2)\pi(\rho_j).
\end{aligned}
\end{equation}
This formulation yields spatially smooth, non-crossing quantile functions while substantially reducing computational complexity compared to the full likelihood approach. Full details of the construction, including covariance estimation, the non-crossing parametrization, and prior specification, are provided in Supplement~\ref{sec:supp_BSQR}.\label{sec:BQR_spe}

\subsection{Marginal modeling of the tail (Univariate extreme value theory)}\label{sec:GEV}
The classical starting point of EVT is the modeling of block maxima. Let $Y_1,\dots,Y_n$ be independent and identically distributed random variables with common distribution function $F$, and define the block maximum $M_n=\max(Y_1,\dots,Y_n)$. If sequences $\{a_n\}$ with $a_n>0$ and $\{b_n\}$ exist such that the normalized maximum $M_n^*=(M_n-b_n)/a_n$ converges in distribution to a non-degenerate limit, then the limiting distribution must belong to the GEV family \parencite{fisher1928limiting,jenkinson1955frequency,coles2001introduction}. The GEV distribution function is given by
\begin{equation}\label{eq:GEV_CDF}
G(z)=
\begin{dcases}
\exp \Bigl[-\{ \max(1+\xi(z-\mu)/\sigma,0)\}^{-1/\xi}\Bigr], & \xi\neq 0, \\
\exp \bigl[-\exp\{-(z-\mu)/\sigma\}\bigr], & \xi=0,
\end{dcases}
\end{equation}
where $\mu\in\mathbb{R}$ is a location parameter, $\sigma>0$ is a scale parameter, and $\xi\in\mathbb{R}$ is a shape parameter controlling tail behavior.

For each grid cell center $\svec\in\Rregion$, we observe daily maxima within each summer season and extract the annual maximum. Let $M_{\svec,t}=\max_{d=1,\ldots,D} X(\svec,d,t)$ denote the annual maximum over all days in year $t$ for the grid cell of which $\svec$ is the center. We further observe $p$ time-varying covariates $\covariates(t)\in\mathbb{R}^p$
common to all sites as in Section~\ref{sec:BQR_spe}. These covariates enter the marginal GEV parameters, yielding a nonstationary spatial GEV model \parencite{coles2001introduction}
\begin{equation}\label{eq:max_model}
    M_{\svec,t}\sim \operatorname{GEV}\bigl(\mu_{\svec}(t),\sigma_{\svec}(t),\xi_{\svec}\bigr).
\end{equation}
We model the location and scale parameters through linear predictors
\begin{align}
\label{eq:loc_gev_function}
    \mu_{\svec}(t)
    &=\mu_{0,\svec}+c_1(t)\mu_{1,\svec}+c_2(t)\mu_{2,\svec}+\dots+c_p(t)\mu_{p,\svec}, \\
\label{eq:scale_gev_function}
    \log(\sigma_{\svec}(t))
    &=\log(\sigma_{0,\svec})+c_1(t)\sigma_{1,\svec}+c_2(t)\sigma_{1,\svec}+\dots+c_p(t)\sigma_{p,\svec},
\end{align}
while the shape parameter $\xi_{\svec}$ is assumed to remain constant over time. The location and scale parameters $\mu_{\svec}(t)$ and $\sigma_{\svec}(t)$ are expressed in \si{\degreeCelsius}, while the shape parameter $\xi_{\svec}$ is dimensionless.

Because nearby locations are exposed to the same regional extreme events, the GEV parameters are expected to vary smoothly over space. To encode this spatial structure while retaining interpretable global effects, we adopt a hierarchical Bayesian formulation based on an
Intrinsic Conditional Autoregressive (ICAR) prior, which induce intrinsic Gaussian Markov random Field (GMRF) structures\parencite{ferreira2024modeling,rue2005gaussian}. This approach is similar to what has been done by \textcite{sang2009hierarchical,dyrrdal2015bayesian,cooley2010spatial}. For each coefficient type $\ell$ (e.g., intercept or covariate slope in $\mu$ or $\log\sigma$, or the shape parameter $\xi$), define the spatial coefficient vector
\begin{equation}\label{eq:gamma_GEV}
    \boldsymbol{\gamma}_{\ell}=(\gamma_{\ell,1},\dots,\gamma_{\ell,n})^\top.
\end{equation}
Rather than assigning an ICAR prior directly to $\boldsymbol{\gamma}_{\Rregion,\ell}$, we decompose each spatial coefficient vector into a global mean plus a spatially structured deviation. We assume this decomposition is specified independently for each coefficient type $\ell$.
Specifically,
\begin{equation}
\label{eq:gamma_decomp}
\boldsymbol{\gamma}_{\Rregion,\ell}=\gamma_{\ell,g}\ind+s_\ell\bu_\ell,
\qquad
\bu_\ell\sim\mathrm{ICAR}(\boldsymbol{A}),
\end{equation}
where $\ind$ denotes the $n$-vector of ones, $\gamma_{\ell,g}\in\mathbb{R}$ is a global parameter, $s_\ell>0$ is a spatial scale parameter, and $\bu_\ell=(u_{\ell,1},\dots,u_{\ell,n})^\top$ is a latent ICAR field defined with respect to the spatial adjacency $n \times n$ matrix $\boldsymbol{A}=(a_{ik})$. Let $\Ngh_{\svec}$ denote the set of neighbors of the grid cell of which $\svec$ is the center and define
\begin{equation}\label{eq:S_matrix}
\boldsymbol{S}=\operatorname{diag}(|\Ngh_{\svec_1}|,\dots,|\Ngh_{\svec_n}|)-\boldsymbol{A},\qquad
a_{ik}=
\begin{dcases}
1,& k\in\Ngh_{\svec_i},\\
0,& \text{otherwise}.
\end{dcases}
\end{equation}
The intrinsic GMRF prior for $\bu_\ell$ has improper density proportional to
$$
\pi(\bu_\ell)\propto \exp\!\left\{-\frac12\,\bu_\ell^\top\boldsymbol{S}_{\Rregion}\bu_\ell\right\},
$$
restricted to the identifiable subspace defined by a sum-to-zero constraint within each connected component of the adjacency graph. Note that no separate precision parameter appears in this expression, as the overall marginal variance of $\boldsymbol{\gamma}_{\ell}$ is controlled by the scale parameter $s_\ell$ of Eq.~\eqref{eq:gamma_decomp}. The quadratic form can be written as
$$
\bu_\ell^\top\boldsymbol{S}_{\Rregion}\bu_\ell
=
\frac{1}{2}\sum_{i=1}^n\sum_{k=1}^n a_{ik}\,(u_{\ell,i}-u_{\ell,k})^2,
$$
which makes explicit that the ICAR prior penalizes squared differences between neighboring grid cells. Under the decomposition above, the implied spatial precision is $1/s_\ell^2$, so that larger values of $s_\ell$ correspond to weaker spatial smoothing.
In the general formulation, the index $j$ may range over an arbitrary number of covariates $p$. However, in our application we restrict attention to an intercept and a single covariate effect ($j=0,1$), corresponding to the intercept and the yearly GMST anomaly. Applying this decomposition to each GEV coefficient yields, for all $\svec\in\Rregion$,
\begin{align*}
\mu_{j,\svec} &= \mu_{j,g} + s_{\mu_j}\,u_{\mu_j,\svec},
& \bu_{\mu_j} &\sim \mathrm{ICAR}(\boldsymbol{A}),
\qquad j=0,1,
\\
\log(\sigma_{0,\svec})& =\log(\sigma_{0,g})+s_{\sigma_0}\,u_{\sigma_0,\svec},
& \bu_{\sigma_0} &\sim \mathrm{ICAR}(\boldsymbol{A}),
\\
\sigma_{1,\svec} &= \sigma_{1,g} + s_{\sigma_1}\,u_{\sigma_1,\svec},
& \bu_{\sigma_1} &\sim \mathrm{ICAR}(\boldsymbol{A}),
\\
\xi_{\svec} &= \xi_g + s_{\xi}\,u_{\xi,\svec},
& \bu_{\xi} &\sim \mathrm{ICAR}(\boldsymbol{A}).
\end{align*}
We assign weakly informative priors to the global and spatial scale parameters,
\begin{align*}
\mu_{0,g} &\sim \Normal(m_0,s_0^2),
&
\mu_{1,g} &\sim \Normal(0,2^2),
\\
\log(\sigma_{0,g}) &\sim \Normal(\log s_0,1^2),
&
\sigma_{1,g} &\sim \Normal(0,1^2),
\\
\xi_g &\sim \Normal(0,0.1^2)\;\mathbbm{1}_{[-0.4,\,0.4]},
\end{align*}
where $m_0$ and $s_0$ denote the empirical mean and standard deviation, respectively, of the observed annual maxima and $\HalfNormal$ is the half normal distribution.  A
schematic summary of the hierarchical construction is provided in Supplement~\ref{sec:supp_GEV}.

Let $\Theta^{\mathrm{GEV}}_{\Rregion}$ denote the collection of all unknown parameters in the spatial GEV model in a region $\Rregion$, namely
\begin{equation}
\label{eq:gev_param_def}
\Theta^{\mathrm{GEV}}_{\Rregion}
=
\left\{
    \mu_{0,\svec}, \mu_{1,\svec}, 
    \sigma_{0,\svec}, \sigma_{1,\svec},
    \xi_{\svec} :
    \svec \in \Rregion
\right\}
\;\cup\;
\left\{
    s_{\psi}, \boldsymbol{u}_{\psi} :
    \psi \in \{\mu_0, \mu_1, \sigma_0, \sigma_1, \xi\}
\right\}.
\end{equation}
Assuming conditional independence of annual maxima across years given the model parameters, the posterior distribution is
\begin{equation}\label{eq:posterior_gev}
\pi(\Theta^{\mathrm{GEV}}_{\Rregion}\mid\{M_{\svec_i,t}\})
\propto
\prod_{i=1}^n\prod_{t=1}^T
f_{\mathrm{GEV}}\!\left(M_{\svec_i,t};
\mu_{\svec_i}(t),\sigma_{\svec_i}(t),\xi_{\svec_i}\right)
\,
\pi(\Theta^{\mathrm{GEV}}_{\Rregion}),
\end{equation}
where $f_{\mathrm{GEV}}(\,\cdot\,;\mu,\sigma,\xi)$ denotes the GEV density and $\pi(\Theta^{\mathrm{GEV}}_{\Rregion})$ is the joint prior implied by the hierarchical specification. Inference is performed using Markov chain Monte Carlo, yielding posterior samples for all marginal GEV parameters and spatial effects.

\subsection{Spatio-temporal modeling} \label{space-time model}
\subsubsection{Model construction} \label{Model construction}
We adopt a copula-based approach to construct processes with flexible extremal dependence structures. For a random vector $X = (X_1,\ldots,X_m)$ with marginals $ X_j \sim F_j $ that are continuous, the $m$-dimensional copula function $C_X $ is defined by
\begin{equation*}
    C_X(u_1,\dots,u_m) = \prob(F_1(X_1)\leq u_1,\dots,F_m(X_m)\leq u_m),
    \qquad u_1,\ldots,u_m \in [0, 1].
\end{equation*}
According to Sklar's theorem \parencite{sklar1959fonctions}, any multivariate joint distribution can be expressed in terms of a copula. Moreover, when the marginal distributions $F_j$, for $j = 1,\dots,m $, are continuous, the copula representation is unique.
In this work, we assume that covariates affect only the marginal distributions $F_j$, while the copula function remains invariant. In particular, climate forcing is allowed to modify the marginal behavior of temperatures through the BSQR and GEV components, but the spatio-temporal dependence structure is assumed to be stationary and independent of the covariates. 

Our objective is to construct a process $ Z(\svec,d)$ that exhibits the desired within-year extremal space-time dependence. For a fixed year $t$, we use the copula associated with $Z(\svec,d) $ to model the dependence structure of $(X(\svec,d,t))_{\svec,d}$. 
Consider the following equation:
\begin{equation}
    \label{eq:Zsd}
    Z(\svec,d) = R(\svec,d)^{\delta} \cdot W(\svec,d)^{1-\delta}, 
     \qquad \delta \in [0,1],
\end{equation}   
where $R(\svec,d)$ and $W(\svec,d)$ denote two independent processes that are asymptotically dependent and asymptotically independent, respectively, and that have unit-Pareto marginals: 
\[
    \prob(R(\svec,d)>r) = \prob(W(\svec,d)>r) = 1/r, 
    \qquad r\ge1. 
\]
The dependence characteristics of the process $Z(\svec,d)$ are determined by the asymptotic dependence or independence of the processes $R(\svec,d)$ and $W(\svec,d)$. If $\delta \leq 1/2$ then the tail of $Z(\svec,d)$ is driven by $W(\svec,d)$ and the process is asymptotically independent, while if $\delta > 1/2$, the tail is driven by $R(\svec,d)$ and the process is asymptotially dependent. The marginal distribution function $Q(z)=\prob(Z(\svec,d)\leq z)$ for $z \ge 1$ is
\begin{equation}
    Q(z) =
\begin{dcases} 
    1-[\delta/(2\delta-1)z^{-1/\delta}-(1-\delta)/(2\delta-1)z^{-1/(1-\delta)}], & \text{if } \delta \neq 1/2, \\
    1-z^{-2}[2\log(z)+1],  & \text{if } \delta = 1/2.
\end{dcases}
\label{eq:marginal_process}
\end{equation}
Additional technical specifications are available in \textcite{dell2024flexible,huser2019modeling}.

In our application, we construct $R(\svec,d)$ by transforming an underlying kind of finite Smith storm profile model $R^*(\svec,d)$ \parencite{smith1990statistical,smith1990max}; see Section~\ref{sec:Strom process} below. For each site $\svec$ and day $d=1,\dots,D$, we define 
\begin{equation}
\label{eq:R_process}
    R(\svec,d)=\frac{1}{1-F_{R^*}(R^*(\svec,d))},
\end{equation}
where $F_{R^*}$ is the marginal distribution function, namely $F_{R^*}(r)=\prob(R^*(\svec,d)\leq r)$.
Similarly, we model $W(\svec,d)$ by transforming an underlying Gaussian field; see Section~\ref{sec:Gaussian process} below. Specifically, let $W^*(\svec,d)$ denote an anisotropic Gaussian process with standard normal margins, and let $\Phi$ be the standard normal cumulative distribution function. For each site $\svec$ and day $d=1,\ldots,D$, we define
\begin{equation}
\label{eq:Wstar_to_W}
    W(\svec,d)
    =
    \frac{1}{1-\Phi \bigl(W^*(\svec,d)\bigr)}.
\end{equation}
These transformations are applied componentwise and map the Gaussian field $W^*(\svec,d)$ and the process $R^*(\svec,d)$ to a unit-Pareto process. 

\subsubsection{Gaussian process}
\label{sec:Gaussian process}
Let $\{W^*(\svec,d):\svec\in \Rregion,d\in \{1,\dots,D\}\}$ be a space-time Gaussian random field. The function $(\svec_1,\svec_2,d_1,d_2)\rightarrow C_{s,d}(\svec_1,\svec_2,d_1,d_2)$ defined on the product space \(\mathbb{R}^2\times \mathbb{R}^2\times\mathbb{R}\times\mathbb{R}\) is called the spatio-temporal covariance function. This function needs to be positive definite, meaning that for every finite set of real coefficients $a_i$ and points $(\svec_i,d_i)$ and as soon as not all coefficients $a_i$ are zero, we have the nonnegativity constraint
\begin{equation*}
    \sum_{i=1}^n\sum_{j=1}^na_ia_jC_{s,d}(\svec_i,\svec_j,d_i,d_j)>0.
\end{equation*}
Under the assumption of second-order stationarity we have without loss of generality that $\E(W^*(\svec,d))=0$  and
\begin{equation}
     \Cov(W^*(\svec_i,d_i),W^*(\svec_j,d_j))=C_{s,d}(\boldsymbol{h},l)
    \quad \text{where} \quad 
    (\boldsymbol{h},l)=(\svec_i-\svec_j,d_i-d_j),
\label{eq:stationarity condition}
\end{equation}
that is, the covariance depends only on the spatial and temporal separation. Finally, a stationary covariance function is isotropic if it is rotation and translation invariant, meaning that $C_{s,d}(\boldsymbol{h},l)=\Tilde{C}_{s,d}(\|\boldsymbol{h}\|,\abs{l})$ with $\|\cdot\|$ the usual Euclidean norm and $\Tilde{C}_{s,d}$ a positive definite function \parencite{porcu2007covariance}.

A covariance function is called separable if it can be written as the product of two covariance functions, that is, $C_{s,d}(\boldsymbol{h},l)=C_{s}(\boldsymbol{h})C_{d}(l)$ where $C_{s}(\cdot)$ denotes the spatial covariance and $C_{d}(\cdot)$ the temporal covariance. Separability is computationally convenient, especially for large spatio-temporal fields. Under the separability assumption we can define the variance–covariance matrix of the process $W^*(\svec,d)$ as $\Sigma=\Sigma_{s}\otimes\Sigma_{d}$, with $\Sigma_{s}$ and $\Sigma_{d}$ the spatial and temporal covariance matrices and $\otimes$ the Kronecker product. Below, we model the spatial and temporal covariance functions separately.

\paragraph{Spatial covariance function}

The assumption of isotropy is often made because it simplifies the spatial model and facilitates parameter estimation. However, this is a restrictive assumption and does not hold in many empirical applications \parencite{deng2008anisotropic}. 
In cases where the covariance of a second-order stationary process depends on the direction, the spatial dependence is referred to as anisotropic. Isotropy implies that correlation decays only with distance regardless of the direction, resulting in circular or spherical isolines on a variogram map. Anisotropy introduces directional variation, which may produce elongated, elliptical patterns of semi-variance. A specific and frequently encountered form, called geometric anisotropy, can often be managed by applying a suitable linear transformation to the spatial coordinates \parencite{schabenberger2017statistical}. 
To correct for geometric anisotropy, one changes the coordinate system with an appropriate linear transformation. If  $\svec$ is the spatial coordinate of a geometric anisotropic process $W^*(\svec,t)$, then $W^*(\Anisomat\svec,t)$ has an isotropic covariance function, where
\begin{equation}
    \Anisomat=\begin{pmatrix}
\cos(\omega) & -\sin(\omega) \\
\sin(\omega) & \phantom{-}\cos(\omega)
\end{pmatrix}\begin{pmatrix}
1 & 0 \\
0 & \eta
\end{pmatrix}.
\label{eq:Anisotropic matrix}
\end{equation}
Hence, geometric anisotropy is corrected by first rotating the coordinate system to align with the principal directions of the elliptical contours, which is governed by the parameter $\omega\,[\si{\radian}]\in[0,\pi]$, followed by scaling the axis to restore circular symmetry governed by the parameter $\eta\geq1$.
  
We consider a stationary but anisotropic spatial correlation function in dimension~2 of the form
\begin{equation}
\label{eq:Cs-general}
C_s(\svec_i,\svec_j;\theta)
=
C\!\left(
\frac{\|\Anisomat(\svec_i-\svec_j)\|}{\theta}
\right),
\end{equation}
where $C$ is a valid isotropic correlation function and $\theta\,[\si{\km}]>0$ is a spatial range parameter.
For our application, we specify $C$ as a Cauchy-type correlation function as in \textcite{dell2024flexible}, yielding
\begin{equation}
\label{eq:Cs2}
C_s(\svec_i,\svec_j;\theta)
=
\frac{1}{
1+\left(
\frac{\|\Anisomat(\svec_i-\svec_j)\|}{\theta}
\right)^2
}.
\end{equation}
With $\Anisomat$ defined as in Eq.~\eqref{eq:Anisotropic matrix}, the parameters $\omega\in[0,\pi]$ and $\eta\in[1,\infty[$ control the orientation and strength of the anisotropy, respectively, while $\theta>0$ determines the spatial range over which the correlation decays.

\paragraph{Temporal covariance function}

The temporal covariance function between time steps is specified using the Matérn class \parencite{matern1960spatial,rue2005gaussian}: for days $d_r$ and $d_q$, we put
\begin{equation}
\label{eq:temporal_matern}
C_d(d_r, d_q)
=
\frac{2^{1 - \nu}}{\Gamma(\nu)}
\left(
\frac{\sqrt{2\nu}\,|d_r - d_q|}{\kappa}
\right)^{\nu}
K_{\nu}\!\left(
\frac{\sqrt{2\nu}\,|d_r - d_q|}{\kappa}
\right)
+
\ind(d_r = d_q),
\end{equation}
where $\nu>0$ is a smoothness parameter, $\kappa\,[\si{\day}]>0$ is a temporal range parameter, $\Gamma(\cdot)$ denotes the Gamma function, and $K_{\nu}(\cdot)$ is the modified Bessel function of the second kind of order~$\nu$. The indicator term ensures that $C_t(d,d)=1$, so that the process has unit marginal variance. The Matérn smoothness parameter $\nu$ is difficult to estimate reliably from data, particularly in spatio-temporal models. In practice, it is common to set $\nu$ to a fixed value; this is what has been done by \textcite{healy2025inference} in the context of extreme spatial temperature events in Ireland. Recently, \textcite{cognot2025spatio} report temporal Matérn smoothness estimates consistently below one for daily temperatures over France, with values clustering around $0.6$ up to $0.9$. Following this empirical evidence and common practice in applied spatio-temporal modeling, we fix $\nu$ to $0.8$ in the data application rather than estimating it.

\subsubsection{Storm process} \label{sec:Strom process}
We construct the latent process $R^*(\svec,d)$ to represent episodic heatwave-like events with spatial extent and temporal persistence. 
Let $K \sim \text{Poisson}(\lambda)$ denote the number of heatwave episodes occurring during a given year. Conditionally on $K$, we let 
\[
\{A_k\}_{k=1}^{K} \overset{\text{iid}}{\sim} \mathrm{Pareto}(1), 
\qquad \mathbb P(A_k > x) = x^{-1}, \quad x \ge 1,
\]
where the variables $A_k$ represent independent event intensities. In addition to the amplitudes $A_k$, we sample for each episode $k$ a pair $(\boldsymbol{C}_k, T_k)$, where the location $\boldsymbol{C}_k$ is sampled uniformly from the set of grid cell centres $\{\svec_1,\dots,\svec_n\}$ and $T_k$ is sampled uniformly from $\{1,\dots,D\}$. 
The variable $\boldsymbol{C}_k$ represents the spatial centre of the heatwave episode, from which the event propagates spatially, while $T_k$ denotes the day at which the episode reaches its peak intensity. The latent process is then defined as
\begin{equation}
R^*(\svec,d)
=
\max_{k=1,\dots,K}
A_k
\exp\!\left(
-\frac{\|\Anisomat(\svec-\boldsymbol{C}_k)\|}{\rho_s}
-\frac{|d-T_k|}{\rho_t}
\right).
\end{equation}
Here $\Anisomat$ denotes the anisotropy matrix defined in Eq.~\eqref{eq:Anisotropic matrix}. The parameter $\rho_s\,[\si{\km}]>0$ controls the spatial decay of the episode away from its centre $\boldsymbol{C}_k$, while $\rho_t\,[\si{\day}]>0$ controls the temporal decay away from the peak day $T_k$. For a fixed spatial location $\svec$ and day $d$, let $F_{R^*(\svec,d)}(x)=\prob\{R^*(\svec,d)\le x\}$ denote the distribution function of the latent process. Using the Poisson construction of the episodic field and the Pareto$(1)$ distribution of the episode intensities, the distribution function admits the explicit form
\begin{equation}
\label{eq:F_Rstar}
F_{R^*(\svec,d)}(x)
=
\begin{cases}
0, & x<0,\\[2mm]
\exp(-\lambda), & x=0,\\[2mm]
\displaystyle 
\exp\!\left\{-\lambda\left(1-\frac{1}{nD}\sum_{i=1}^{n}\sum_{t=1}^{D}
\left(1-\frac{\exp\!\left(-\frac{\|\Anisomat(\svec-\svec_i)\|}{\rho_s}-\frac{|d-t|}{\rho_t}\right)}{x}\right)_+\right)\right\},
& x>0 ,
\end{cases}
\end{equation}
where $(u)_+=\max(u,0)$, $n$ denotes the number of spatial grid cells $\{\svec_1,\dots,\svec_n\}$, and $D$ the number of days in the season. The derivation of this expression is provided in the Supplement~\ref{sec:supp_storm_cdf}. Given this distribution function we can construct our process $R(\svec,d)$ as in Eq.~\eqref{eq:R_process}.

\subsubsection{Estimation} \label{Estimation}

For a given region $\Rregion$, we let the parameter that controls the space-time dependence be denoted by
\begin{equation}
\label{eq:depParams}
    \paramsDEP
    :=
    (\delta,\theta,\omega, \eta,\kappa,\lambda,\rho_s,\rho_t)^{\top}.
\end{equation}
The estimation of $\paramsDEP$ is complex: we were unable to establish the likelihood function, and even if it had been possible, the maximization problem would likely have been intractable. Instead, we used likelihood-free parameter estimation with the neural Bayes estimator \parencite{sainsbury2024likelihood}. This estimation framework has been used by \textcite{vihrs2022using,sainsbury2024likelihood,dell2024flexible,richards2024neural} to estimate parameters of spatial processes when conventional methods, such as maximum likelihood estimation or pseudo-likelihood techniques, are unfeasible. The concept is straightforward and involves treating the estimation issue as a prediction challenge.  The only prerequisite for the approach to work is the ability to simulate from the data-generating process to produce simulated data for model training and validation.

The method consists of establishing a prior distribution for the parameters and then conducting extensive random sampling over the parameter space. Following that, a dataset is created from the process for each parameter setting. A Neural Network (NN) is then trained to learn the map between the data to the parameters that have generated the data. Training the NN is a costly task; but, once completed, obtaining estimates from a dataset may be executed rapidly.  This is the reason why \textcite{zammit2024neural} designate it as an amortized estimator. In contrast to the method of \textcite{sainsbury2024likelihood}, we employed the approach proposed by \textcite{dell2024flexible,vihrs2022using}, so we did not input the data directly into the NN but rather utilized certain statistics. Selecting which statistics are appropriate instead of letting the NN learning them brings some subjectivity to the process. However, this can massively reduce the memory needed to train the neural network.

To train the NN to estimate the parameters of the underlying spatiotemporal process, we first transformed the value of $Z(\svec,d)$ to the uniform scale $U(\svec,d)$ by using Eq.~\eqref{eq:marginal_process}, and then we computed several summary statistics that characterize extremal dependence. These include spatial extremogram grids at multiple lags \parencite{dell2024flexible}, an auto-tail dependence function (ATDF) \parencite[Chapter~13]{reiss2007statistical}, distance-binned spatial extremograms \parencite{davis2009extremogram} summarized by quantile profiles and interquartile ranges (IQR) and a madogram map \parencite{cooley2006variograms}. In practice, the observed data consist of $T$ independent replicates of the spatio-temporal process $Z(\svec,d,t)$ $t\in\{1,\dots,T\}$, corresponding to distinct summer seasons (years).

All summary statistics defined in this section are computed on the copula scale $U(\svec,d,t)$. To train the neural Bayes estimator, we can work directly with $U(\svec,d,t)$ because the marginal distribution of the latent process $Z(\svec,d,t)$ is known. Because we simulate $Z(\svec,d,t)$ explicitly, the transformation in Eq.~\eqref{eq:marginal_process} is available in closed form.
In contrast, for the observed temperature anomalies $X(\svec,d,t)$ the marginal distribution varies with year through the covariates, and $U(\svec,d,t)$ must be constructed in a way that removes this temporal nonstationarity. We achieve this by mapping $X(\svec,d,t)$ to the copula scale using the estimated conditional quantile function from the BSQR model (Section~\ref{sec:BQR_spe}): for each $X(\svec,d,t)$ we define
\begin{equation}\label{eq:uniform_transformation}
    \hat U(\svec,d,t)
    =
    \hat q^{-1}\!\bigl(X(\svec,d,t)\mid \covariates(t),\svec\bigr),
\end{equation}
where $\hat q(\tau\mid \covariates(t),\svec)$ denotes the fitted conditional quantile function of Section~\ref{sec:BQR_spe}.
Equivalently, $\hat U(\svec,d,t)$ is the estimated conditional CDF value of $X(\svec,d,t)$ given $\covariates(t)$, so that the transformed field has approximately uniform margins and can be compared across years.
All dependence summaries (the $\chi$-grids, ATDF, extremogram quantile profiles, IQR curves, and the thresholded madogram map) are then computed by replacing $U(\svec,d,t)$ with $\hat U(\svec,d,t)$.

For the $\chi$-grids, we evaluate the pairwise tail dependence coefficient $\taildep{\svec_i}{\svec_k}{h;u}$ defined as:
\begin{equation}\label{eq:chi}
    \taildep{\svec_i}{\svec_k}{h;u}
    =
    \prob\bigl\{
    U(\svec_k,d+h,t)>u
    \;\big|\;
    U(\svec_i,d,t)>u
    \bigr\}.
\end{equation}
which measures the probability of an extreme event at location $\svec_k$ on day $d+h$ given an extreme event at location $\svec_i$ on day $d$. We evaluate this quantity empirically over probability thresholds $u \in \{0.90, 0.95, 0.99\}$ with the following estimator. Define the binary indicator of exceedance:
$$
I_{\svec,u}(t,d) = \ind\{U(\svec,d,t) > u\}.
$$
Then, for locations $\svec_i$ and $\svec_k$ and lag $h \geq 0$. we estimate the tail dependence coefficient via:
\begin{equation}\label{eq:chi_est}
\hat{\chi}_{\svec_i,\svec_k}(h;u)
=
\frac{
\displaystyle \sum_{t=1}^T \sum_{d=1}^{D-h}
I_{\svec_i,u}(t,d)\, I_{\svec_k,u}(t,d+h)
}{
\displaystyle \sum_{t=1}^T \sum_{d=1}^{D-h}
I_{\svec_i,u}(t,d)
}.
\end{equation}
To reduce dimensionality, we aggregate pairwise estimates according to the
Euclidean distance between locations.
Let $\|\svec_i-\svec_k\|$ denote the distance between locations $\svec_i$ and
$\svec_k$, and define distance bins
$$
B_l = [r_l, r_l+100\text{ km}), \qquad l=1,\dots,L,
$$
where $r_l = 100(l-1)$ km.
For a given lag $h$ and threshold $u$, the bin-averaged tail dependence
coefficient is defined as
\begin{equation}
\label{eq:chi_bin}
\bar{\chi}(r_l,h;u)
=
\frac{1}{\lvert B_l \rvert }
\sum_{\substack{i<k:\\ \|\svec_i-\svec_k\|\in B_m}}
\hat{\chi}_{\svec_i,\svec_k}(h;u),
\end{equation}
where $\lvert B_l \rvert$ denotes the number of location pairs whose distance falls in bin
$B_l$. Repeating this process for lags $h = 0, \dots, 10$ yields a two-dimensional matrix ($\chi$-grid) representing extremal dependence across space and time. For numerical stability, we apply a logarithmic transformation to the averaged estimates, computing $\log\left(\bar{\chi}(r_l,h;u)+\varepsilon\right)$, with $\varepsilon = 10^{-6}$. The use of such grids has also been proposed by \textcite{dell2024flexible} in a similar setting. 
An illustration of the estimated $\chi$-grids are visualized in Supplement~\ref{sec:chi grid} Figure~\ref{fig:tail_dependence_perspective}.

To provide more information about the temporal persistence of extremes, we compute the
auto-tail dependence function (ATDF) for lags $h = 0,\dots,10$.
For each spatial location $\svec$, the ATDF at lag $h$ is estimated as
\begin{equation}
\label{eq:ATDF}
\widehat{\mathrm{ATDF}}_{\svec}(h)
=
\frac{
\displaystyle \sum_{t=1}^T \sum_{d=1}^{D-h}
I_{\svec,u}(t,d)\, I_{\svec,u}(t,d+h)
}{
\displaystyle \sum_{t=1}^T \sum_{d=1}^{D-h}
I_{\svec,u}(t,d)
},
\end{equation}
which estimates the conditional probability
$\prob\{U(\svec,d+h,t)>u \mid U(\svec,d,t)>u\}$. The final ATDF curve is obtained by averaging across all spatial locations $\svec$ in the region. We have
\begin{equation}\label{eq:atdf_est}
    \widehat{\mathrm{ATDF}}_{\Rregion}(h)=\frac{1}{n_{\Rregion}}\sum_{i=1}^{n_\Rregion}\widehat{\mathrm{ATDF}}_{\svec_i}(h),
\end{equation}
where $n_\Rregion$ denotes the number of grid cells in region $\Rregion$. For numerical stability we apply a logarithmic transformation, computing $\log\left(\widehat{\mathrm{ATDF}}_{\Rregion}(h)+\varepsilon\right)$.

We further characterize the spatial heterogeneity of extremal dependence by
computing a quantile profile of the spatial extremogram at lag $h=0$.
For a fixed exceedance threshold $u=0.95$, we first estimate the pairwise tail
dependence coefficients $\hat{\chi}_{\svec_i,\svec_k}(0;u)$ using Eq.~\eqref{eq:chi_est} for all pairs of locations $(\svec_i,\svec_k)$ in the
region $\Rregion$. These estimates are then grouped into distance bins of fixed width
(\SI{100}{\kilo\metre}) according to the spatial separation $\|\svec_i-\svec_k\|$. Let $B$ denote the total number of distance bins, and define
$$
\mathcal{C}_b
=
\Bigl\{
\hat{\chi}_{\svec_i,\svec_k}(0;u)
:
\|\svec_i-\svec_k\| \in \text{bin } b
\Bigr\},
\qquad b=1,\dots,B.
$$
Within each bin $b$, we compute the empirical $\alpha$-quantiles of
$\mathcal{C}_b$, namely
\begin{equation}
\label{eq:bin-quantiles}
\hat{q}_{\alpha}(b)
=
\inf\Bigl\{
x :
\frac{1}{|\mathcal{C}_b|}
\sum_{c\in\mathcal{C}_b} \ind(c \le x)
\ge \alpha
\Bigr\},
\qquad
\alpha \in \{0.1,\,0.5,\,0.9\}.
\end{equation}
This yields a $3\times B$ summary describing the median decay of extremal
dependence with distance together with its lower and upper variability envelopes.
For numerical stability, we apply a logarithmic
transformation to the bin-wise quantiles,
\[
\log \bigl(\hat{q}_{\alpha}(b) + \varepsilon\bigr).
\]
As a compact summary of dispersion, we additionally report the interquartile
range (IQR),
\begin{equation}
\label{eq:IQR}
\widehat{\mathrm{IQR}}(b) = \hat{q}_{0.75}(b) - \hat{q}_{0.25}(b),
\end{equation}
which quantifies the variability of extremal dependence across spatial pairs at comparable separations. For numerical stability, a small constant $\varepsilon$ is
added prior to applying a logarithmic transformation to all binwise summaries $\log\left(\widehat{\mathrm{IQR}}(b)+\varepsilon\right)$.

To characterize directional features and spatial anisotropy in the dependence structure of extremes, we compute a thresholded madogram map based on the field $U(\svec,d)$. Let $u=0.95$ denote a high probability threshold so that only high values contribute to the dependence measure. For any pair of spatial locations $\svec_i=(s_{i,x},s_{i,y})$ and $\svec_k=(\svec_{k,x},\svec_{k,y})$, consider the displacement vector $\boldsymbol{h}=\svec_k-\svec_i=(\svec_{k,x}-\svec_{i,x},\svec_{k,y}-\svec_{i,y})=(h_x,h_y)$.
For a spatial lag vector $\hvec=(h_x,h_y)$, the empirical extremal thresholded madogram is
defined as
\begin{multline}
\label{eq:extremal_madogram}
\hat{\nu}(\hvec) 
=
\frac{1}{TD}
\sum_{t=1}^{T}
\sum_{d=1}^{D}
\Biggl\{ \frac{1}{2\,N(\hvec,d,t)} \\
\sum_{i,k: \svec_k-\svec_i \in \mathcal B(\hvec)}
\bigl|U(\svec_k,d,t)-U(\svec_i,d,t)\bigr|
\;\ind\!\bigl\{U(\svec_k,d,t)\geq u ,U(\svec_i,d,t) \geq u \bigr\} \Biggr\},
\end{multline}
where $\mathcal B(\hvec)$ denotes a bin of displacement vectors centered at $\hvec$, and $N(\hvec,d,t)$ is the number of spatial pairs such that $\svec_k-\svec_i \in \mathcal{B}(\hvec)$ and $U(\svec_k,d,t)\geq u$ and $U(\svec_i,d,t) \geq u$. The lag-vector bins $\mathcal B(\hvec)$ are defined on a regular two-dimensional grid in $(h_x,h_y)$ with bin width \SI{100}{\kilo\metre} in both directions and maximum extent \SI{1000}{\kilo\metre}. A pair $(\svec_i,\svec_k)$ contributes to
the bin centered at $\hvec$ if both components of $\svec_k-\svec_i$ fall within $\pm50$~km of $(h_x,h_y)$. This construction is common in geostatistics \parencite[Chapter~2]{chiles2012geostatistics} and know as the variogram map; we only adjusted it to a thresholded madogram to be coherent with EVT. The resulting madogram map summarizes the average absolute difference between
high values of $U$ at pairs of locations separated by a given spatial
displacement. Isotropic dependence would produce approximately symmetric
patterns in this map, while elongated or directional features indicate spatial anisotropy in the structure of extremes. For stability purposes, the madogram values are log-transformed $\log\left(\hat{\nu}(\hvec)+ \varepsilon\right)$.

\subsection{Return periods and uncertainty quantification} \label{sec:Return periods}
In extreme value analysis, the return period $\period$ and the return level $\zlevel$ are key metrics for quantifying the rarity of extreme events. The return period $\period$ represents the average time interval between events whose magnitude is at least $\zlevel$. Mathematically, if $M$ denotes the annual maximum of the variable of interest (e.g., temperature anomalies), the return period associated with a level $\zlevel$ is  
$T(\zlevel) = 1/\prob(M > \zlevel).$ 
The return level $\zlevel$ is defined as the magnitude that is expected to be exceeded on average once every $\period$ years, i.e., it satisfies  
$\prob(M > \zlevel) = 1/\period.$ 
In the simple univariate case, one can fit a GEV distribution to the annual maxima. The estimated return level associated with a return period of $\period$ years is given by $\zlevel = G^{-1}(1 - 1/\period)$ with $G(\cdot)$ defined as Eq.~\eqref{eq:GEV_CDF}.
However, when the analytical form of $G$ is not available, for instance, when using complex stochastic models, the relationship between $\period$ and $\zlevel$ can be obtained through numerical approaches such as Monte Carlo simulation.

To return simulated values from the latent process $Z(\svec, d,t)$ to the original physical scale for a given year $t$, we apply a marginal back-transformation based on the nonstationary fitted GEV parameters of annual maxima at each location $\svec$. Full derivations of this transformation are provided in Supplement~\ref{sec:back_transform_deriv}. Specifically, we first transform the realizations of $Z(\svec, d,t)$ to a uniform scale using Eq.~\eqref{eq:marginal_process}, and then back-transform the uniform variates $U(\svec,d,t)$ to the original scale using
\begin{equation}
X(\svec,d,t) =
\begin{cases} 
q(U(\svec,d,t) \mid \covariates(t),\svec), &
U(\svec,d,t) <\tau_{\svec,t}, \\[1mm]
\Tthres + \dfrac{\tilde{\sigma}_{\svec,t}}{\xi_{\svec}}
\left[
\left(\dfrac{1-U(\svec,d,t)}{1-\tau_{\svec,t}}\right)^{-\xi_{\svec}} - 1
\right], 
& U(\svec,d,t)\ge\tau_{\svec,t},\ \xi_{\svec}\neq 0, \\[1mm]
\Tthres - \tilde{\sigma}_{\svec,t}
\log\!\left(\dfrac{1-U(\svec,d,t)}{1-\tau_{\svec,t}}\right),
& U(\svec,d,t)\ge\tau_{\svec,t},\ \xi_{\svec}=0.
\end{cases}
\label{eq:backtransform_full}
\end{equation}
Here, $q(\cdot \mid \covariates(t),\svec)$ is the BSQR model described in Section~\ref{sec:BQR_spe} and $\tau_{\svec,t}=q^{-1}(\Tthres \mid \covariates(t),\svec)$ is obtained by inverting it. The quantity
$\tilde{\sigma}_{\svec,t}
= \sigma_{\svec}(t)+\xi_{\svec}\{\Tthres-\mu_{\svec}(t)\}$ involves the
nonstationary GEV parameters $\mu_{\svec}(t)$, $\sigma_{\svec}(t)$, and
$\xi_{\svec}$ defined in Eq.~\eqref{eq:max_model}, while the threshold $\Tthres$ is given in Eq.~\eqref{eq:threshold}.

Now that we can simulate from our random field, we can generate independent replicates and use a Monte Carlo technique to approximate return periods. For each replicate $j$, we are interested in a certain property $\prop$ that summarizes aspects of the simulated field over the year e.g., the duration of an event, the frequency of occurrence, or the intensity. Given a property $\prop$, we fix a return level $\zlevel^{(\prop)}$ and approximate the corresponding return period empirically from the simulated data. Specifically, we generate a large number $J$ of replicates. Let $\propval_j$ denote the value of property $\prop$ in replicate $j$. The empirical approximation of the return period associated with the threshold $\zlevel^{(\prop)}$ is then given by
\begin{equation}
    \label{eq:RP_estimates}
    \hat{T}_J^{(\prop)}(\zlevel^{(\prop)})
    \;=\;
        \frac{J}{\sum_{j=1}^J \ind{ \{ \propval_j > \zlevel^{(\prop)} } \} }.
\end{equation}
The relative error of our Monte Carlo approximation is given by $1/\sqrt{J p}$, where $p$ is the probability of exceedance that we are trying to approximate.

To quantify the uncertainty in the return period $\hat{T}_J^{(\prop)}(\zlevel^{(\prop)})$ that arises from the estimation procedure, we propose to use the following approach. 
For a given region $\Rregion$ let the parameter set be denoted by
$
\Theta_{\Rregion} = \left(\paramsBSQR,\paramsGEV,\paramsDEP\right)^{\top},
$
where $\paramsBSQR$ are the parameters of the BSQR as defined in Eq.~\eqref{eq:quantile_reg_param_def} in Supplement~\ref{sec:supp_BSQR}, $\paramsGEV$ are the nonstationary GEV parameters as in Eq.~\eqref{eq:gev_param_def}, and $\paramsDEP$ are the space-time dependence parameters of the process $Z(\svec,t)$ as defined in Eq.~\eqref{eq:depParams}. As a full bootstrap strategy is computationally infeasible, we adopted a twofold resampling strategy: (i) a parametric bootstrap for the space-time dependence parameters as suggested by \textcite{zammit2024neural,dell2024flexible,sainsbury2025neural}, (ii) a posterior sampling for the marginal GEV parameters from the posterior distribution as described in Section~\ref{sec:GEV}, Eq.~\eqref{eq:posterior_gev}, and a posterior sampling for the BSQR coefficient as described in Section~\ref{sec:BQR_spe}, Eq.~\eqref{eq:posterior_bsqr_short}.
From these procedures, we can generate $B$ triples of parameters
\begin{equation}
    \Theta_{\Rregion}^{(b)} = (\paramsBSQRboot,\paramsGEVboot, \paramsDEPboot)^{\top}, 
\qquad b = 1, \dots, B.
\label{eq:param_boot}
\end{equation}
For each parameter replicate $b$, we simulate $J'$ data replicates ($J'$ possibly smaller than $J$) from the model; we then approximate the return period $\hat{T}_{N,b}^{(P)}(\zlevel^{(\prop)})$ via the previous Monte Carlo method. 
The empirical distribution of these estimates is then used to assess uncertainty, with pointwise $1-\alpha$ confidence intervals derived from the empirical quantiles.

\section{Data application}\label{sec:Data}
\subsection{Data acquisition and preprocessing}
The data application is based on climate model simulations from the Coupled Model
Intercomparison Project Phase~6 (CMIP6). As described in Section~2, we focus on
simulations produced by the Meteorological Research Institute Earth System Model
version~2 (MRI--ESM2).

For both the factual and counterfactual climates, we use daily maximum
near-surface air temperature (TASMAX). In all experiments, we rely on the
\texttt{r1i1p1f1} ensemble member to ensure consistency across scenarios. The
factual climate is represented by the historical experiment extended with the
SSP2--4.5 scenario beyond 2014, while the counterfactual climate is represented
by the historical natural-forcing-only (\texttt{hist-nat}) experiment, as detailed
in Section~2.1. All TASMAX fields are processed according to the methodology
described in Section~2.

\begin{figure}[t]
  \centering
  \includegraphics[width=0.85\textwidth]{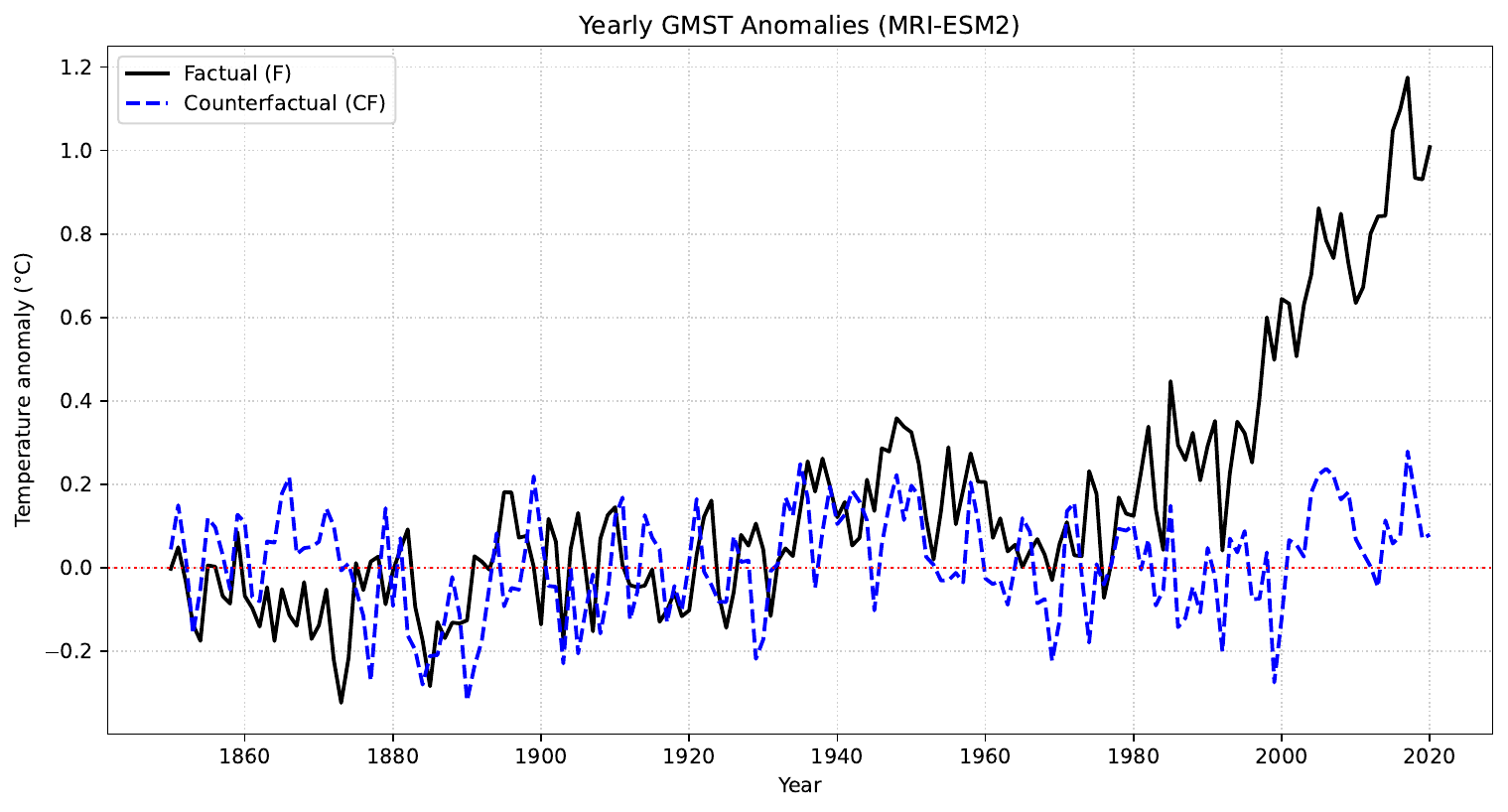}
  \caption{
    Yearly GMST anomalies from MRI--ESM2,
    computed relative to the 1850--1949 baseline. The factual series includes
    anthropogenic and natural forcings, while the counterfactual series
    includes natural forcings only. Both series are based on the
    \texttt{r1i1p1f1} ensemble member.
  }
  \label{fig:gmst}
\end{figure}

To capture large-scale climatic forcing, we use GMST anomalies as a covariate in the marginal models. GMST is
computed from monthly near-surface air temperature (TAS) fields over all 12 months of each year, provided by
MRI--ESM2, again using the \texttt{r1i1p1f1} ensemble member for both factual and
counterfactual simulations. Monthly TAS fields are aggregated to a global mean.
The resulting global mean series are then averaged to yearly means. Yearly GMST anomalies are computed relative to the same early-industrial baseline
period (1850--1949) used for the TASMAX anomalies, separately for the factual and
counterfactual climates. For the factual climate, the GMST series is truncated in
2020 to match the temporal availability of the counterfactual simulation.
The resulting GMST anomaly series serves as the sole covariate in the BSQR model and the Bayesian spatial GEV model introduced in Sections~\ref{sec:BQR_spe} and~\ref{sec:GEV}
, where it drives
nonstationary changes in the marginal distribution of daily temperature
anomalies. The temporal evolution of this covariate in the factual and
counterfactual climates is shown in Figure~\ref{fig:gmst}.

\begin{figure}[t]
  \centering
  \includegraphics[scale=0.5]{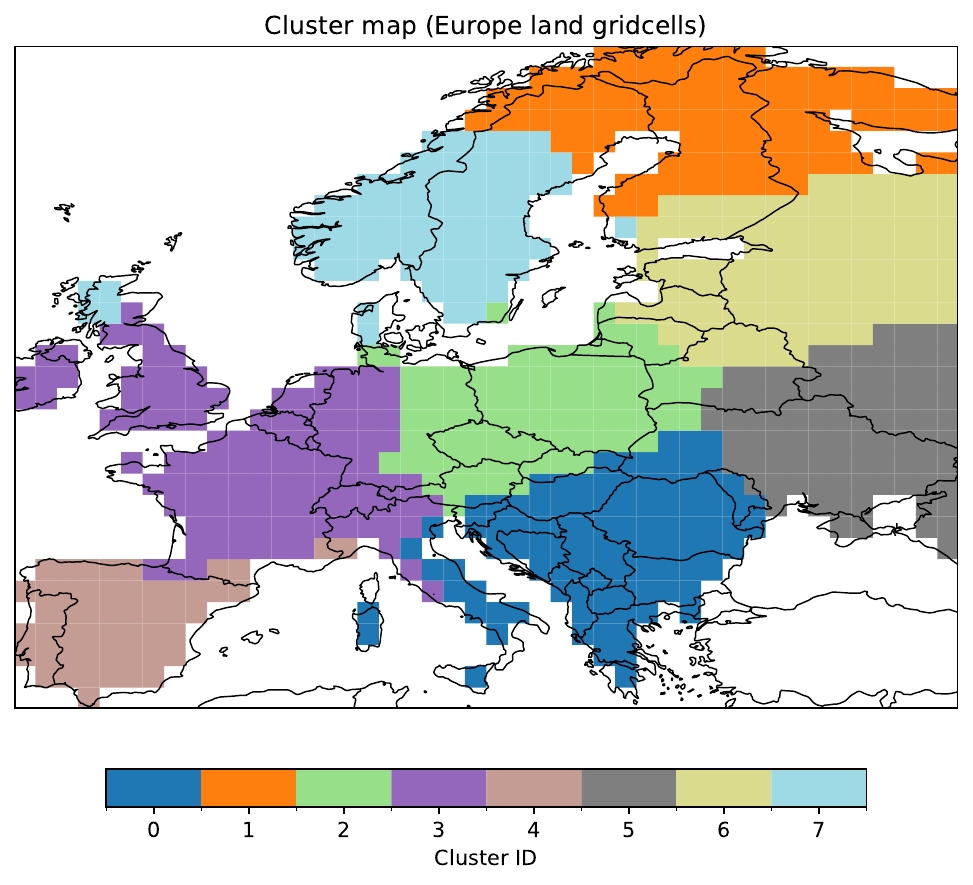}
  \caption{
    Spatial clustering of Europe based on extremal dependence of summer (JJA)
    temperature anomalies in the counterfactual climate. The partition was
    obtained using a $k$-medoids algorithm with the chi-based dissimilarity in Eq.~\eqref{eq:dissimilarity}
    computed at threshold $u=0.8$, yielding $k=8$ clusters.
  }
  \label{fig:eu_clim_clusters}
\end{figure}

Europe spans a wide range of climate regimes,
and fitting a single spatio-temporal model over the entire domain is unlikely to
be realistic. In particular, both the marginal behavior and the dependence
structure of extreme temperatures can vary substantially across regions (e.g.,
Mediterranean versus Scandinavian climates). This motivates a regionalization step prior to model fitting, presented in Supplement~\ref{sec:Clustering}. We focus on Cluster~3, as shown in Figure~\ref{fig:eu_clim_clusters}.

\subsection{BSQR application}\label{sec:BSQR_app}

We apply the BSQR methodology described in
Section~\ref{sec:BQR_spe} to estimate spatially smooth conditional quantile
functions of daily summer temperature anomalies. In the BSQR model, the
conditional quantile function at level $\tau\in(0,1)$ is expressed as a
Bernstein-polynomial expansion; in the application we set the number of basis
functions to $M=15$ following \textcite{reich2011bayesian}. The covariate entering the model is yearly GMST anomalies.
To match the BSQR formulation, we rescale this covariate to the unit interval by
a min--max transformation. Specifically, letting $g_t$ denote the yearly GMST
anomaly at year $t$, we define
\[
\tilde g_t=\frac{g_t-g_{\min}}{g_{\max}-g_{\min}},
\qquad
g_{\min}=\min\{g_t^{\text{F}},g_t^{\text{CF}}\},\quad t\in \{1,\dots,T\},\quad g_{\max}=4\si{\degreeCelsius},
\]
where $g_t$, $g_{\min}$ are expressed in \si{\degreeCelsius}. We choose $g_{\max}=4  \si{\degreeCelsius}$ as this is the maximum value we allow for extrapolation, so that $\tilde g_t\in[0,1]$ for the range of values encountered in the present application.

A key role of the fitted BSQR model is to remove temporal nonstationarity in the
margins prior to dependence modeling. Using the estimated conditional quantile
function $\hat q_{\svec,t}(\tau)$, we transform daily anomalies in both the
factual and counterfactual worlds to the copula scale via the estimated
conditional distribution function,
\[
\widehat{U}(\svec,d,t)
=
\hat F_{\svec,t}\bigl(X(\svec,d,t)\bigr)
=
\hat q_{\svec,t}^{-1}\bigl(X(\svec,d,t)\bigr).
\]
A diagnostic check of this uniformization step for Cluster~3 is provided
Supplement~\ref{sec:bsqr-uniform-diagnostic} Figure~\ref{fig:uniform_check_cluster3}. The empirical distributions are close to the uniform
distribution on $[0,1]$, supporting the adequacy of the marginal standardization.
Deviations are nevertheless visible in the lower and upper tails, indicating that
the BSQR model captures the central part of the conditional distribution more
accurately than the extremes.

\begin{figure}[!htbp]
  \centering
  \captionsetup[sub]{font=footnotesize, skip=1pt}

  % % -------- Row 1 --------
  % \begin{subfigure}[t]{0.37\textwidth}
  %   \centering
  %   \includegraphics[width=\textwidth]{figures/cluster3_T95_CF_threshold.pdf}
  %   \caption{Empirical counterfactual threshold $\TthresEmp$ (95th percentile of CF anomalies).}
  % \end{subfigure}\hfill
  % \begin{subfigure}[t]{0.38\textwidth}
  %   \centering
  %   \includegraphics[width=\textwidth]{figures/cluster3_tau_CF_GMST0.pdf}
  %   \caption{Probability level $\tau_{\svec}$ of $\TthresEmp$ under the counterfactual BSQR fit at GMST $=0^\circ$C.}
  % \end{subfigure}

  % \vspace{-0.1em}

  % -------- Row 2 --------
  \begin{subfigure}[t]{0.38\textwidth}
    \centering
    \includegraphics[width=\textwidth]{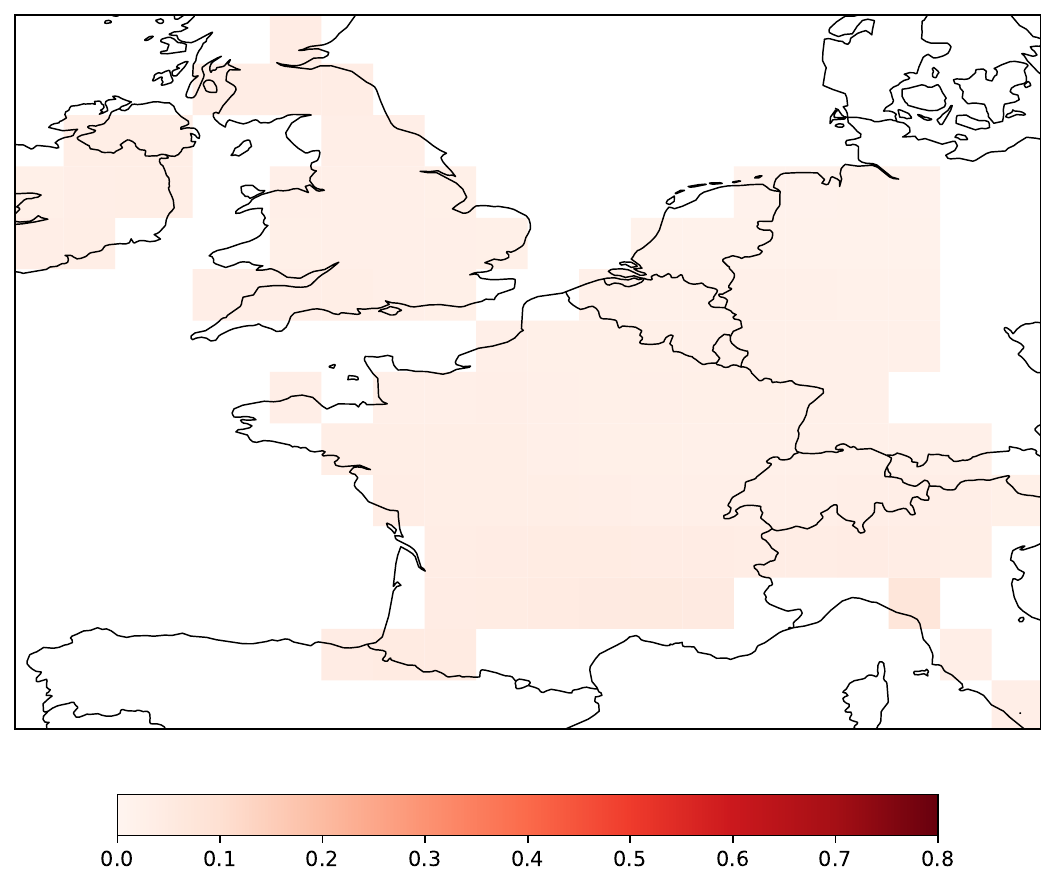}
    \caption{
    Exceedance probability change for GMST $=1^\circ$C
    }
  \end{subfigure}\hfill
  \begin{subfigure}[t]{0.38\textwidth}
    \centering
    \includegraphics[width=\textwidth]{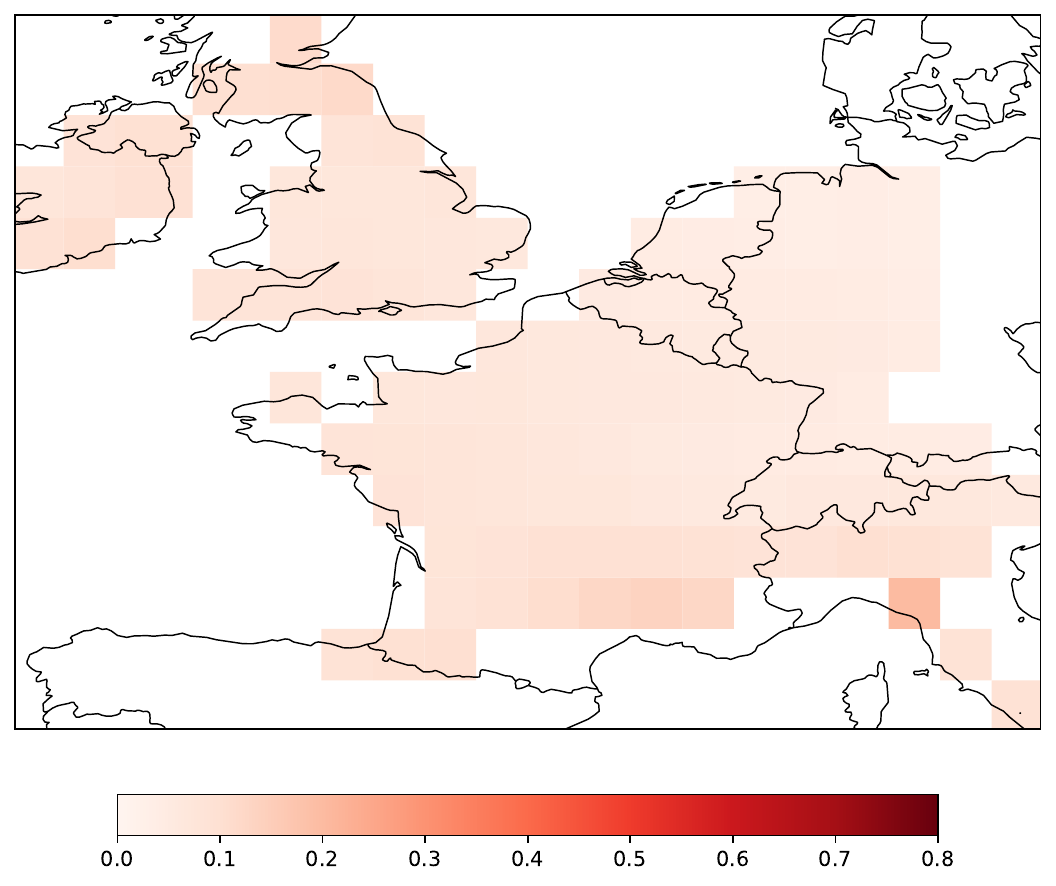}
    \caption{Same as (a), but for GMST $=2^\circ$C.}
  \end{subfigure}

  \vspace{-0.1em}

  % -------- Row 3 --------
  \begin{subfigure}[t]{0.38\textwidth}
    \centering
    \includegraphics[width=\textwidth]{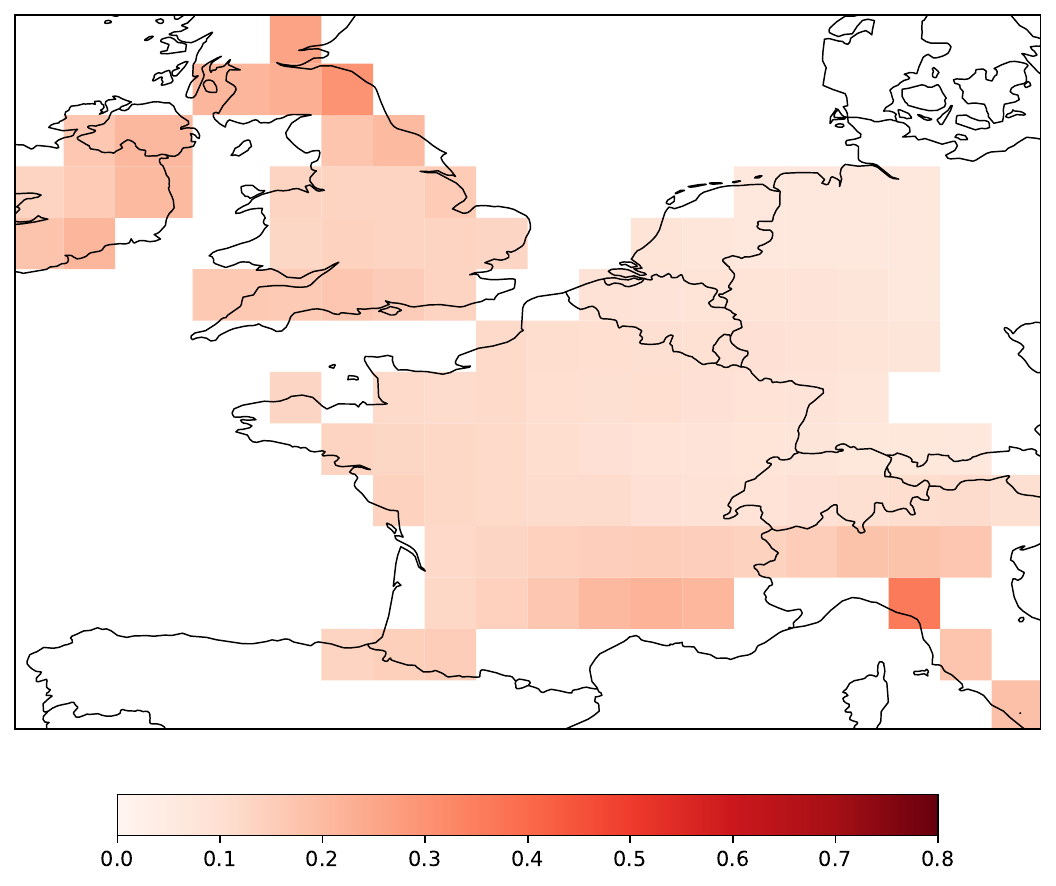}
    \caption{Same as (a), but for GMST $=3^\circ$C.}
  \end{subfigure}\hfill
  \begin{subfigure}[t]{0.38\textwidth}
    \centering
    \includegraphics[width=\textwidth]{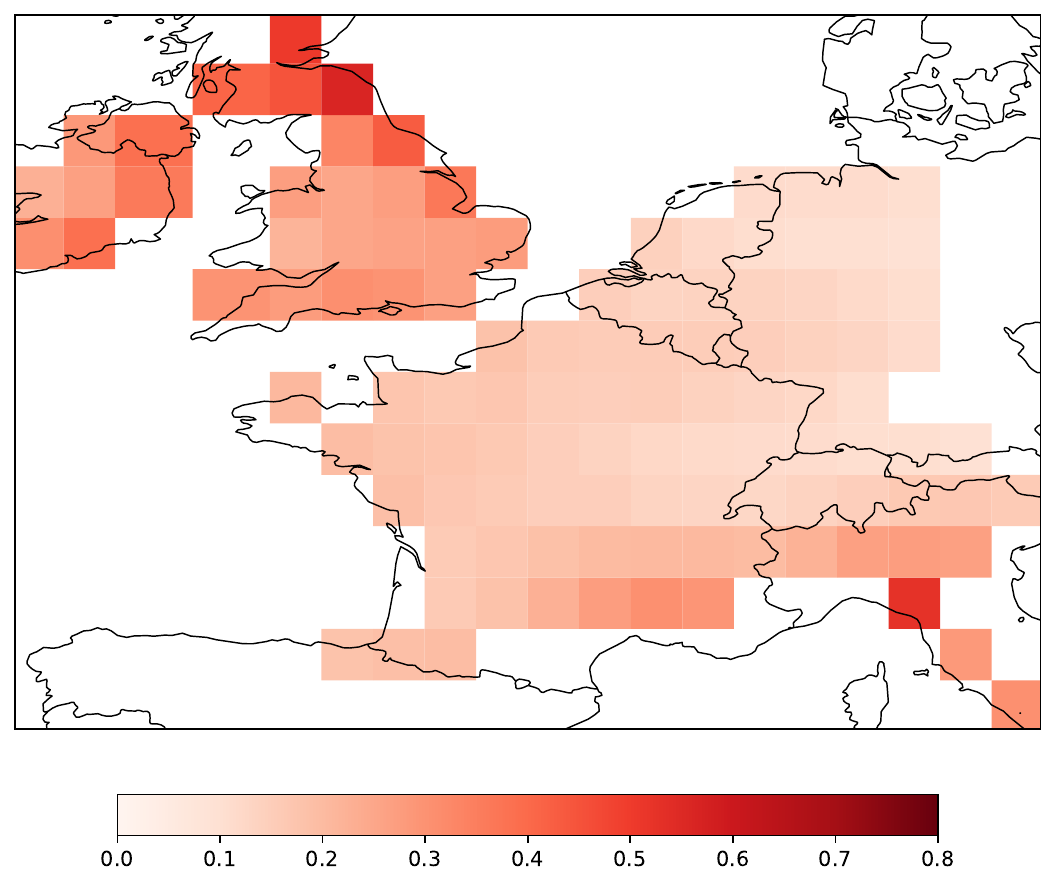}
    \caption{Same as (a), but for GMST $=4^\circ$C.}
  \end{subfigure}
  \caption{
    (a)--(d) Change in exceedance probability conditional on GMST anomalies.
  }
  \label{fig:threshold_map}
\end{figure}

Another key contribution of the BSQR model is its ability to quantify how a fixed counterfactual heatwave threshold shifts in the factual distribution of the temperature distribution under global warming. Specifically, we consider the counterfactual threshold $\TthresEmp$ defined in Eq.~\eqref{eq:threshold_empirical} and examine how its associated probability level
\[
\tau_{\svec,t}
=
q^{-1}\!\left(\Tthres \mid \covariates(t), \svec \right)
\]
evolves as a function of GMST. This provides a characterization of how temperatures that were extreme in the counterfactual climate are repositioned within the temperature distribution at each grid cell under factual warming. The results are presented in Figure~\ref{fig:threshold_map}. Panels (a)--(d) present the exceedance probability change
\[
  \bigl[1-q^{-1}_{\mathrm{F}}(\TthresEmp \mid \mathrm{GMST}=g,\svec)\bigr]
  -
  \bigl[1-q^{-1}_{\mathrm{CF}}(\TthresEmp \mid \mathrm{GMST}=0,\svec)\bigr]
\] 
conditional on GMST anomalies $g$ of $\SI{1}{\celsius}$ up to $\SI{4}{\celsius}$.

\subsection{GEV application}\label{sec:GEV_app}

We apply the spatially varying nonstationary GEV model in
Section~\ref{sec:GEV} to annual summer maxima of daily temperature anomalies
over Cluster~3. Annual maxima are extracted independently for each grid cell
and year, and the GEV parameters are modeled as functions of the rescaled GMST
covariate, with spatial structure introduced through intrinsic Gaussian Markov
random field priors. The model is fitted separately for the counterfactual
and factual climates using Bayesian inference.

\begin{figure}[!htbp]
  \centering
  \captionsetup[sub]{font=footnotesize, skip=1pt}

  % -------- Row 1: mu --------
  \begin{subfigure}[t]{0.48\textwidth}
    \centering
    \includegraphics[width=\textwidth]{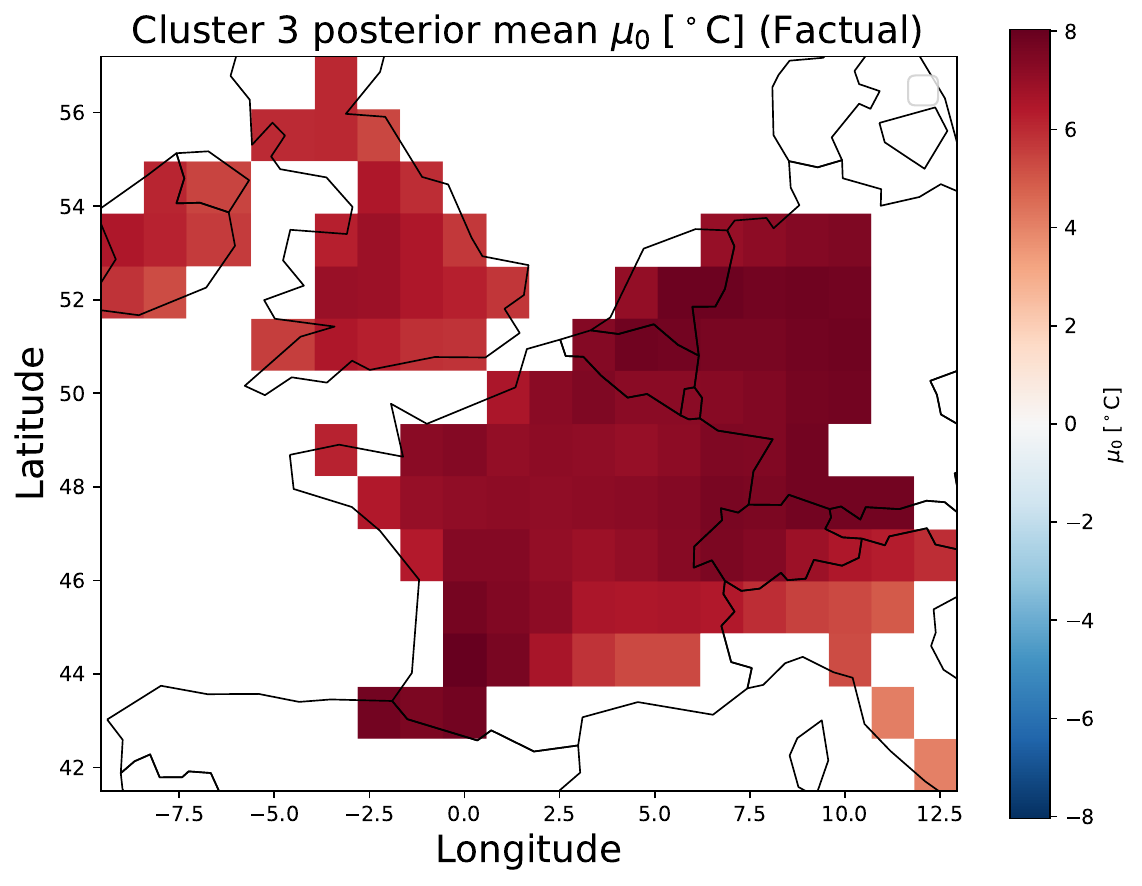}
    \caption{$\mu_0\,[\si{\degreeCelsius}]$}
  \end{subfigure}\hfill
  \begin{subfigure}[t]{0.48\textwidth}
    \centering
    \includegraphics[width=\textwidth]{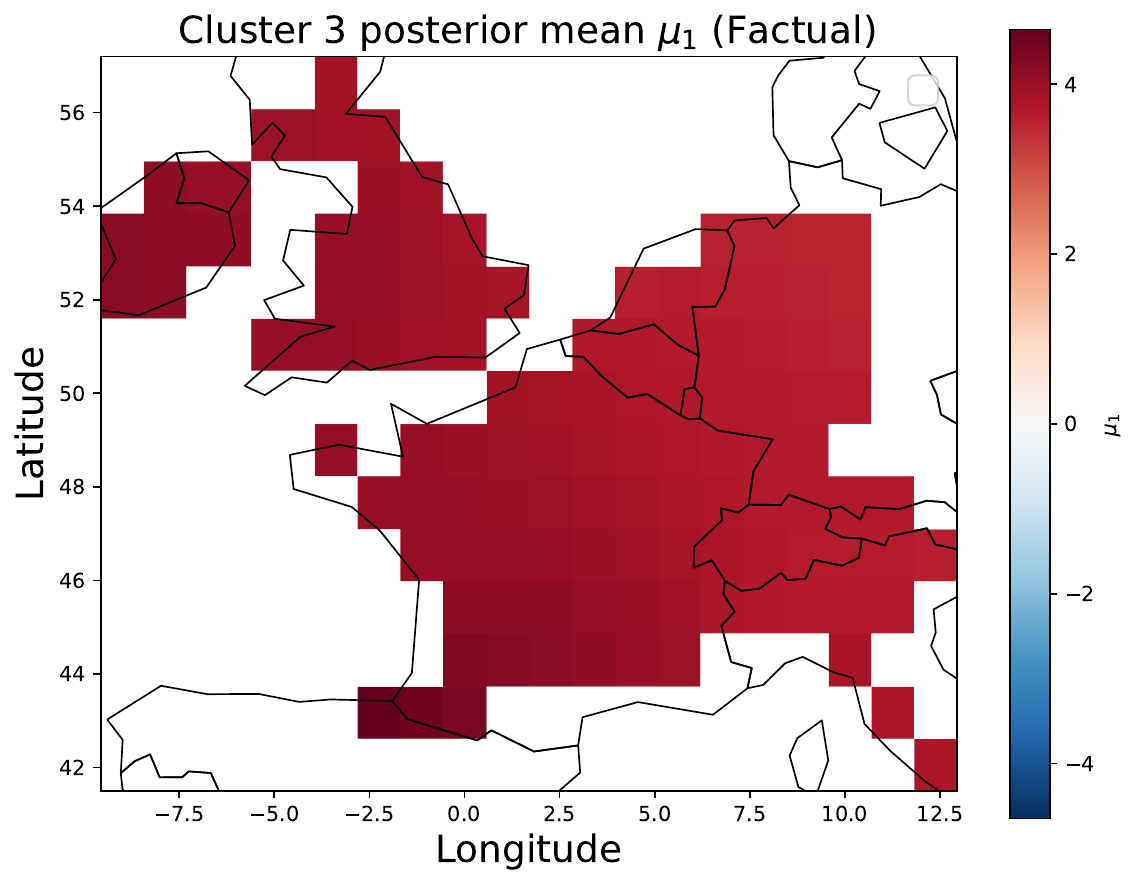}
    \caption{$\mu_1$}
  \end{subfigure}

  \vspace{-0.2em}

  % -------- Row 2: sigma --------
  \begin{subfigure}[t]{0.48\textwidth}
    \centering
    \includegraphics[width=\textwidth]{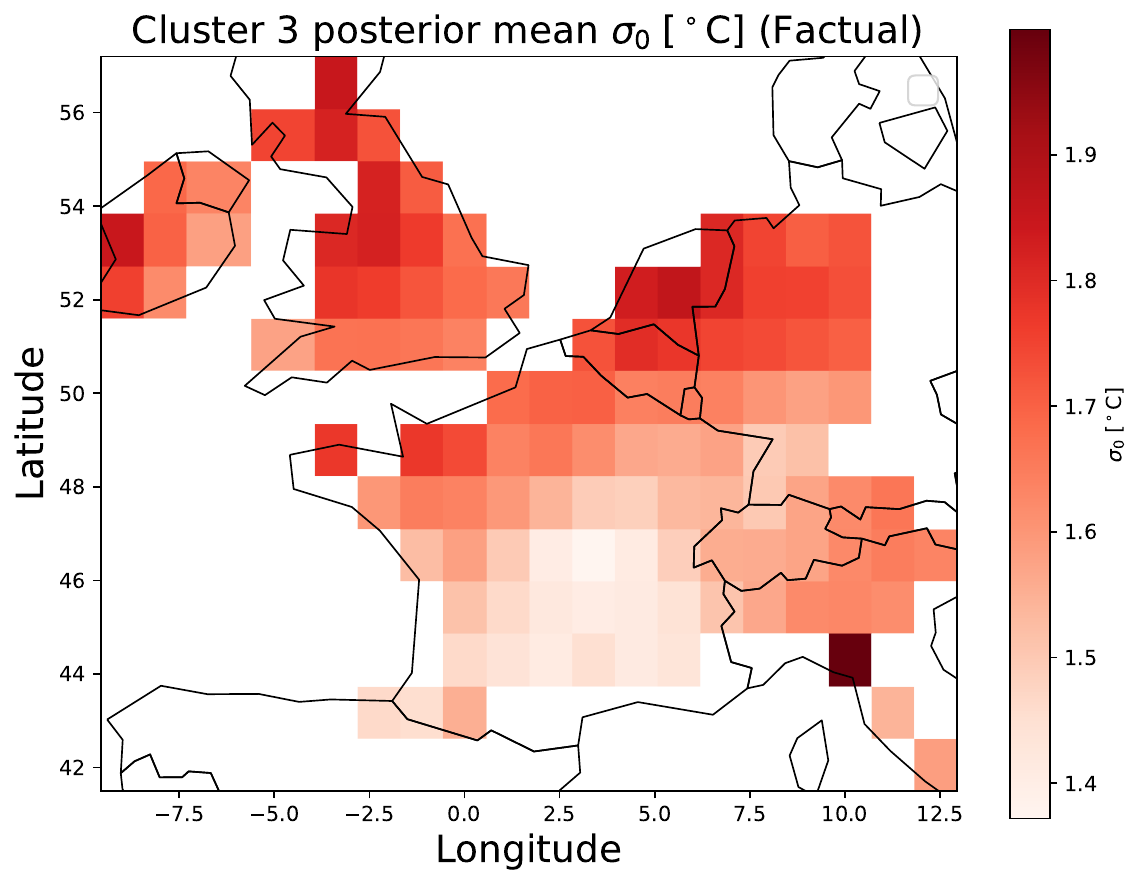}
    \caption{$\sigma_0\,[\si{\degreeCelsius}]$}
  \end{subfigure}\hfill
  \begin{subfigure}[t]{0.48\textwidth}
    \centering
    \includegraphics[width=\textwidth]{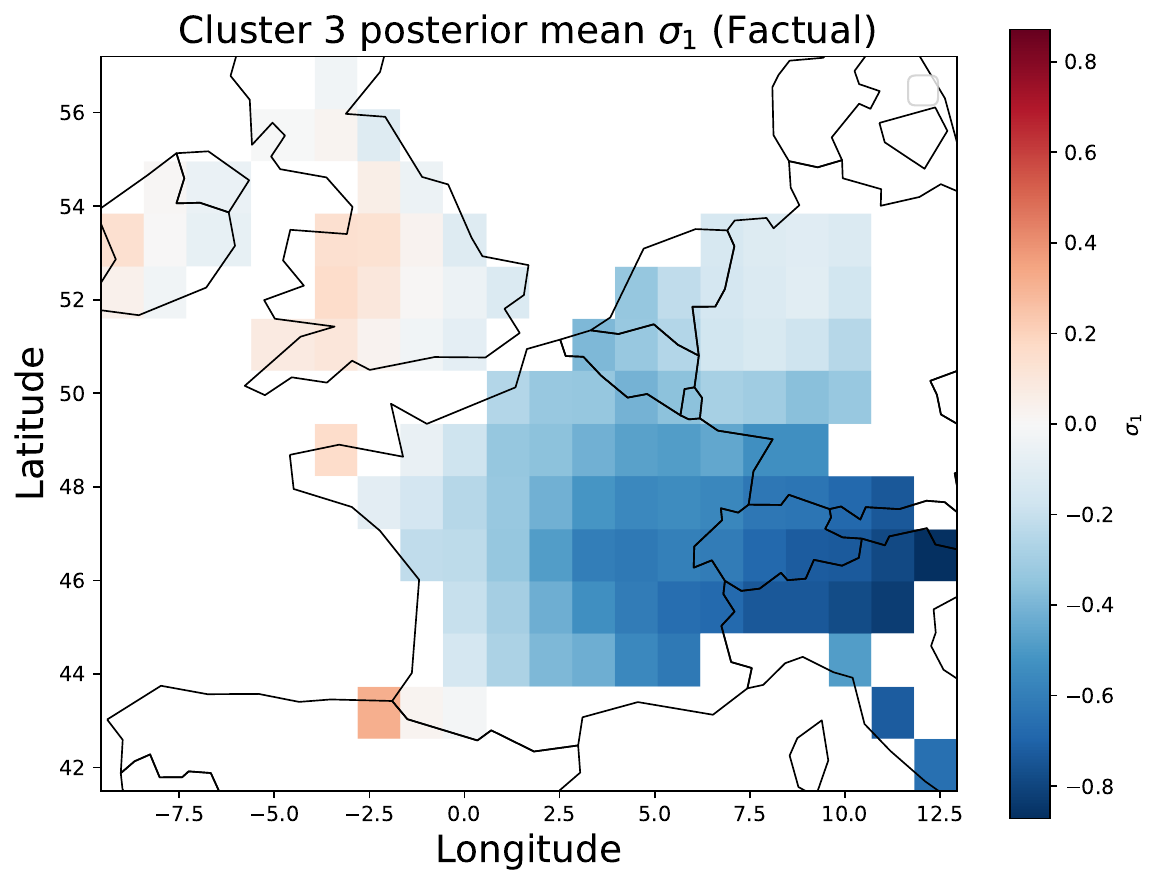}
    \caption{$\sigma_1$}
  \end{subfigure}

  \vspace{-0.2em}

  % -------- Row 3: xi --------
    \begin{subfigure}[t]{0.5\textwidth}
      \centering
      \includegraphics[width=\textwidth]{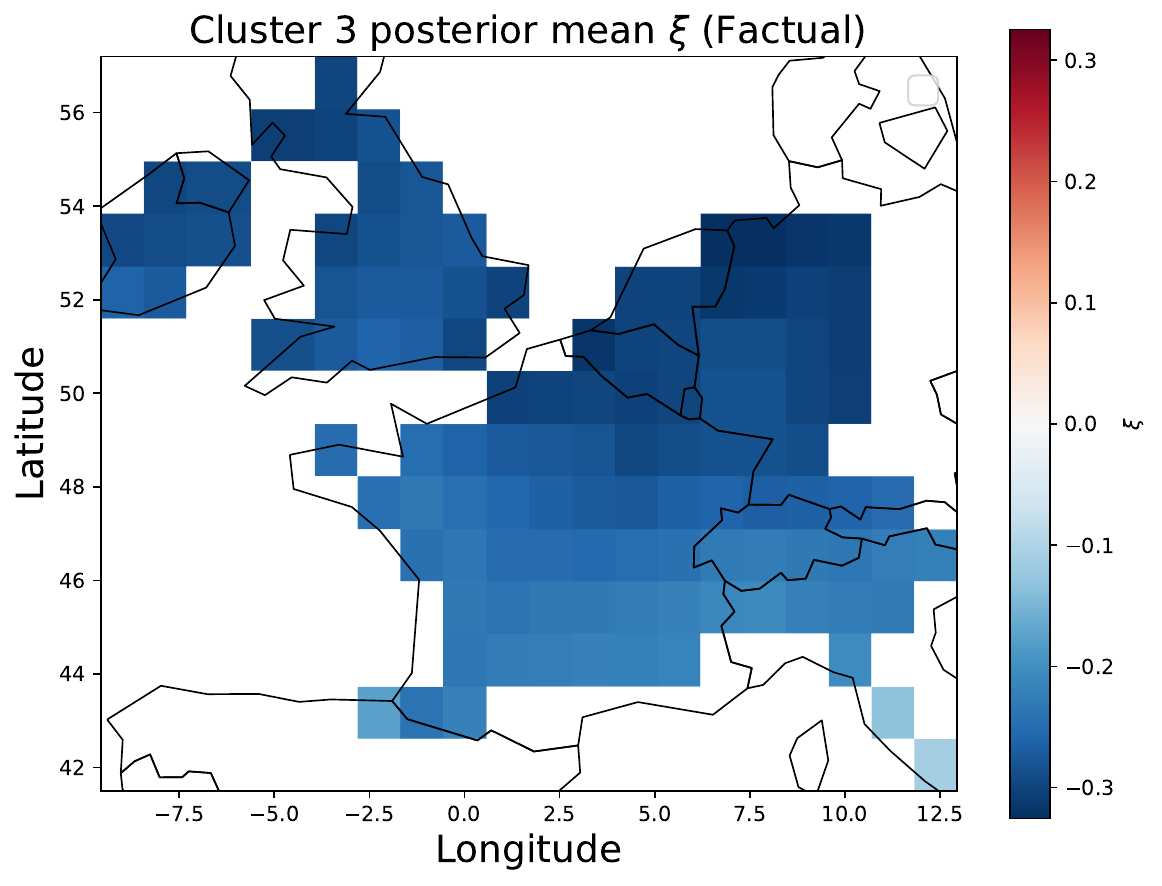}
      \caption{$\xi$}
    \end{subfigure}

  \caption{
    Posterior mean maps of key GEV parameters over Cluster~3 under the factual climate.
  }
  \label{fig:gev_maps_main}
\end{figure}

Figure~\ref{fig:gev_maps_main} displays posterior mean estimates of the GEV
parameters that are most directly relevant for attribution over Cluster~3, for factual climate. Specifically, we report the spatial patterns of the covariate effects in the location and scale parameters, $\mu_1$ and $\sigma_1$, which quantify the sensitivity of extreme temperatures to GMST anomalies. We find that as GMST anomalies increase, extremes tend to become more intense, as indicated by the positive values of $\mu_1$. On the other hand, the variability appears to decrease with increasing GMST, as suggested by $\sigma_1$. This behavior is also reported by \textcite{auld2023changes}, who found that the effect of global warming on the scale parameter is heterogeneous across clusters. However, for clusters similar to ours, our results are consistent with their findings. Caterpillar plots by grid cell are shown in Figure~\ref{fig:gev_caterpillar_mu}, Figure~\ref{fig:gev_caterpillar_sigma} and Figure~\ref{fig:gev_caterpillar_xi} in the Supplement~\ref{sec:supp_gev_maps}, where the uncertainty around the parameters is also illustrated. See Supplement~\ref{sec:supp_gev_mcmc} (Figure~\ref{fig:gev_traceplots}) and Supplement~\ref{sec:supp_gev_ppc} (Figure~\ref{fig:qq_ppc_gev}) for MCMC diagnostics and posterior predictive checks.

\subsection{Neural Bayes estimator for Cluster 3}\label{sec:NBE_cluster3}

For Cluster~3, we trained a neural Bayes estimator on a large number of simulated datasets generated from the proposed spatio-temporal model. Each training sample consisted of the collection of summary statistics described above, namely the $\chi$-grids, the thresholded madogram map, the spatial extremogram spread summaries, the ATDF curve in Eq.~\eqref{eq:atdf_est}, and the IQR-by-distance curve in Eq.~\eqref{eq:IQR}. 
The corresponding target output was the dependence parameter vector
\[
\paramsDEP
=
(\delta,\theta,\omega,\eta,\kappa,\lambda,\rho_s,\rho_t)^\top.
\]
For the simulation of the training datasets, we specified independent prior distributions for each dependence parameter component. We adopted uniform priors over ranges chosen to cover plausible values of the model parameters while ensuring sufficient variability in the simulated data. Specifically, we used
$$
\begin{aligned}
\delta\,[\text{--}]&\sim \mathcal{U}(0,1), & \theta [\si{\km}] &\sim \mathcal{U}(50,1000), \\
\kappa [\si{\day}] &\sim \mathcal{U}(1,30), & \eta\,[\text{--}] &\sim \mathcal{U}(1,5), \\
\lambda \,[\text{--}] &\sim \mathcal{U}(0,5), & \omega\,[\si{\radian}] &\sim \mathcal{U}(0,\pi), \\
\rho_s[\si{\km}] &\sim \mathcal{U}(10,400), & \rho_t[\si{\day}] &\sim \mathcal{U}(0,15).
\end{aligned}
$$
with all parameters sampled independently. These ranges were selected to span a broad yet realistic domain given the cluster size for the spatio-temporal dependence structure, allowing the neural Bayes estimator to learn across diverse regimes of extremal dependence.

The NN was implemented as a multi-input architecture combining convolutional branches for image-like summaries and one-dimensional convolutional branches for curve-like summaries. More precisely, the $\chi$-grids, madogram map, and spatial spread summaries were processed through two-dimensional convolutional layers, whereas the ATDF and IQR summaries were processed through one-dimensional convolutional layers. The features extracted from all branches were then concatenated and passed through fully connected layers to predict the eight model parameters jointly. The network was trained on \num{299000} simulated samples of $171$ years, while \num{1000} samples were used for validation during training. Furthermore, \num{1000} simulated samples were kept as an independent holdout set to assess predictive accuracy.

To evaluate the estimator, we compared the predicted and true parameter values on the holdout set. Figure~\ref{fig:cluster3_scatter_params} in Supplement~\ref{sec:sup_plot_NBE} displays scatterplots of predicted versus true values for all eight parameters. Points concentrated around the diagonal indicate accurate recovery of the parameter. Figure~\ref{fig:cluster3_boxplot_params} in Supplement~\ref{sec:sup_plot_NBE} complements this analysis by showing boxplots of the prediction errors, defined as predicted minus true value, for each parameter. We observe that the estimation accuracy depends on the value of $\delta$. Recall $Z(\svec,d) = R(\svec,d)^{\delta} \cdot W(\svec,d)^{1-\delta}$ in Eq.~\eqref{eq:Zsd}. When $\delta$ is small, the data are primarily driven by the process $W(\svec,d)$, and consequently the parameters associated with the component $R(\svec,d)$, namely $\lambda$, $\rho_s$, and $\rho_t$, are not well identified. Conversely, when $\delta$ is close to~$1$, the data are mainly governed by the process $R(\svec,d)$, and the parameters related to $W(\svec,d)$ become difficult to estimate. In addition, the value of $\eta$ in Eq.~\eqref{eq:Anisotropic matrix} also impacts identifiability. When $\eta$ is close to~$1$, the process exhibits little to no anisotropy, which makes the anisotropy angle $\omega$ poorly identifiable.

We report posterior estimates for all parameters in the factual world. Table~\ref{tab:dep_est} presents the true parameter values alongside the corresponding 95\% credible intervals derived from the parametric bootstrap based on 1000 draws.
\begin{table}[ht]
\centering
\caption{Parameter estimates of dependence over Cluster~3 with 95\% credible intervals for Factual world}
\begin{tabular}{llc}
\hline
 Parameter & Estimated & 95\% CI  \\
\hline
 $\delta\,[\si{-}]$   & 0.4494 & (0.4155, 0.4691)  \\
 $\theta\,[\si{\km}]$   & 661.9112 & (638.4598, 718.5402)  \\
 $\kappa\,[\si{\day}]$   & 20.9955 & (20.0814, 23.8779)  \\
 $\omega\,[\si{\radian}]$   & 2.3335 & (2.2359, 2.3671)  \\
 $\eta\,[\si{-}]$     & 2.0366 & (1.9770, 2.2302) \\
 $\lambda\,[\si{-}]$  & 2.0668 & (1.4143, 3.1116)  \\
 $\rho_s\,[\si{\km}]$   & 300.9300 & (261.5675, 352.3192) \\
 $\rho_t\,[\si{\day}]$   & 1.1806 & (0.6109, 1.5983)\\
\hline
\label{tab:dep_est}
\end{tabular}
\end{table}

\begin{figure}[htbp]
    \centering
    \begin{subfigure}[t]{0.48\textwidth}
        \centering
        \includegraphics[width=\textwidth]{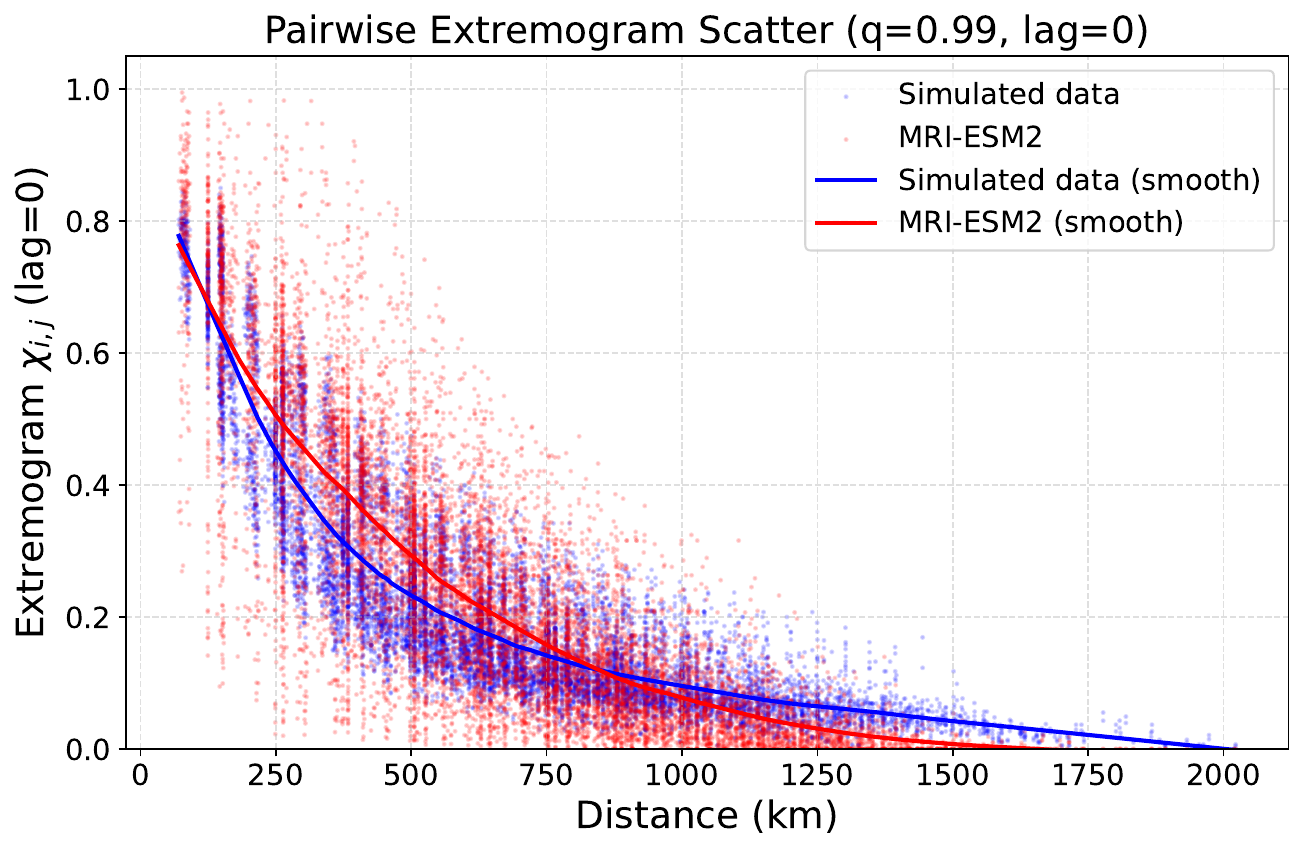}
        \caption{Lag 0}
    \end{subfigure}
    \hfill
    \begin{subfigure}[t]{0.48\textwidth}
        \centering
        \includegraphics[width=\textwidth]{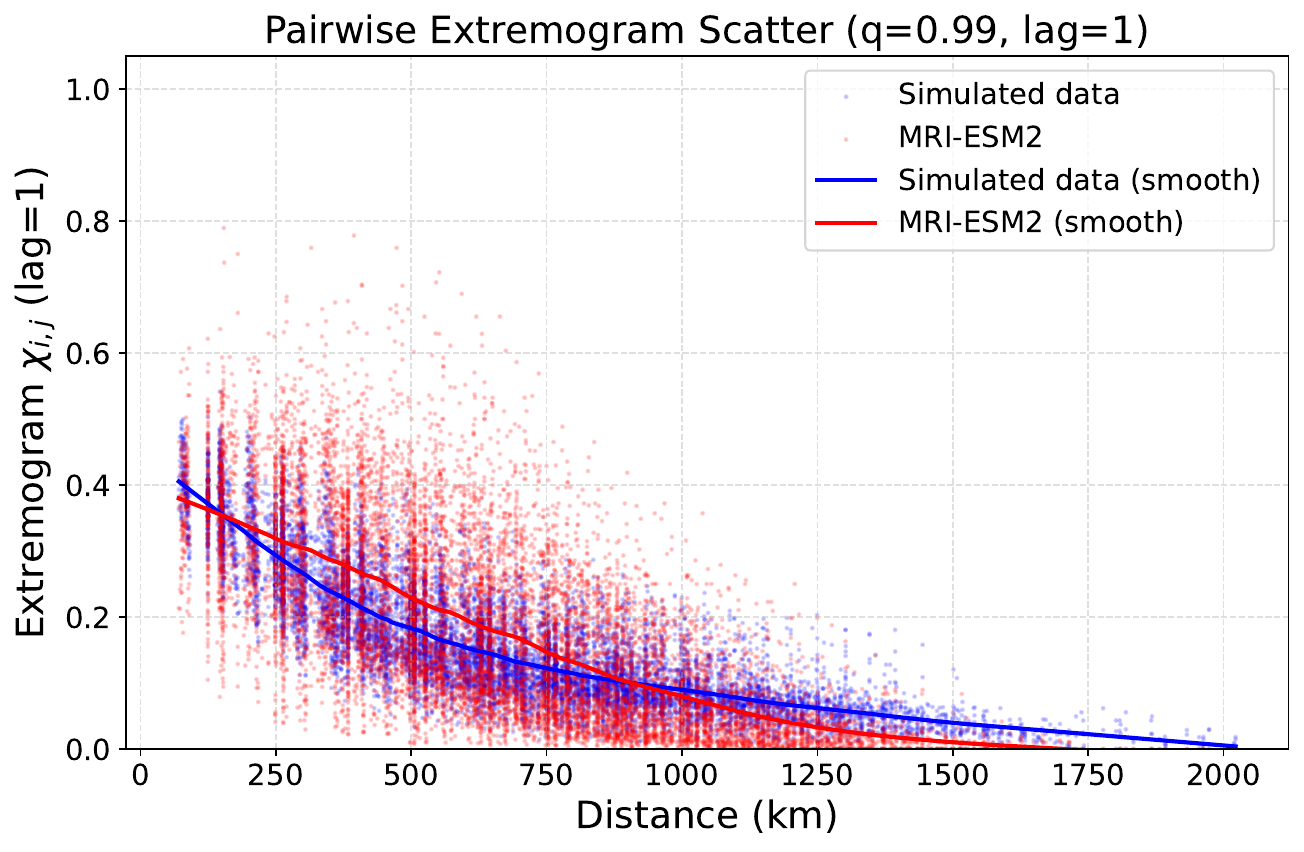}
        \caption{Lag 1}
    \end{subfigure}
    \begin{subfigure}[t]{0.48\textwidth}
        \centering
        \includegraphics[width=\textwidth]{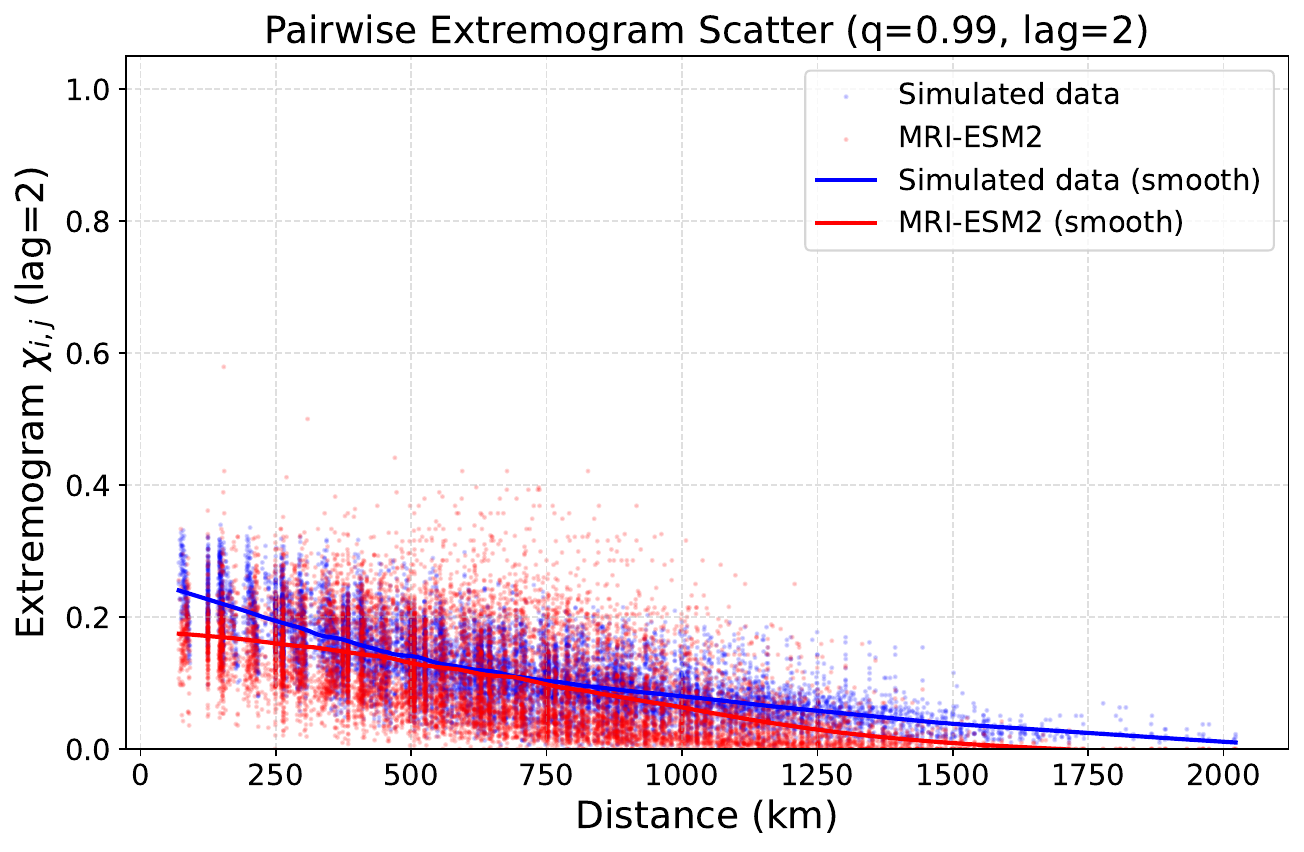}
        \caption{Lag 2}
    \end{subfigure}
    \hfill
    \begin{subfigure}[t]{0.48\textwidth}
        \centering
        \includegraphics[width=\textwidth]{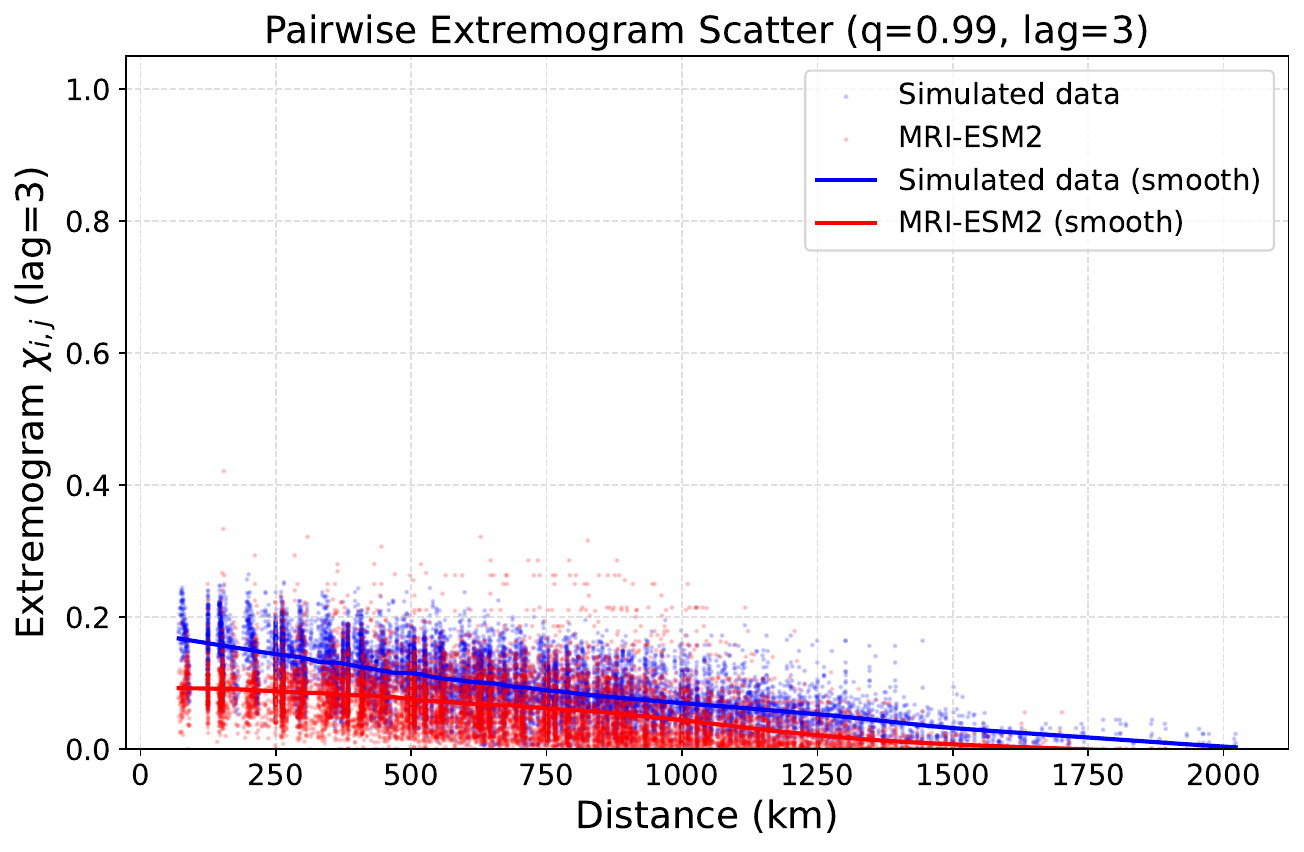}
        \caption{Lag 3}
    \end{subfigure}

    \caption{
    Pairwise extremogram plots for Cluster~3 at temporal lags $0$ to $3$ for 171 years of $Z(\svec,d)$ vs the observed factual MRI-ESM2 data. Each panel shows the empirical extremogram as a function of spatial distance.
    The estimated dependence parameters are 
    $\hat{\delta}=0.45$, 
    $\hat{\theta}=661.98$, 
    $\hat{\kappa}=20.99$, 
    $\hat{\omega}=2.33$, 
    $\hat{\eta}=2.04$, 
    $\hat{\lambda}=2.07$, 
    $\hat{\rho}_s=301.01$, 
    and $\hat{\rho}_t=1.18$.
    }
    \label{fig:extremogram_cluster3_lags0_3}
\end{figure}

To assess whether our statistical model captures the correct extremal dependence structure, we simulate data from the statistical model and visually compare it with data generated by the MRI-ESM2 climate model. Figure~\ref{fig:extremogram_cluster3_lags0_3} illustrates this comparison for a threshold of $u = 0.99$ and four lag times. We observe that our statistical model slightly underestimates extremal dependence at short distances. Moreover, although anisotropy is included, the climate model data exhibit greater variability than what is captured by our statistical model.

\subsection{Data simulation, return period estimation \& attribution}
\label{sec:data_sim}

\begin{figure}[!htbp]
    \centering
    \includegraphics[width=0.4\linewidth]{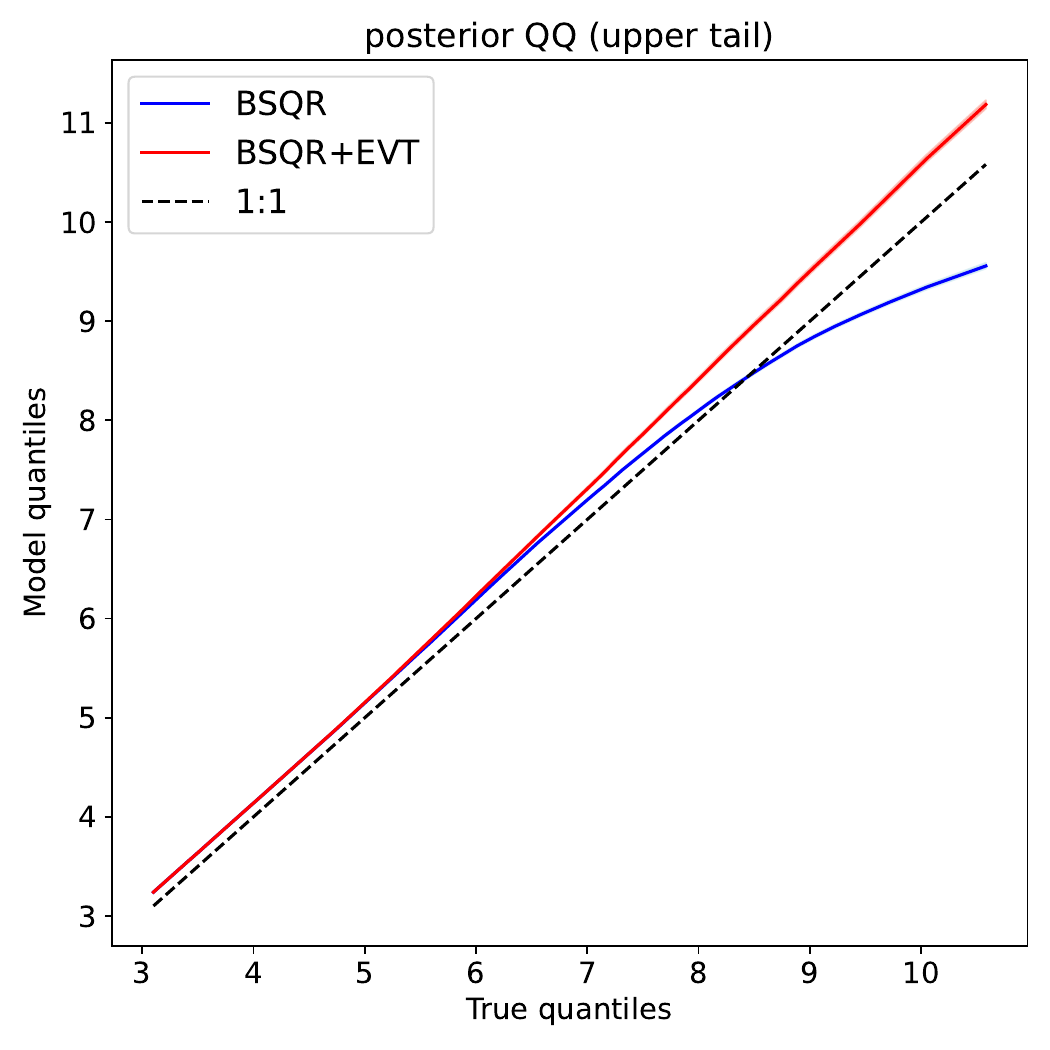}
    \caption{
    Pooled upper-tail QQ plot of observed MRI-ESM2 anomalies versus simulations from the fitted model. The BSQR-only transformation underestimates upper quantiles, whereas the hybrid BSQR+EVT approach is slightly above the $1{:}1$ line at the highest quantiles. Which reflects its ability to extrapolate beyond the observed range.
    }
    \label{fig:BSQR vs BSQR+EVT}
\end{figure}

Given the parameters of the copula $Z(\svec,d)$ and the parameters of the EVT and BSQR models, we can now use our simulation framework as described in Section~\ref{sec:Return periods}. We first illustrate empirically why it is necessary to combine the BSQR model for the bulk of the data with the EVT model for the tail of the distribution at each location. Figure~\ref{fig:BSQR vs BSQR+EVT} shows a pooled upper-tail quantile-quantile (QQ) comparison between observed (MRI-ESM2) and simulated temperature anomalies. The empirical quantiles of the factual data from the climate model MRI-ESM2, $x^{\text{MRI-ESM2,F}}(\svec,d,t)$ with $t\in\{1,\dots,171\}$, are compared to the corresponding quantiles of simulated fields obtained from the copula model. More precisely, we generate realizations $z(\svec,d,t)$ of the latent process $Z(\svec,d)$, which are transformed into uniform variables $u(\svec,d,t)$ according to Eq.~\eqref{eq:marginal_process}, using the parameters estimated by the Neural Bayes estimator. These uniform realizations are then backtransformed to the anomaly scale using two approaches. First, using only the BSQR model, we obtain $x^{\text{BSQR}}(\svec,d,t) = \hat{q}\big(u(\svec,d,t)\mid \covariates(t),\svec\big),$
where $\hat{q}$ denotes the estimated conditional quantile function. Second, using the hybrid approach described in Eq.~\eqref{eq:backtransform_full}, we obtain $x^{\text{BSQR+EVT}}(\svec,d,t)$, which combines the BSQR model for the bulk with a EVT model for the upper tail. The QQ plot focuses on high quantile levels and pools all locations and times in order to emphasize tail behavior. The BSQR only transformation clearly underestimates the upper quantiles, as shown by its systematic deviation below the $1{:}1$ line. In contrast, the hybrid BSQR+EVT transformation corrects the underestimation of the upper tail and reaches the appropriate magnitude of extreme values. In the most extreme quantiles, it exceeds the $1{:}1$ line, reflecting its ability to extrapolate beyond the range of the observed data.

\begin{figure}[htbp]
    \centering
    \begin{subfigure}[b]{0.48\textwidth}
        \centering
        \includegraphics[width=\textwidth]{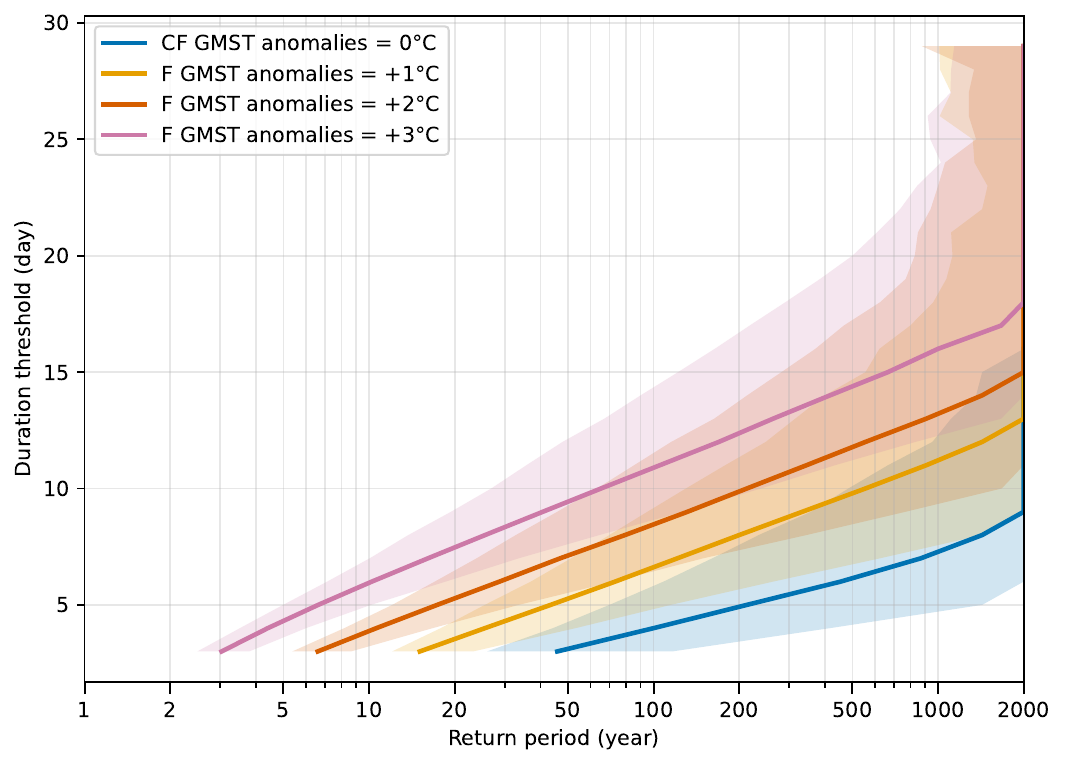}
        \caption{Duration}
    \end{subfigure}
    \hfill
    \begin{subfigure}[b]{0.48\textwidth}
        \centering
        \includegraphics[width=\textwidth]{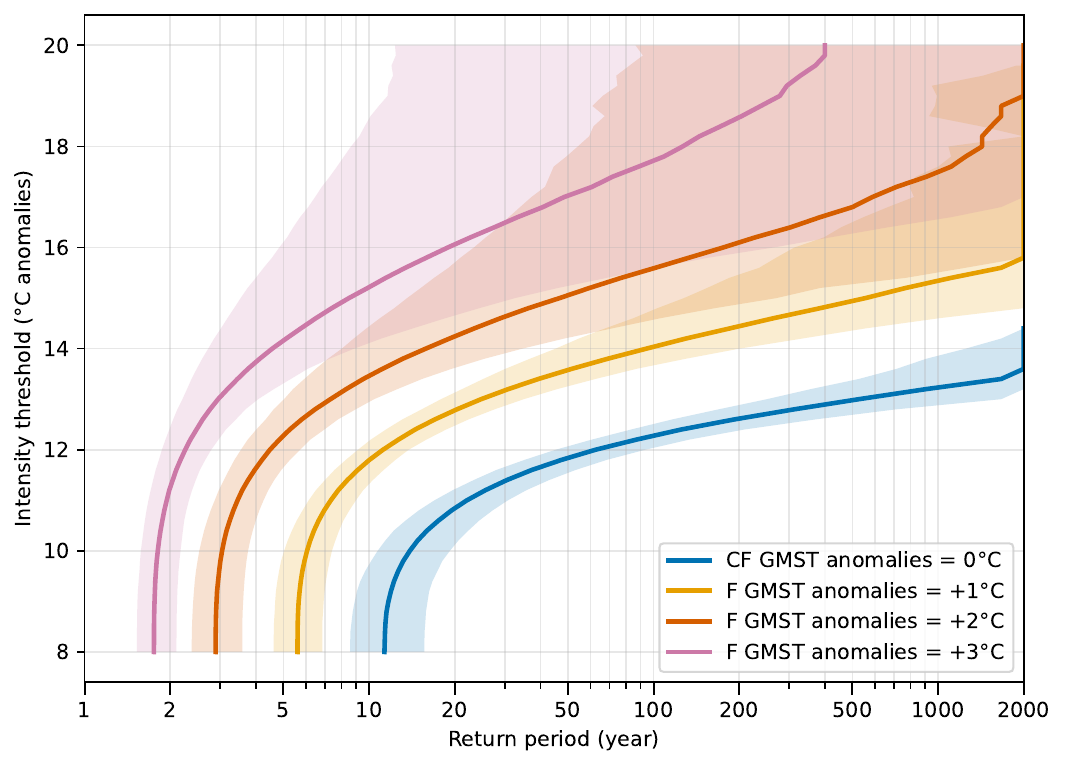}
        \caption{Intensity}
    \end{subfigure}
        \caption{Return period plots under different GMST anomaly scenarios. 
    Panel (a) shows the return period plot for the duration of heatwave episodes, 
    while panel (b) shows the return period plot for the maximum intensity 
    observed during a heatwave episode.
    F and CF stands for factual and counterfactual  climate. 
}
    \label{fig:cluster_3_return_period}
\end{figure}

Given the estimated model parameters, empirical return levels can now be computed using Monte Carlo simulation. For each parameter triple defined in Eq.~\eqref{eq:param_boot}, we simulated \num{10000} years of data and applied the back-transformation described in Eq.~\eqref{eq:backtransform_full}. Figure~\ref{fig:cluster_3_return_period} presents the return periods associated with the duration of heatwave episodes and with the maximum intensity observed during a heatwave episode under GMST anomaly scenarios of $\SI{0}{\celsius}$ up to $\SI{3}{\celsius}$.

% Once the exceedance probabilities have been estimated, attribution metrics such as the attributable risk (AR) can be computed (Eq.~\eqref{eq:AR}). Figure~\ref{fig:AR plot}a shows the AR for the maximum heatwave intensity, expressed as temperature anomalies of \SI{10}{\celsius}, \SI{12}{\celsius}, and \SI{14}{\celsius}. The AR is consistently negative, and the corresponding confidence intervals do not cross zero, indicating a statistically significant increase in the probability of such events due to anthropogenic forcing. For event duration (Fig.~\ref{fig:AR plot}b), expressed in days, the AR is also negative across all warming scenarios for a duration of \SI{4}{day}. As global mean surface temperature (GMST) increases, the AR becomes more negative and the associated uncertainty decreases. For longer durations (\SI{6}{day} and \SI{8}{day}), the \SI{95}{\percent} confidence interval includes zero under the \SI{1}{\celsius} GMST anomaly scenario, indicating a lack of statistical significance. However, for the \SI{2}{\celsius} and \SI{3}{\celsius} scenarios, the AR remains negative with confidence intervals excluding zero, again suggesting that anthropogenic forcing has increased the probability of such events.

\section{Discussion \& Conclusion}\label{sec:Discussion}
Our main objective was to develop a complete statistical pipeline to quantify the extent at which anthropogenic forcing has influenced the properties of heatwaves, including their duration and intensity. To address this question, we modeled temperature fields in both space and time within a nonstationary framework by combining marginal models with a flexible dependence model. The proposed approach integrates several modern statistical tools into a unified framework that allows the simulation of realistic spatio-temporal temperature fields. In particular, recent work has emphasized the importance of using sub-asymptotic models to accurately capture extremal dependence, as misspecification of the dependence structure can substantially affect attribution results \parencite{huser2025modeling,li2025importance}. Although the estimation procedure is computationally demanding, once the model is fitted it enables practitioners to generate coherent spatio-temporal temperature fields and to investigate a wide range of heatwave characteristics.

Our results based on the MRI-ESM2 climate model are consistent with findings reported in the attribution literature \parencite{ipcc2021,engdaw2023attribution,quilcaille2025systematic}, namely that anthropogenic forcing has increased both the severity and the intensity of heatwaves. However, our approach differs from most previous studies in that we explicitly model heatwaves as spatio-temporal events rather than relying on scalar temperature indices or heatwave indicators. By simulating full temperature fields, the proposed framework makes it possible to study a broad set of heatwave properties, including their spatial extent, persistence, and intensity, within a coherent probabilistic framework.

Despite these promising results, several areas of improvement remain. First, the diagnostic analysis presented in Section~\ref{sec:NBE_cluster3} suggests that the proposed copula slightly underestimates the tail dependence. Over large spatial domains such as the one considered in this study, we observe substantial variability in the tail dependence that cannot be fully captured by a simple anisotropic structure. One possible improvement would be to introduce non-stationary covariance structures; however, such an extension would substantially increase the number of parameters and the computational complexity of the model. A second limitation concerns the assumption that the tail dependence regime—whether asymptotically dependent or independent—is constant across the entire study region. This assumption may be unrealistic in practice. More flexible models, such as those proposed by \textcite{hazra2025efficient} or \textcite{shi2026spatial}, allow different dependence regimes to coexist within the same spatial domain, although these approaches are currently restricted to purely spatial settings. Finally, the dependence structure was assumed to be stationary over time. An interesting extension would be to allow the mixture parameter $\delta$ to vary with covariates, following ideas similar to those proposed by \textcite{maume2024spatio}.

Future work could therefore focus on improving the dependence model and extending the proposed framework to incorporate more flexible copula structures. Another natural direction would be to apply the methodology to ensembles of climate models in order to assess the robustness of the results and to evaluate whether consistent conclusions are obtained across different climate simulations.  In addition, applying the framework to observational datasets would provide an important benchmark. The methodology could also be extended to other spatio-temporally coherent processes, such as droughts, and to different regions of the world.

\printbibliography

\clearpage
\appendix

\section*{Supplementary Material for ``\scititle''}
\label{sec:supplement}

% Restart and prefix supplement numbering
\setcounter{section}{0}
\setcounter{subsection}{0}
\setcounter{figure}{0}
\setcounter{table}{0}
\setcounter{equation}{0}

\renewcommand{\thesection}{S\arabic{section}}
\renewcommand{\thesubsection}{S\arabic{section}.\arabic{subsection}}
\renewcommand{\thefigure}{S\arabic{figure}}
\renewcommand{\thetable}{S\arabic{table}}
\renewcommand{\theequation}{S\arabic{equation}}

\section{Supplementary Methods}
\label{sec:supp_methods}
\subsection{Bayesian Spatial Quantile Regression}
\label{sec:supp_BSQR}
This section presents the Bayesian hierarchical model corresponding to the \emph{approximate method} of \textcite[Sec.~3]{reich2011bayesian}. We adopt the BSQR model to characterize the conditional distribution of temperature anomalies $X(\svec,d,t)$ given covariates $\covariates(t)=(c_1(t),\dots,c_{p}(t))^\top$. The covariates are transformed such that $c_1(t)=1$ is the intercept and $c_j(t)\in[0,1]$ for $j\ge 2$. In practice, this transformation can be performed using a min–max rescaling or by applying the inverse-CDF–based normalization. In principle, the covariates could also depend on the day $d$, allowing for higher temporal resolution (e.g., monthly or daily climate indicators). However, in our application the only used large-scale covariate is the GMST, which is observed at monthly resolution and varies slowly over time. Because GMST primarily reflects long-term climate forcing rather than day-to-day variability, we treat it as a yearly covariate in the model. This simplification reduces model complexity while remaining consistent with the temporal scale at which the forcing evolves. The model allows covariate effects to vary smoothly across quantile level $\tau$ and spatial location $\svec$ as we expect nearby locations to be exposed to the same regional events, while enforcing non-crossing quantile curves. The approximation replaces the full data likelihood with a Gaussian likelihood for site-specific quantile regression coefficient estimates, while retaining the spatial quantile-process prior of \textcite[Sec.~2]{reich2011bayesian}. Let $q(\tau\mid\covariates(t),\svec)$ denote the conditional $\tau$-th quantile of $X(\svec,d,t)$. Following \textcite{reich2011bayesian}, we assume
\begin{equation}
q(\tau\mid\covariates(t),\svec)
=
\covariates(t)^\top \paramsQR(\tau,\svec),
\end{equation}
where $\paramsQR(\tau,\svec)=(\beta^{\mathrm{QR}}_1(\tau,\svec),\dots,\beta^{\mathrm{QR}}_p(\tau,\svec))^\top$ are spatially varying quantile coefficient functions.
The approximation starts by fitting for each spatial location $\svec\in \Rregion$, a classical quantile regression at quantile levels $\tau_1,\ldots,\tau_K$.
For each $\tau_{k}$ and site $\svec$, the Koenker--Bassett estimator \parencite{koenker2005quantile} is
\begin{equation}
\label{eq:KB}
    \paramsQRest(\tau_{k},\svec) 
    = \bigl(
        \hat\beta_{1}^{\mathrm{QR}}(\tau_{k},\svec),
        \ldots,
        \hat\beta_{p}^{\mathrm{QR}}(\tau_{k},\svec)
    \bigr)^{\top}
    = \operatornamewithlimits{\arg\min}_{\paramsQR}\sum_{t}\sum_{d}\rho_{\tau_{k}}\bigl(x(\svec,d,t)-\covariates(t)^{\top}\paramsQR\bigr),
\end{equation}
where $\rho_{\tau}(u)=u \cdot (\tau-\ind\{u<0\})$ is the pinball loss.  
These estimators are consistent for the true quantile coefficients and possess a known asymptotic covariance structure. Specifically, for any two quantile levels $\tau_{k}$ and $\tau_{l}$, we have
\begin{equation}
\Cov\!\left(
\sqrt{DT}\,\paramsQRest(\tau_{k},\svec),\;
\sqrt{DT}\,\paramsQRest(\tau_{l},\svec)
\right)
\approx H(\tau_{k},\svec)^{-1} J(\tau_{k},\tau_{l},\svec) H(\tau_{l},\svec)^{-1}. 
\label{eq:asymCov}
\end{equation}
where $H(\tau,\svec)$ and $J(\tau_{k},\tau_{l},\svec)$ are the standard quantile regression sandwich $p \times p$ matrices defined in \textcite{koenker2005quantile}.
We define the stacked vector of estimated coefficients
\begin{equation}
\label{eq:beta_hat_stack}
\paramsQRest(\svec)
=    \left(
        \paramsQRest(\tau_{1},\svec)^\top,
        \ldots,
        \paramsQRest(\tau_{K},\svec)^\top
    \right)^\top
\in\mathbb{R}^{pK},
\end{equation}
and $\Cov(\paramsQRest(\svec))=\Sigma(\svec)$ is the $pK \times pK$ covariance matrix, where the elements are given by Eq.~\eqref{eq:asymCov}. Let $\hat\Sigma(\svec)\in\mathbb{R}^{pK\times pK}$ denote the estimated asymptotic covariance matrix of $\paramsQRest(\svec)$ obtained from standard quantile regression theory. Its $((j,k),(j',\ell))$ entry is
\begin{equation}
\label{eq:Sigma_hat_entry}
\hat\Sigma_{(j,k),(j',\ell)}(\svec)
=
\hat H(\tau_k,\svec)^{-1}\,
\hat J(\tau_k,\tau_\ell,\svec)\,
\hat H(\tau_\ell,\svec)^{-1},
\end{equation}
for $j,j'=1,\ldots,p$ and $k,\ell=1,\ldots,K$, where
\begin{align}
\label{eq:H_hat}
\hat H(\tau,\svec)
&=
\frac{1}{DT}
\sum_{t,d}
\covariates(t)\covariates(t)^\top\,
\hat f_{\svec}\!\left(\covariates(t)^\top \paramsQRest(\tau,\svec)\right),
\\[0.5em]
\label{eq:J_hat}
\hat J(\tau_k,\tau_\ell,\svec)
&=
\left(\tau_k \wedge \tau_\ell - \tau_k \tau_\ell\right)\,
\frac{1}{DT}
\sum_{t,d}
\covariates(t) \covariates(t)^\top,
\end{align}
with $\hat f_{\svec}(\cdot)$ a nonparametric density estimate at location $\svec$ evaluated at the fitted quantile $\covariates(t)_i^\top \paramsQRest(\tau,\svec)$.

In the second step, the quantile-regression coefficients $\paramsQRest(\svec)$ are assumed to follow a normal distribution,
\begin{equation}
\label{eq:approx_likelihood}
\paramsQRest(\svec)
\sim
\mathcal N\!\bigl(\paramsQR(\svec),\,\Sigma(\svec)\bigr),
\end{equation}
independently over $\svec\in\Rregion$. Where $\Sigma(\svec)$ will be replaced by its estimator $\hat{\Sigma}(\svec)$ defined in Eq.~\eqref{eq:Sigma_hat_entry}. For each covariate $j=1,\ldots,p$ and location $\svec$, the quantile-regression coefficient function is modeled using a Bernstein polynomial expansion,
\begin{equation}
\label{eq:bernstein_beta}
\beta^{\text{QR}}_j(\tau,\svec)
=
\sum_{m=1}^{M} B_m(\tau)\,\alpha_{jm}(\svec),
\qquad
B_m(\tau)
=
\binom{M}{m}\tau^m(1-\tau)^{M-m},
\qquad \tau\in[0,1].
\end{equation}
We choose $M=15$ basis functions. Following \textcite{reich2011bayesian}, 
moderate values of $M$ are sufficient to accurately approximate smooth quantile curves, 
and their simulation results were shown to be robust to the choice of $M$.
The vector $\paramsQR(\svec)$ in Eq.~\eqref{eq:approx_likelihood} is obtained by evaluating $\beta^{\text{QR}}_j(\tau,\svec)$ at $\tau_1,\ldots,\tau_K$ and stacking over $j$ as in Eq.~\eqref{eq:beta_hat_stack}. Define the increment parameters $\{\delta_{jm}(\svec)\}$ by
\begin{align}
\label{eq:delta_def}
\delta_{j1}(\svec) &= \alpha_{j1}(\svec), \\[0.25em]
\delta_{jm}(\svec) &= \alpha_{jm}(\svec) - \alpha_{j,m-1}(\svec),
\qquad m=2,\ldots,M,
\end{align}
so that
\begin{equation}
\label{eq:alpha_from_delta}
\alpha_{jm}(\svec)
=
\sum_{\ell=1}^{m}\delta_{j\ell}(\svec),
\qquad m=1,\ldots,M.
\end{equation}
Next, let us introduce unconstrained latent variables $\delta^{*}_{jm}(\svec)$. For $m=2,\ldots,M$, define the constrained increments by
\begin{equation}
\label{eq:delta_constraint}
\delta_{jm}(\svec)=
\begin{cases}
\delta^{*}_{jm}(s),
& \text{if }
\delta^{*}_{1m}(\svec)
+\displaystyle\sum_{i=2}^{p}
\ind \!\left\{\delta^{*}_{im}(\svec)<0\right\}\,
\delta^{*}_{im}(\svec)
\ge 0, \\[0.75em]
0, & \text{otherwise},
\end{cases}
\end{equation}
and for $m=1$ set $\delta_{j1}(\svec)=\delta^{*}_{j1}(\svec)$. Under the assumption that the intercept covariate $c_1(t)$ equals $1$ and that $c_j(t)\in[0,1]$ for $j=2,\ldots,p$, this construction ensures that the conditional quantile function is nondecreasing in $\tau$ for all $\covariates(t)$; see \textcite[Section~2.2 Eq.~(11)]{reich2011bayesian}.

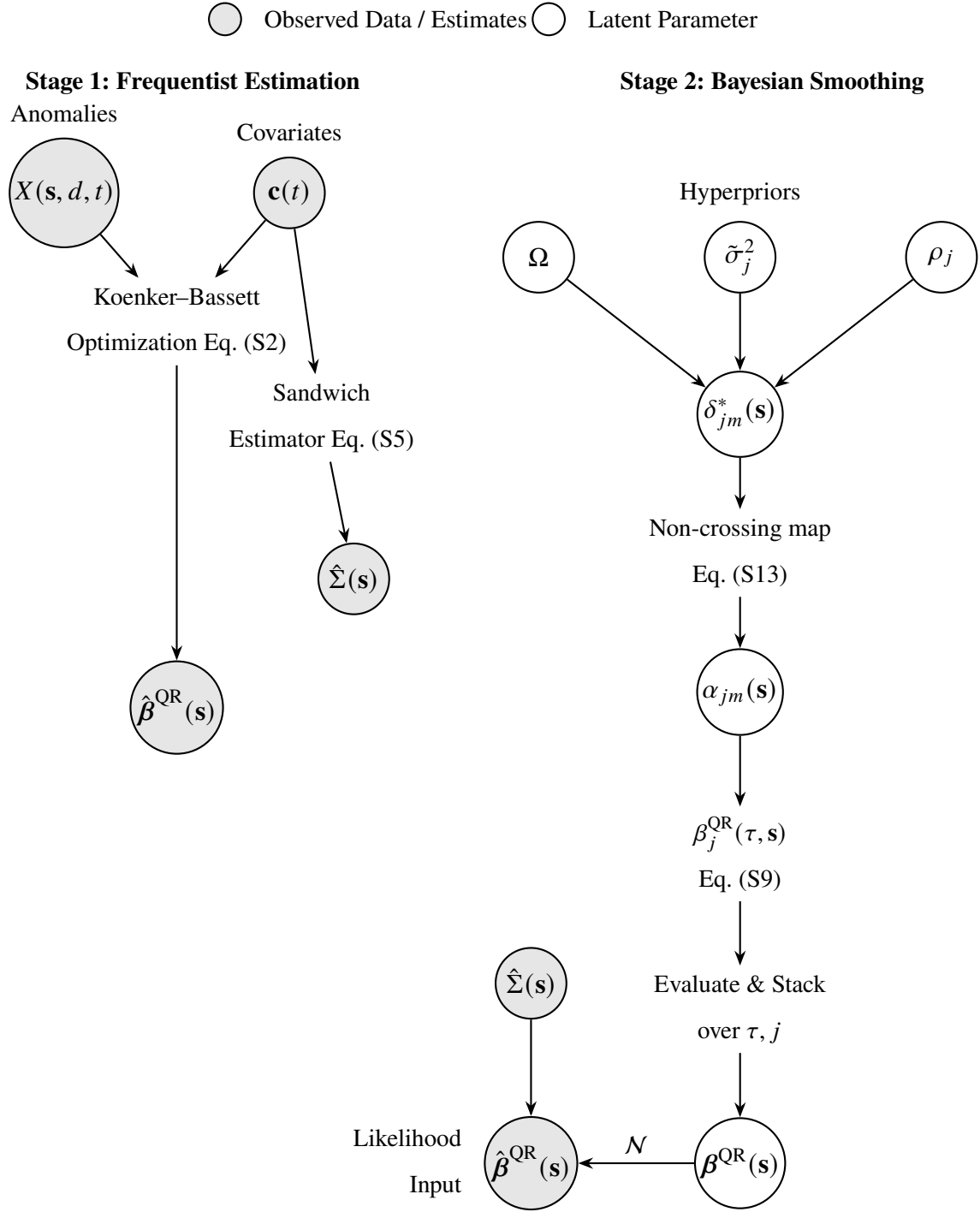
\begin{figure}[htbp]
\centering
\begin{tikzpicture}[
  node distance=1.1cm and 2.0cm,
  >=Stealth,
  % ---- Styles ----
  latent/.style={draw, circle, minimum size=11mm, inner sep=1pt, thick},
  obs/.style={draw, circle, fill=gray!20, minimum size=11mm, inner sep=1pt, thick}, 
  det/.style={align=center, font=\small}, 
  plate/.style={draw, dashed, inner sep=12pt, color=gray!60, thick},
  arrow/.style={->, thick},
  lab/.style={font=\small},
  stagelab/.style={font=\small\bfseries}
]

  % =========================
  % Top Legend
  % =========================
  \node[obs, minimum size=5mm] (k1) at (-5, 3.2) {};
  \node[right=2mm of k1, lab] {Observed Data / Estimates};
  
  \node[latent, minimum size=5mm, right=4.5cm of k1] (k2) {};
  \node[right=2mm of k2, lab] {Latent Parameter};

  % % =========================
  % % Vertical Separator (Simple Line)
  % % =========================
  % \draw[dashed, thick, gray!40] (-1.3, 2.5) -- (-1.3, -14);

  % =========================
  % Stage Labels
  % =========================
  \node[stagelab] (stage1title) at (-5.5, 2.2) {Stage 1: Frequentist Estimation};
  \node[stagelab] (stage2title) at (3.5, 2.2) {Stage 2: Bayesian Smoothing};

  % =========================
  % Stage 1: Left
  % =========================
  \node[obs] (X_data) at (-7.5, 0.5) {$X(\svec, d, t)$};
  \node[obs] (C_data) at (-4, 0.5) {$\covariates(t)$};
  \node[lab, above=1mm of X_data] {Anomalies};
  \node[lab, above=1mm of C_data] {Covariates};

  \node[det] (KB_est) at (-5.75, -1.5) {Koenker--Bassett \\ Optimization Eq.~\eqref{eq:KB}};
  \node[det] (Sandwich) at (-3.5, -3) {Sandwich \\ Estimator Eq.~\eqref{eq:Sigma_hat_entry}};

  % Outputs of Stage 1
  \node[obs] (Shat_S1) at (-3, -5.5) {$\hat{\Sigma}(\svec)$};
  \node[obs] (bhat_S1) at (-5.75, -7.5) {$\paramsQRest(\svec)$};

  \draw[arrow] (X_data) -- (KB_est);
  \draw[arrow] (C_data) -- (KB_est);
  \draw[arrow] (C_data) -- (Sandwich);
  \draw[arrow] (KB_est) -- (bhat_S1);
  \draw[arrow] (Sandwich) -- (Shat_S1);

  % =========================
  % Stage 2: Right
  % =========================
  % Hyperpriors
  \node[latent] (sig) at (3, -0.5) {$\tilde{\sigma}_j^2$};
  \node[latent, left=of sig] (Omega) {$\Omega$};
  \node[latent, right=of sig] (rho) {$\rho_j$};
  \node[lab, above=1mm of sig] {Hyperpriors};

  % Latent Process
  \node[latent, below=1.2cm of sig] (dstar) {$\delta^{*}_{jm}(\svec)$};

  % Construction
  \node[det, below=0.8cm of dstar] (map) {Non-crossing map \\ \small Eq.~\eqref{eq:delta_constraint}};
  \node[latent, below=0.8cm of map] (alpha) {$\alpha_{jm}(\svec)$};

  % Quantile layer
  \node[det, below=1.1cm of alpha] (beta_fun) {$\beta^{\mathrm{QR}}_{j}(\tau,\svec)$ \\ \small Eq.~\eqref{eq:bernstein_beta}};
  \node[det, below=1cm of beta_fun] (eval) {Evaluate \& Stack \\ \small over $\tau, j$};
  \node[latent, below=1cm of eval] (beta_stack) {$\paramsQR(\svec)$};

  % RECREATED NODES IN STAGE 2
  % These represent the Stage 1 outputs being used as Stage 2 inputs
  \node[obs, left=1.8cm of beta_stack] (bhat_S2) {$\paramsQRest(\svec)$};
  \node[obs, above=1.5cm of bhat_S2] (Shat_S2) {$\hat{\Sigma}(\svec)$};
  \node[lab, left=2mm of bhat_S2, align=right] {Likelihood \\ Input};

  % =========================
  % Arrows in Stage 2
  % =========================
  \draw[arrow] (Omega) -- (dstar);
  \draw[arrow] (sig) -- (dstar);
  \draw[arrow] (rho) -- (dstar);
  
  \draw[arrow] (dstar) -- (map);
  \draw[arrow] (map) -- (alpha);
  \draw[arrow] (alpha) -- (beta_fun);
  \draw[arrow] (beta_fun) -- (eval);
  \draw[arrow] (eval) -- (beta_stack);

  % The Likelihood link within Stage 2
  \draw[arrow] (beta_stack) -- node[above, lab, pos=0.5] {$\Normal$} (bhat_S2);
  \draw[arrow] (Shat_S2) -- (bhat_S2);

\end{tikzpicture}
\caption{Schematic of the two-stage BSQR model. Stage 1 (left) independently estimates the quantile coefficients and their covariance at each location $\svec$. Stage 2 (right) re-incorporates these estimates as observed data (shaded nodes) to perform spatial smoothing via a Bernstein polynomial-based Gaussian process prior.}
\label{fig:bsqr_diagram}
\end{figure}

To introduce spatial smoothness, for each pair $(j,m)$, the latent variables are assumed to follow independent Gaussian spatial processes:
\begin{equation}
\label{eq:gp_prior_delta}
\{\delta^{*}_{jm}(\svec): \svec \in\mathcal \Rregion\}
\sim
GP\!\left(
\bar\delta_{jm}(\Omega),\;
\tilde{\sigma}_j^2 \exp\!\left(-\frac{\|\svec-\svec'\|}{\rho_j}\right)
\right).
\end{equation}
Where $\bar\delta_{jm}(\Omega)$ is chosen to center the quantile process on a parametric base distribution $f_0(y \mid \Omega)$. Following \textcite{reich2011bayesian}, we take $f_0$ to be a skew-normal distribution with parameter vector $\Omega = (\psi_1,\psi_2,\psi_3)^{\top}$. For the intercept term ($j=1$), the prior mean coefficients $\bar\delta_{1m}(\Omega)$ are selected so that the implied quantile function is centered on the corresponding skew-normal quantile function $q_0(\tau \mid \Omega)$.
Specifically,
$$
q_0(\tau\mid\Omega)
\approx
\sum_{m=1}^{M} B_m(\tau)\,\bar\alpha_{1m}(\Omega),
\qquad
\bar\alpha_{1m}(\Omega)=\sum_{\ell=1}^{m}\bar\delta_{1\ell}(\Omega).
$$
The increment parameters $\{\bar\delta_{1m}(\Omega)\}$ are obtained by solving the constrained ridge regression problem with $\lambda=1$ following \textcite{reich2011bayesian}
\begin{equation}
\label{eq:centering_ridge}
(\bar\delta_{11}(\Omega),\ldots,\bar\delta_{1M}(\Omega))
=
\operatornamewithlimits{\arg\min}_{\{d_m\}}
\sum_{k=1}^{K}
\left[
q_0(\tau_k\mid\Omega)
-
\sum_{m=1}^{M}
B_m(\tau_k)
\left(
\sum_{\ell=1}^{m} d_\ell
\right)
\right]^2
+
\sum_{m=1}^{M} d_m^2,
\end{equation}
subject to $d_m\ge 0$ for $m\ge 2$, and we set $\bar\delta_{1m}(\Omega)=d_m$. For all non-intercept terms ($j=2,\ldots,p$), we set
$$
\bar\delta_{jm}(\Omega)=0,
\qquad j=2,\ldots,p; m=1,\ldots,M,
$$
shrinking covariate effects toward no effect in the absence of strong evidence from the data. For each covariate $j=1,\ldots,p$, the spatial variance and range parameters in Eq.~\eqref{eq:gp_prior_delta} are assigned independent priors
\begin{align}
\label{eq:hyper_sigma_rho}
\tilde{\sigma}_j^2 &\sim \mathrm{Inv\text{-}Gamma}(a_\sigma,b_\sigma), \\
\rho_j &\sim \mathrm{Gamma}(a_\rho,b_\rho),
\end{align}
and the base-distribution parameters $\Omega$ receive a prior $\Omega\sim\pi_\Omega(\Omega)$, independent of $\{\tilde{\sigma}_j^2,\rho_j\}_{j=1}^p$. A schematic summary of the hierarchical construction is provided in Figure~\ref{fig:bsqr_diagram}.

Let $\Theta^{\mathrm{QR}}_{\Rregion}$ denote the collection of all unknown parameters,
\begin{equation}
\label{eq:quantile_reg_param_def}
\Theta^{\mathrm{QR}}_{\Rregion}
=
\Big\{
\delta^{*}_{jm}(\svec) : j=1,\ldots,p,\; m=1,\ldots,M,\; \svec \in \Rregion
\Big\}
\;\cup\;
\Big\{
\tilde{\sigma}_j^2,\rho_j : j=1,\ldots,p
\Big\}
\;\cup\;
\{\Omega\}.
\end{equation}
For each spatial location $\svec$, the Bernstein polynomial representation in
Eq.~\eqref{eq:bernstein_beta} induces a finite-dimensional vector of quantile
regression coefficients evaluated at the grid
$\tau_1,\ldots,\tau_K$.
Specifically, we define
\begin{equation}
\label{eq:beta_QR_stack_def}
\boldsymbol{\beta}^{\mathrm{QR}}(\svec)
:=
\left(
\beta^{\mathrm{QR}}_1(\tau_1,\svec),\ldots,\beta^{\mathrm{QR}}_1(\tau_K,\svec),
\;\ldots,\;
\beta^{\mathrm{QR}}_p(\tau_1,\svec),\ldots,\beta^{\mathrm{QR}}_p(\tau_K,\svec)
\right)^\top
\in \mathbb{R}^{pK},
\end{equation}
where $\beta^{\mathrm{QR}}_j(\tau,\svec)$ is given by
Eq.~\eqref{eq:bernstein_beta}. The posterior distribution of $\Theta_{\mathrm{QR}}$ given the estimated quantile regression coefficients $\paramsQRest$ is
\begin{equation}
\label{eq:posterior_bsqr}
\begin{aligned}
\pi\left(\Theta^{QR}_{\Rregion} \mid \{\paramsQRest(\svec)\}_{\svec \in \Rregion}\right)
\;\propto\;&
\prod_{\svec \in \Rregion}
\Normal\!\left(
\paramsQRest(\svec) \mid
\boldsymbol\beta^{\mathrm{QR}}(\svec),
\hat\Sigma(\svec)
\right)
\\[0.75em]
&\times
\prod_{j=1}^{p}
\prod_{m=1}^{M}
GP\!\left(
\delta^{*}_{jm}(\cdot)\mid
\bar\delta_{jm}(\Omega),
\tilde{\sigma}_j^2,
\rho_j
\right)
\pi_\Omega(\Omega)
\prod_{j=1}^{p}
\pi(\tilde{\sigma}_j^2)\pi(\rho_j).
\end{aligned}
\end{equation}
This posterior induces spatially smooth, noncrossing quantile functions while borrowing strength across spatial locations and quantile levels. From Eq.~\eqref{eq:posterior_bsqr}, we see that the posterior distribution of the model parameters is conditioned on frequentist quantile regression estimates, rather than on the original data directly. This constitutes the key step of the approximate approach proposed by \textcite{reich2011bayesian}, and leads to a substantial reduction in computational complexity. For a fully Bayesian formulation, we refer the reader to \textcite[Section~2]{reich2011bayesian}.

\subsection{GEV method}
\label{sec:supp_GEV}
\begin{figure}[htbp]
\centering
\begin{tikzpicture}[
  node distance=1.0cm and 2.0cm,
  >=Stealth,
  % ---- Styles ----
  latent/.style={draw, circle, minimum size=11mm, inner sep=1pt, thick},
  obs/.style={draw, circle, fill=gray!20, minimum size=11mm, inner sep=1pt, thick},
  det/.style={align=center, font=\small},
  plate/.style={draw, dashed, inner sep=15pt, color=gray!60, thick},
  arrow/.style={->, thick},
  lab/.style={font=\small},
  stagelab/.style={font=\small\bfseries}
]
  % =========================
  % Top Legend
  % =========================
  \node[obs, minimum size=5mm] (k1) at (-4, 2.8) {};
  \node[right=2mm of k1, lab] {Observed Data};
  
  \node[latent, minimum size=5mm, right=3.5cm of k1] (k2) {};
  \node[right=2mm of k2, lab] {Latent Parameter};
  
  % =========================
  % Stage Labels
  % =========================
  \node[stagelab] (stage1title) at (-5.5, 2) {Data \& Likelihood};
  \node[stagelab] (stage2title) at (3.5, 2) {Hierarchical Spatial Prior};

  % =========================
  % Stage 1: Observed (Left)
  % =========================
  \node[obs] (M_st) at (-5.5, -4.5) {$M_{\svec,t}$};
  \node[obs, above=3.5cm of M_st] (c_t) {$\covariates(t)$};
  \node[lab, below=2mm of M_st] {Annual Maxima};
  \node[lab, above=2mm of c_t] {Global Covariates};

  % =========================
  % Stage 2: Parameters (Right)
  % =========================
  % Hyperpriors
  \node[latent] (gamma_g) at (2, 0) {$\gamma_{\ell,g}$};
  \node[latent, right=of gamma_g] (s_ell) {$s_\ell$};
  \node[lab, above=1mm of gamma_g] {Global Mean};
  \node[lab, above=1mm of s_ell] {Spatial Scale};

  % Latent ICAR Field
  \node[latent, below=1.2cm of gamma_g, xshift=1.5cm] (u_ell) {$\bu_\ell$};
  \node[lab, right=2mm of u_ell] {$\bu_\ell \sim \mathrm{ICAR}(\boldsymbol{A})$};

  % Decomposition Relation
  \node[det, below=1cm of u_ell] (decomp) {$\boldsymbol{\gamma}_{\ell} = \gamma_{\ell,g}\ind + s_\ell \bu_\ell$ \\ \small (Spatial Coeff. Vector)};

  \node[det, below=1.2cm of decomp] (coeffs) {
    $\mu_{j,\svec}, \sigma_{j,\svec}, \xi_{\svec}$ \\ 
    \small (Elements of $\boldsymbol{\gamma}_\ell$ at site $\svec$)
  };

  % Final Predictors
  \node[det, below=1.2cm of coeffs] (gev_params) {
    $\mu_{\svec}(t) = \sum_j c_j(t) \mu_{j,\svec}$ \\ 
    $\log\sigma_{\svec}(t) = \sum_j c_j(t) \sigma_{j,\svec}$\\
    $\xi_{\svec}=\xi_{\svec}$
  };

  % =========================
  % Arrows
  % =========================
  \draw[arrow] (gamma_g) -- (decomp);
  \draw[arrow] (s_ell) -- (decomp);
  \draw[arrow] (u_ell) -- (decomp);
  
  \draw[arrow] (decomp) -- (coeffs);
  \draw[arrow] (coeffs) -- (gev_params);

  % Interaction with covariates
  \draw[arrow] (c_t) -- (gev_params);
  
  % Final Likelihood
  \draw[arrow] (gev_params) -- node[above, lab, pos=0.3, sloped] {GEV Likelihood} (M_st);

\end{tikzpicture}
\caption{Schematic representation of the spatial GEV framework. Spatial dependence is modeled via ICAR priors on latent fields $\mathbf{u}_\ell$, forming spatially varying coefficients $\boldsymbol{\gamma}_\ell$. These coefficients are combined with global covariates $\mathbf{c}(t)$ through linear predictors to define the time-dependent GEV distribution at each location $\boldsymbol{s}$.}
\label{fig:gev_diagram}
\end{figure}
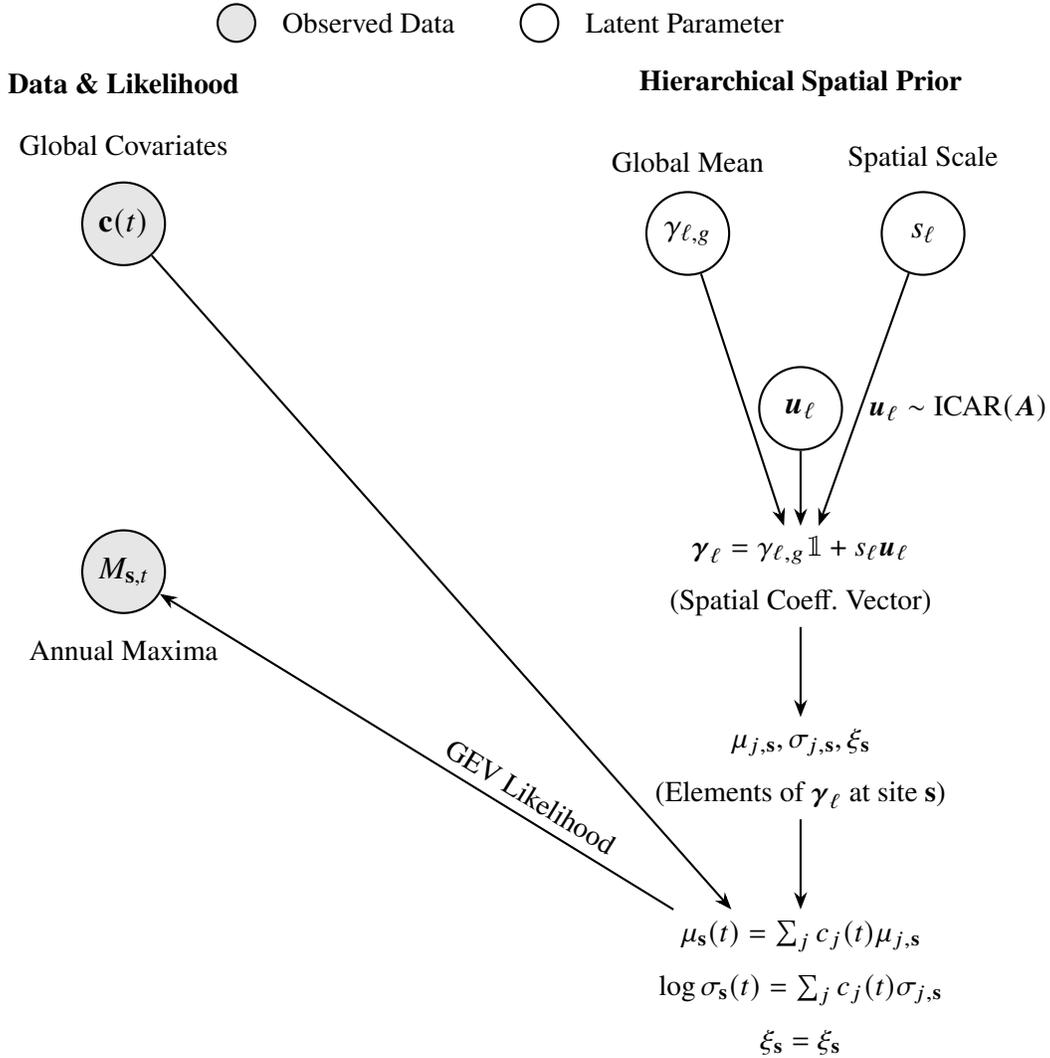

This section provides a schematic overview of the hierarchical Bayesian GEV model described in the main text. Figure~\ref{fig:gev_diagram} summarizes the full model structure, linking observed annual maxima $M_{\svec,t}$ and global covariates $\covariates(t)$ to the GEV parameters. Spatial variability in the GEV coefficients is introduced through the decomposition $\boldsymbol{\gamma}_\ell = \gamma_{\ell,g}\ind + s_\ell \bu_\ell$, where $\bu_\ell$ follows an ICAR prior. These spatially varying coefficients are then combined with covariates to define the location, scale, and shape parameters of the GEV distribution at each site.

\subsection{Marginal distribution of $R^*(\svec,d)$}\label{sec:supp_storm_cdf}
For a fixed location $\svec$ and day $d$, the process $R^*(\svec,d)$ is the maximum over the contributions of the $K$ episodes. Let
\[
H_k(\svec,d)
=
A_k
\exp\!\left(
-\frac{\|\Anisomat(\svec-\boldsymbol{C}_k)\|}{\rho_s}
-\frac{|d-T_k|}{\rho_t}
\right)
\]
denote the contribution of episode $k$ at $(\svec,d)$, so that
\[
R^*(\svec,d)=\max_{k=1,\dots,K} H_k(\svec,d).
\]
Let $F_{\svec,d}$ denote the distribution function of $R^*(\svec,d)$. 
Conditionally on $K$, the variables $\{H_k(\svec,d)\}_{k=1}^K$ are independent and identically distributed, which yields
\[
\mathbb P\!\left(R^*(\svec,d)\le x \mid K\right)
=
\mathbb P\!\left(H_1(\svec,d)\le x\right)^K .
\]
Since $K \sim \mathrm{Poisson}(\lambda)$, taking expectation with respect to $K$ gives
\[
F_{\svec,d}(x)
=
\mathbb P\!\left(R^*(\svec,d)\le x\right)
=
\exp\!\bigl\{-\lambda\bigl(1-p_{\svec,d}(x)\bigr)\bigr\},
\]
where
\[
p_{\svec,d}(x)
=
\mathbb P\!\left(H_1(\svec,d)\le x\right).
\]
Using the uniform distribution of $(\boldsymbol{C}_1,T_1)$ over $\{\svec_1,\dots,\svec_n\}\times\{1,\dots,D\}$, we obtain
\[
p_{\svec,d}(x)
=
\frac{1}{nD}
\sum_{i=1}^{n}\sum_{t=1}^{D}
\mathbb P\!\left(
A_1
\exp\!\left(
-\frac{\|\Anisomat(\svec-\svec_i)\|}{\rho_s}
-\frac{|d-t|}{\rho_t}
\right)
\le x
\right).
\]
Because $A_1 \sim \mathrm{Pareto}(1)$, the inner probability can be written explicitly as
\[
\mathbb P(A_1 w \le x)
=
\left(1-\frac{w}{x}\right)_+,
\qquad
w=
\exp\!\left(
-\frac{\|\Anisomat(\svec-\svec_i)\|}{\rho_s}
-\frac{|d-t|}{\rho_t}
\right),
\]
where $(u)_+=\max(u,0)$. Consequently,
\[
p_{\svec,d}(x)
=
\frac{1}{nD}
\sum_{i=1}^{n}\sum_{t=1}^{D}
\left(
1-
\frac{
\exp\!\left(
-\frac{\|\Anisomat(\svec-\svec_i)\|}{\rho_s}
-\frac{|d-t|}{\rho_t}
\right)
}{x}
\right)_+ .
\]

\subsection{Chi grid}\label{sec:chi grid}
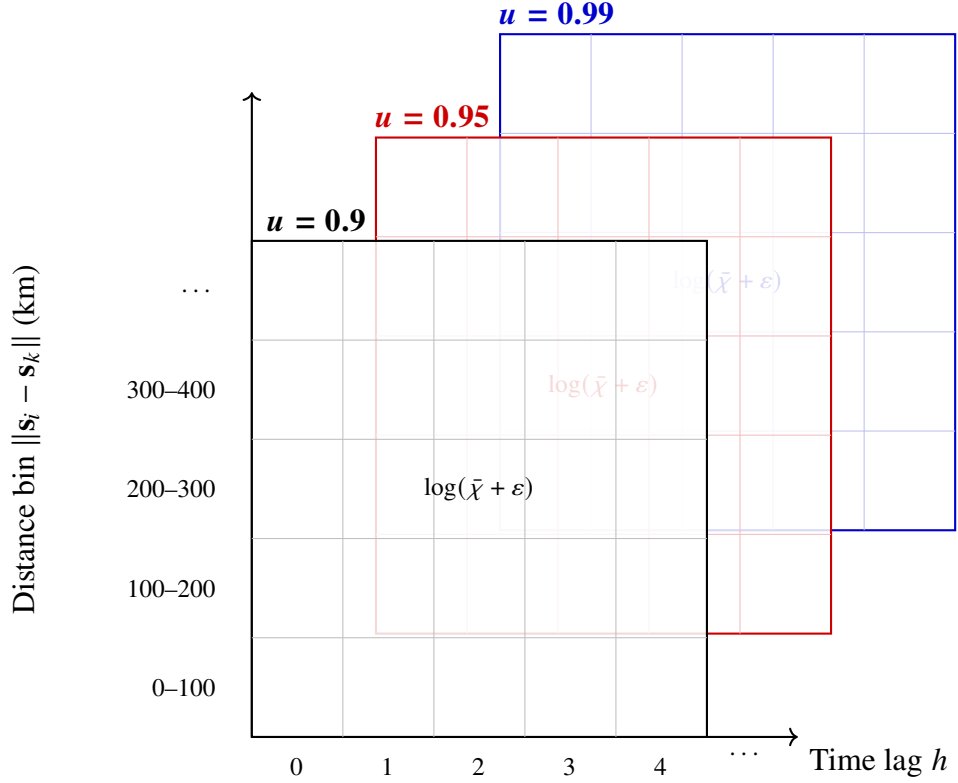
\begin{figure}[H]
    \centering
    \begin{tikzpicture}[x={(1.1cm,0cm)}, y={(0.6cm,0.5cm)}, z={(0cm,1.2cm)}, scale=1.1]

        % Extended Z-Axis (Vertical)
        \draw[thick,->] (0,0,0) -- (0,0,6.5);
        \node[rotate=90, anchor=south] at (-2.2,0,3) {Distance bin $\|\svec_i - \svec_k\|$ (km)};

        % X-Axis (Horizontal)
        \draw[thick,->] (0,0,0) -- (6,0,0) node[anchor=north west] {Time lag $h$};

        % Function to draw a grid plane
        \newcommand{\drawgridplane}[3]{ % #1: y-offset (depth), #2: color, #3: u-label
            \begin{scope}[shift={(0,#1,0)}]
                % Fill for visibility
                \fill[white, opacity=0.85] (0,0,0) -- (5,0,0) -- (5,0,5) -- (0,0,5) -- cycle;
                % Border
                \draw[thick, #2] (0,0,0) -- (5,0,0) -- (5,0,5) -- (0,0,5) -- cycle;
                % Internal Grid Lines
                \foreach \i in {1,...,4} {
                    \draw[#2!25, thin] (\i,0,0) -- (\i,0,5); 
                    \draw[#2!25, thin] (0,0,\i) -- (5,0,\i); 
                }
                % The internal formula (universal)
                \node[#2, scale=0.8] at (2.5,0,2.5) {$\log(\bar{\chi} + \varepsilon)$};
                % The external Threshold label on the left edge
                \node[anchor=east, #2, font=\boldmath] at (1.1,0.5,5) {$u = #3$};
            \end{scope}
        }
        % Draw planes from back to front
        \drawgridplane{5}{blue!80!black}{0.99} 
        \drawgridplane{2.5}{red!80!black}{0.95} 
        \drawgridplane{0}{black}{0.9}          

        % Ticks for Time Lag h (0 to 4)
        \foreach \h in {0,1,2,3,4}
            \draw  (\h,0.9,-0.5) node[below, scale=0.8] {\h};
        
        % Manual dots for the end of the axis
        \node[below, scale=0.8] at (5,0.8,-0.4) {\dots};
        
        % Ticks for Distance Bins (Clearly separated from the axis title)
        \node[anchor=east, scale=0.8] at (-0.3,0,0.5) {0--100};
        \node[anchor=east, scale=0.8] at (-0.3,0,1.5) {100--200};
        \node[anchor=east, scale=0.8] at (-0.3,0,2.5) {200--300};
        \node[anchor=east, scale=0.8] at (-0.3,0,3.5) {300--400};
        \node[anchor=east, scale=0.8] at (-0.3,0,4.5) {\dots};

    \end{tikzpicture}
\caption{
Visual representation of the estimated tail dependence grids $\log(\bar{\chi} + \varepsilon)$ for thresholds $u \in \{0.9, 0.95, 0.99\}$. 
Each grid layer represents a two-dimensional matrix where rows correspond to spatial distance bins and columns to discrete temporal lags $h$. 
Each cell represents the bin-averaged tail dependence estimate $\log(\bar{\chi} + \varepsilon)$ for a specific distance-lag combination.
\label{fig:tail_dependence_perspective}
}
\end{figure}

This section illustrates one of the summary statistics used as input to the neural network. Figure~\ref{fig:tail_dependence_perspective} shows the tail dependence grids based on the transformed statistic $\log(\bar{\chi} + \varepsilon)$, computed across spatial distance bins and temporal lags for different thresholds $u$. Each grid encodes the dependence structure of extremes over space and time, and serves as a compact representation of extremal dependence patterns used for model training.

\subsection{Back-transformation}\label{sec:back_transform_deriv}

This section provides the detailed derivation of the marginal back-transformation presented in Section~3.4 of the main document. 
We assume throughout that the shape parameter $\xi_{\svec}$ is constant in time (i.e., does not depend on $t$).  Furthermore, we assume that the temporal process $\{X(\svec,d,t)\}$ admits an extremal index $\extremalI_{\svec} \in (0,1]$, characterizing the clustering behavior of extremes. In addition, we impose Leadbetter’s $D(u_n)$ and $D'(u_n)$ mixing conditions to control the dependence structure of exceedances and ensure the validity of the asymptotic arguments for block maxima.
Starting from the standard tail decomposition, for a large value $x(\svec,d,t)$ exceeding a high marginal threshold $\Tthres$ of $X(\svec,d,t)$, we decompose
$$
\prob(X(\svec,d,t) > x(\svec,d,t)) 
= 
\prob\left(X(\svec,d,t) > \Tthres\right) 
\, 
\prob\left(X(\svec,d,t) > x(\svec,d,t) \mid X(\svec,d,t) > \Tthres\right).
$$
Let $M_{\svec,t} = \max_{d=1,\ldots,D} X(\svec,d,t)$ denote the annual maximum. 
Under the above mixing conditions, the series $(M_{\svec,t})_t$ has an extremal index $\extremalI_{\svec}$ and converges in distribution to a GEV distribution with parameters $(\mu_{\svec,t},\sigma_{\svec,t},\xi_{\svec})$:
$$
G_{\svec,t}(x) =
\begin{dcases}
\exp\!\Big[-\big(1+\xi_{\svec}\tfrac{x-\mu_{\svec,t}}{\sigma_{\svec,t}}\big)^{-1/\xi_{\svec}}\Big], 
& \xi_{\svec} \neq 0,\quad 1+\xi_{\svec}\tfrac{x-\mu_{\svec,t}}{\sigma_{\svec,t}} > 0, \\
\exp\!\Big[-\exp\!\big(-\tfrac{x-\mu_{\svec,t}}{\sigma_{\svec,t}}\big)\Big], 
& \xi_{\svec} = 0.
\end{dcases}
$$
Since $M_{\svec,t}$ is the maximum of $D$ dependent replicates, 
$$
\prob(M_{\svec,t} \le x(\svec,d,t)) 
= F_{\svec,t}(x(\svec,d,t))^{\extremalI_{\svec}D} 
\approx 
\exp\!\Big[-\extremalI_{\svec}D\{1-F_{\svec,t}(x(\svec,d,t))\}\Big],
$$
where the last approximation uses the first-order tail expansion $\log(1-y)\approx -y$ valid when $y=1-F_{\svec,t}(x) \to 0$.
Using the GEV limit relation 
$$
F_{\svec,t}(x)^{\extremalI_{\svec} D} \approx G_{\svec,t}(x),
$$
we obtain the tail approximation
$$
1 - F_{\svec,t}(x(\svec,d,t)) 
\approx 
\frac{-\log G_{\svec,t}(x(\svec,d,t))}{\extremalI_{\svec}D}
=
\begin{dcases}
\dfrac{\big(1+\xi_{\svec}\tfrac{x(\svec,d,t)-\mu_{\svec,t}}{\sigma_{\svec,t}}\big)^{-1/\xi_{\svec}}}{\extremalI_{\svec}D}, 
& \xi_{\svec} \neq 0, \\
\dfrac{\exp\!\big(-\tfrac{x(\svec,d,t)-\mu_{\svec,t}}{\sigma_{\svec,t}}\big)}{\extremalI_{\svec}D}, 
& \xi_{\svec} = 0.
\end{dcases}
$$
Hence, the conditional tail of the parent distribution is
\begin{align*}
\prob \left( X(\svec,d,t) > x(\svec,d,t) \mid X(\svec,d,t) > \Tthres \right)
&= 
\frac{1-F_{\svec,t}(x(\svec,d,t))}{1-F_{\svec,t}(\Tthres)} \\
&\approx
\begin{dcases}
\left(
\dfrac{
1+\xi_{\svec}\tfrac{x(\svec,d,t)-\mu_{\svec,t}}{\sigma_{\svec,t}}
}{
1+\xi_{\svec}\tfrac{\Tthres-\mu_{\svec,t}}{\sigma_{\svec,t}}
}
\right)^{-1/\xi_{\svec}}, 
& \xi_{\svec} \neq 0, \\
\exp\left(-\dfrac{x(\svec,d,t)-\Tthres}{\sigma_{\svec,t}}\right), 
& \xi_{\svec} = 0.
\end{dcases}
\end{align*}
Introducing the generalized Pareto scale parameter
$$
\tilde{\sigma}_{\svec,t} = \sigma_{\svec,t} + \xi_{\svec}(\Tthres - \mu_{\svec,t}),
$$
the conditional tail takes the standard GPD form:
$$
\prob \left( X(\svec,d,t) > x(\svec,d,t) \mid X(\svec,d,t) > \Tthres \right)
\approx
\begin{dcases}
\left( 1 + \xi_{\svec} \dfrac{x(\svec,d,t)-\Tthres}{\tilde{\sigma}_{\svec,t}} \right)^{-1/\xi_{\svec}}, 
& \xi_{\svec} \neq 0, \\[1mm]
\exp\left(-\frac{x(\svec,d,t)-\Tthres}{\tilde{\sigma}_{\svec,t}}\right), 
& \xi_{\svec} = 0.
\end{dcases}
$$
For a given year $t$ and simulated uniform value $u(\svec,d)\in[\tau_{\svec,t},1)$, where 
$$
\tau_{\svec,t}=q^{-1}_{\svec,t}(\Tthres),
$$
the inverse transformation is obtained by solving
$$
1-u(\svec,d)
= (1-\tau_{\svec,t}) \cdot
\begin{dcases}
\left( 1 + \dfrac{\xi_{\svec}(x(\svec,d,t)-\Tthres)}{\tilde{\sigma}_{\svec,t}} \right)^{-1/\xi_{\svec}}, 
& \xi_{\svec} \neq 0, \\[1mm]
\exp\left(-\frac{x(\svec,d,t)-\Tthres}{\tilde{\sigma}_{\svec,t}}\right), 
& \xi_{\svec} = 0.
\end{dcases}
$$
Solving for $x(\svec,d,t)$ gives the explicit back-transformation:
$$
x(\svec,d,t) =
\begin{dcases} 
\Tthres + \dfrac{\tilde{\sigma}_{\svec,t}}{\xi_{\svec}}
\left[
\left(\dfrac{1-u(\svec,d)}{1-\tau_{\svec,t}}\right)^{-\xi_{\svec}} - 1
\right], 
& u(\svec,d)\ge\tau_{\svec,t},\ \xi_{\svec}\neq 0, \\[1mm]
\Tthres - \tilde{\sigma}_{\svec,t}
\log\!\left(\dfrac{1-u(\svec,d)}{1-\tau_{\svec,t}}\right),
& u(\svec,d)\ge\tau_{\svec,t},\ \xi_{\svec}=0.
\end{dcases}
$$

\clearpage

\section{Supplementary Data Application}
\label{sec:supp_application}
\subsection{Clustering for European heatwaves}\label{sec:Clustering}

To this end, we propose a clustering strategy that identifies spatial subregions
exhibiting similar patterns of concomitant extremes, in the spirit of
\textcite{kiriliouk2020climate} and related approaches
\parencite{bernard2013clustering}. We compute clusters using only the
counterfactual climate simulations, which are assumed to be stationary by
construction.

We first construct pseudo-observations on the uniform scale via
within-location ranks. Denoting by $DT$ the number of available summer days and
by $\operatorname{rank}\{X(\svec,d,t)\}$ the rank of $X(\svec,d,t)$ among
$\{X(\svec,1,1),\dots,X(\svec,D,T)\}$, we define
\[
    \widehat{U}(\svec,d,t) 
    = \frac{\operatorname{rank}\{X(\svec,d,t)\}}{DT},
    \qquad d=1,\dots,D, \; t=1,\dots,T,
\]
so that $\widehat{U}(\svec,d,t)$ has approximately uniform margins on $[0,1]$.

Extremal dependence between two locations $\svec_i$ and $\svec_k$ is then quantified
through a high-threshold tail dependence coefficient at level $u\in(0,1)$:
\[
\taildepest{\svec_i}{\svec_k}{u}
=
\frac{1}{DT(1-u)}
\sum_{t=1}^{T}\sum_{d=1}^{D
}
\ind\!\bigl\{\widehat{U}(\svec_i,d,t)>u,\;\widehat{U}(\svec_k,d,t)>u\bigr\}.
\]
This quantity is an empirical analogue of the conditional exceedance probability
used in \textcite{kiriliouk2015nonparametric} and yields a simple dependence
measure based on simultaneous extremes at identical time instances. Based on $\taildepest{\svec_i}{\svec_k}{u}$, we follow \textcite{kiriliouk2020climate} and define the
dissimilarity
\begin{equation}
\label{eq:dissimilarity}
d(\svec_i,\svec_k\mid u)
=
\frac{1-\taildep{\svec_i}{\svec_k}{u}}{2\bigl(3-\taildep{\svec_i}{\svec_k}{u}\bigr)},
\end{equation}
which is symmetric, equals zero when $\svec_i=\svec_k$, and increases as extremal
dependence weakens.

Given the resulting pairwise dissimilarity matrix, we obtain a
partition of the spatial domain using a $k$-medoids algorithm. In particular, we
consider partitioning around medoids and its scalable variants, and select
the number of clusters $k$ using a silhouette criterion computed from the
precomputed dissimilarities. To reduce sensitivity to initialization, the
silhouette criterion can be evaluated over multiple random restarts, and a stable
choice of $k$ is retained. The final output is a cluster label for each land grid
cell and a set of representative medoid locations.

In the application, we fix the exceedance threshold at $u = 0.8$. This choice is
necessarily somewhat arbitrary and reflects a classical bias-variance
trade-off. Based on this choice, the clustering procedure selects $k = 8$ spatial clusters.
The resulting partition of Europe into regions with similar extremal dependence
properties is displayed in Figure~\ref{fig:eu_clim_clusters}. The clusters we obtained are highly similar to those obtained by \textcite{stefanon2012heatwave} and those used by \textcite{auld2023changes} in his analysis of annual maximum temperatures in Europe.

In the remainder of this work, subsequent analyses focus on Cluster~3.
This cluster predominantly covers Western Europe and parts of Central Europe,
including regions such as France, the Benelux countries, and neighboring areas.

\subsection{Diagnostic of the BSQR Uniform Transformation}
\label{sec:bsqr-uniform-diagnostic}
\begin{figure}[H]
  \centering
  \begin{subfigure}[t]{0.49\textwidth}
    \centering
    \includegraphics[width=\textwidth]{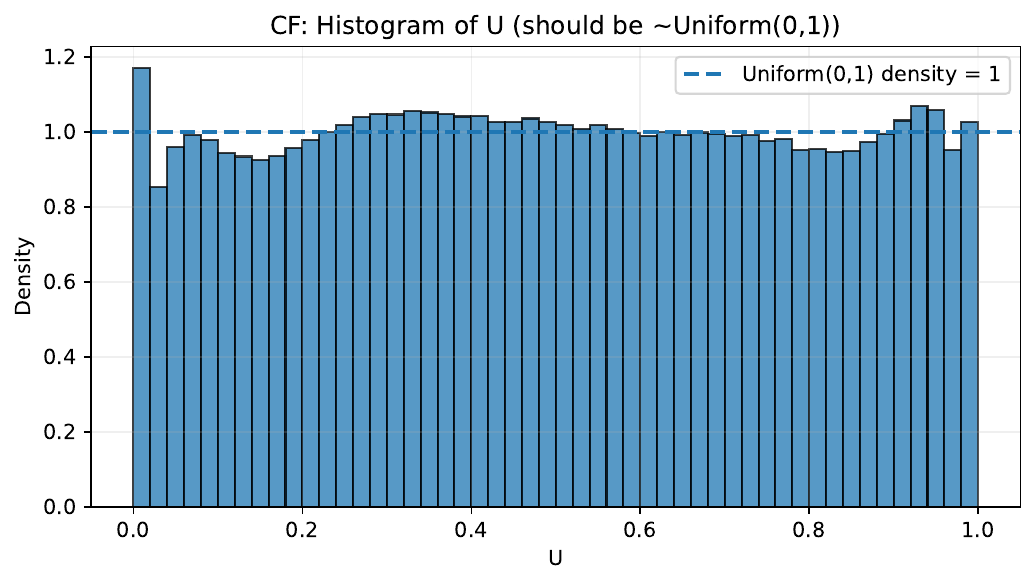}
    \caption{Counterfactual (CF).}
    \label{fig:uniform_check_cf_c3}
  \end{subfigure}\hfill
  \begin{subfigure}[t]{0.49\textwidth}
    \centering
    \includegraphics[width=\textwidth]{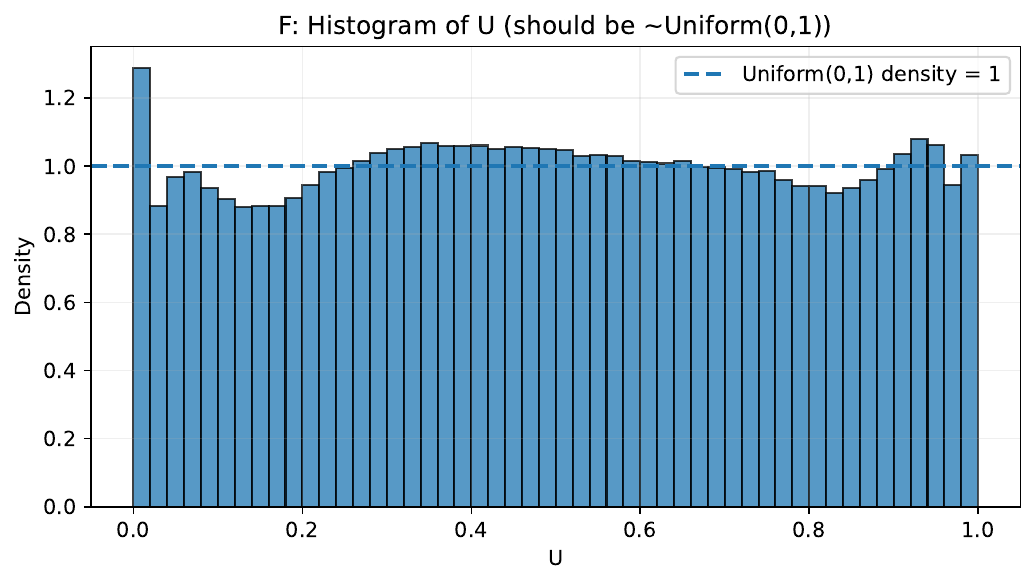}
    \caption{Factual (F).}
    \label{fig:uniform_check_f_c3}
  \end{subfigure}
  \caption{
    Diagnostic check of the BSQR uniformization step for Cluster~3.
    Histograms of $\widehat{U} (\svec,d,t)=\hat F_{\svec,t}(X(\svec,d,t))$ (months JJA) should be
    approximately uniform on $[0,1]$ if the marginal nonstationarity is
    adequately removed. The fit is generally satisfactory, with remaining
    discrepancies most visible in the tails.
  }
  \label{fig:uniform_check_cluster3}
\end{figure}

This section provides a diagnostic check of the marginal transformation used in the BSQR framework. Figure~\ref{fig:uniform_check_cluster3} displays histograms of the transformed variables $\widehat{U}(\svec,d,t)=\hat F_{\svec,t}(X(\svec,d,t))$ for both counterfactual and factual settings. Under a correct specification of the marginal model, these transformed values should follow a uniform distribution on $[0,1]$. The figure confirms that the transformation performs adequately overall, with minor deviations primarily observed in the tails.

\subsection{MCMC convergence diagnostics}\label{sec:supp_gev_mcmc}
\begin{figure}[H]
  \centering
  \captionsetup[sub]{font=footnotesize, skip=1pt}

  \begin{subfigure}[t]{0.48\textwidth}
    \centering
    \includegraphics[width=\textwidth]{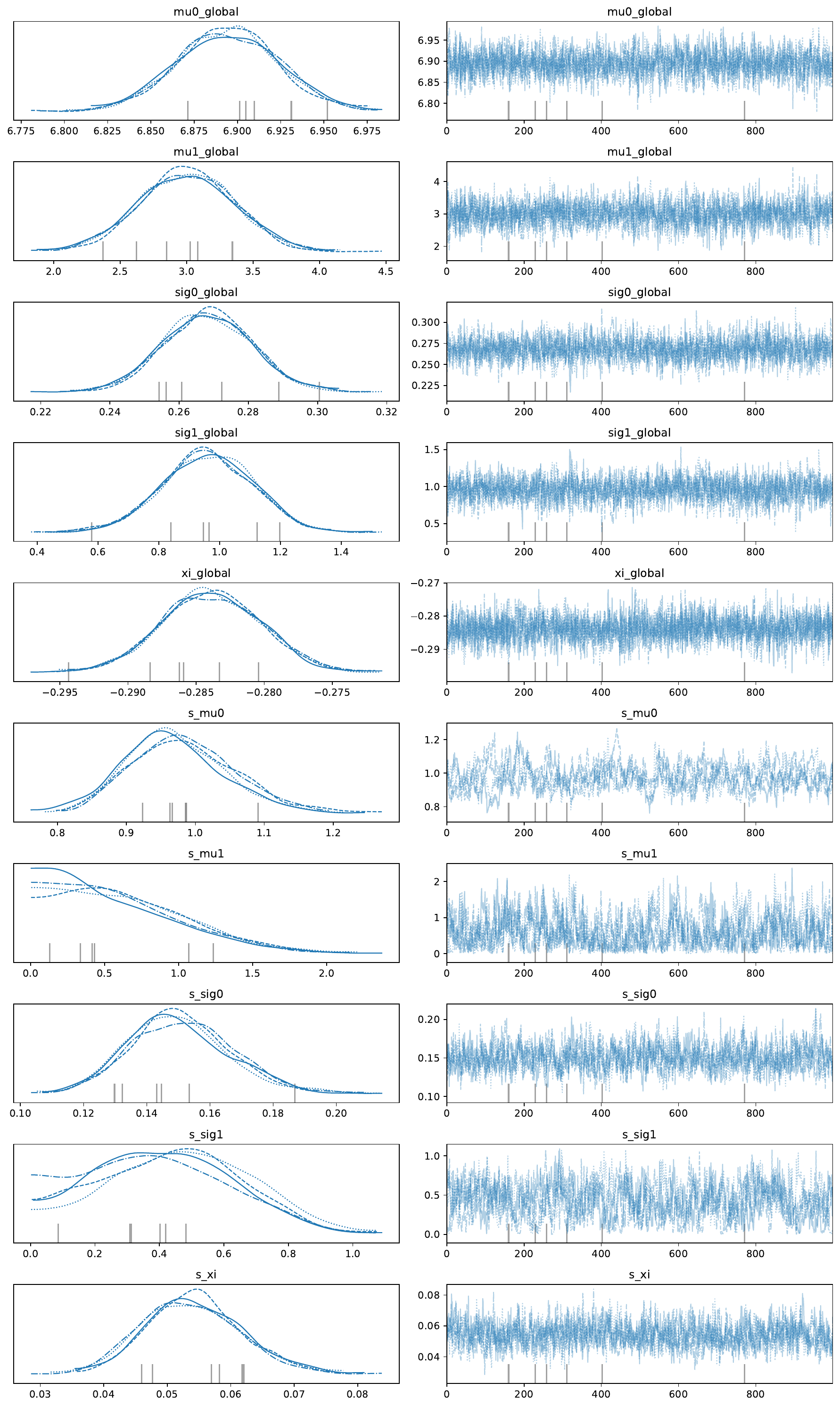}
    \caption{Counterfactual (CF) climate.}
    \label{fig:trace_gev_cf}
  \end{subfigure}\hfill
  \begin{subfigure}[t]{0.48\textwidth}
    \centering
    \includegraphics[width=\textwidth]{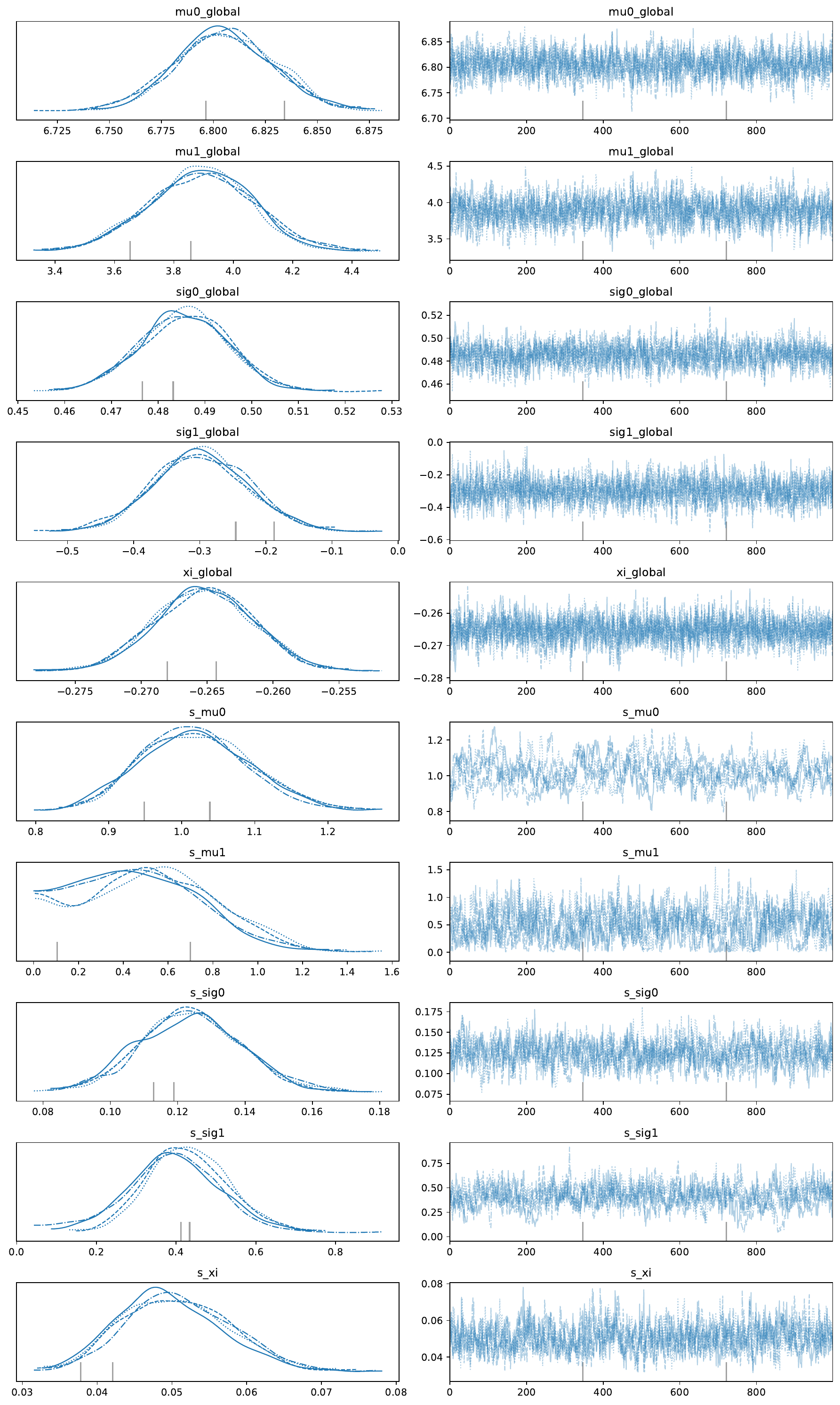}
    \caption{Factual (F) climate.}
    \label{fig:trace_gev_f}
  \end{subfigure}

  \caption{
    Trace plots for selected global and spatial hyperparameters of the GEV
    model under the counterfactual (CF) and factual (F) climates. The chains show
    satisfactory mixing and no apparent trends, supporting convergence of the
    Markov chain Monte Carlo algorithm.
  }
  \label{fig:gev_traceplots}
\end{figure}

This section presents convergence diagnostics for the Bayesian GEV model. Figure~\ref{fig:gev_traceplots} shows trace plots of selected global and spatial hyperparameters under both counterfactual and factual climates. The chains exhibit good mixing and no visible trends, indicating satisfactory convergence of the Markov chain Monte Carlo algorithm.

\subsection{Posterior predictive checks}\label{sec:supp_gev_ppc}

We assess the adequacy of the fitted nonstationary spatial GEV model using a
posterior predictive Q-Q diagnostic based on standardized residuals. For each
grid cell $\svec_i$ and year $t$, recall that
$$
M_{\svec_i,t} \sim \mathrm{GEV}\bigl(\mu_{\svec_i}(t),\sigma_{\svec_i}(t),\xi_{\svec_i}\bigr),
\qquad
\mu_{\svec_i}(t)=\covariates(t)^\top\boldsymbol{\mu}_{\svec_i},
\qquad
\log\sigma_{\svec_i}(t)=\covariates(t)^\top\boldsymbol{\sigma}_{\svec_i},
$$
with $\xi_{\svec_i}$ constant over time. Let
$$
V_{\svec_i,t}
=
\begin{dcases}
\dfrac{1}{\xi_{\svec_i}}
\log\!\left(1+\xi_{\svec_i}\,\dfrac{M_{\svec_i,t}-\mu_{\svec_i}(t)}{\sigma_{\svec_i}(t)}\right),
& \xi_{\svec_i}\neq 0,\\[10pt]
\dfrac{M_{\svec_i,t}-\mu_{\svec_i}(t)}{\sigma_{\svec_i}(t)},
& \xi_{\svec_i}=0,
\end{dcases}
$$
denote the GEV residual transformation, which maps
$M_{\svec_i,t}$ to a scale where the theoretical quantiles are those of a Gumbel
distribution. We pool these transformed residuals over all available
$(\svec_i,t)$ within Cluster~3 and compare the empirical residual quantiles to
theoretical Gumbel quantiles
$$
t_q = -\log\!\{-\log(p_q)\}, \qquad p_q = \frac{q}{N+1}, \quad q=1,\dots,N,
$$
where $N=n_c*T$ is the number of pooled residuals with $n_c$ the number of grid cells in Cluster~3.

To construct a posterior predictive envelope, we draw parameter samples from the
posterior distribution and, for each draw, simulate replicated maxima
$M^{\mathrm{rep}}_{\svec_i,t}$ from the corresponding GEV model. The replicated
data are transformed into residuals using the same mapping, yielding a
replicated Q-Q curve. Repeating this procedure produces an ensemble of
replicated Q-Q curves, from which we compute a pointwise 95\% posterior predictive
envelope and the posterior median replicated curve.

\begin{figure}[!htbp]
  \centering
  \captionsetup[sub]{font=footnotesize, skip=1pt}

  \begin{subfigure}[t]{0.48\textwidth}
    \centering
    \includegraphics[width=\textwidth]{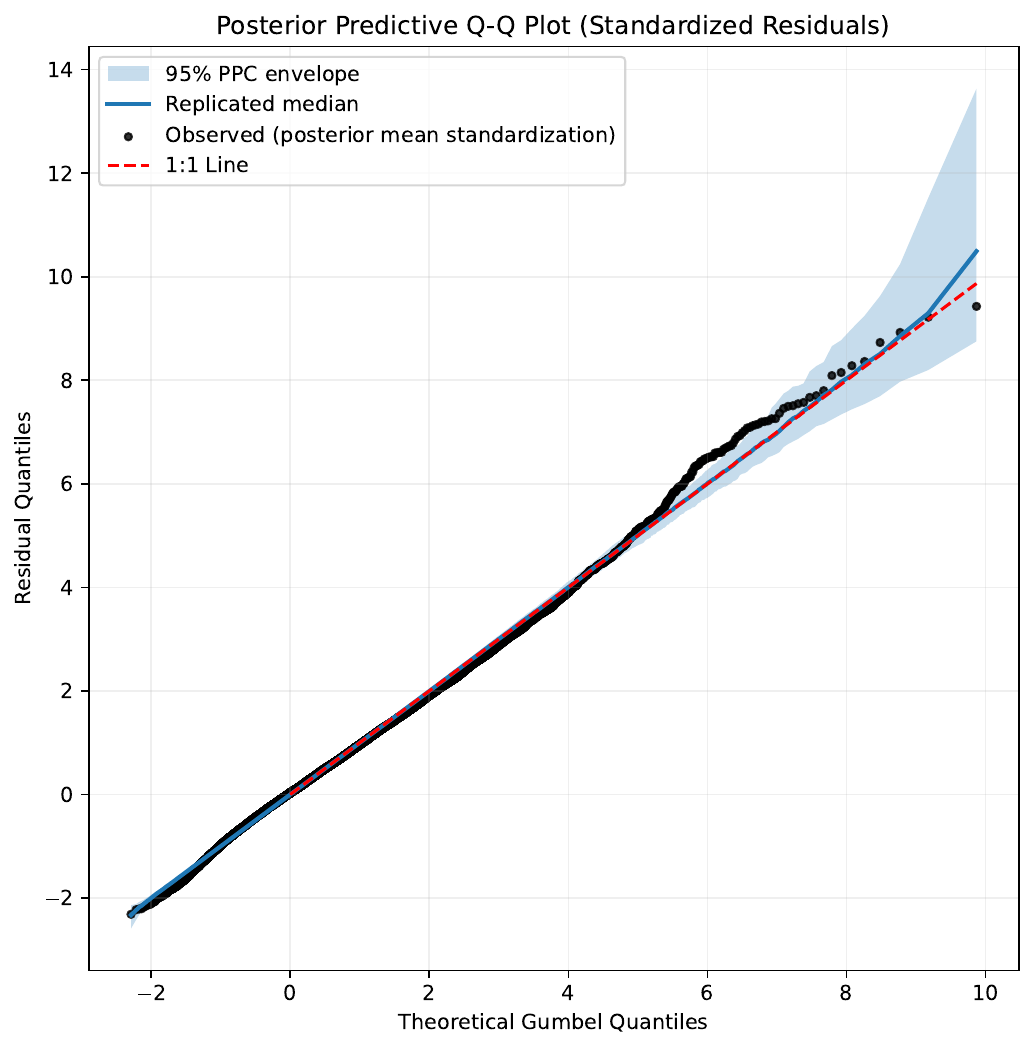}
    \caption{Counterfactual (CF) climate.}
    \label{fig:qq_ppc_gev_cf}
  \end{subfigure}\hfill
  \begin{subfigure}[t]{0.48\textwidth}
    \centering
    \includegraphics[width=\textwidth]{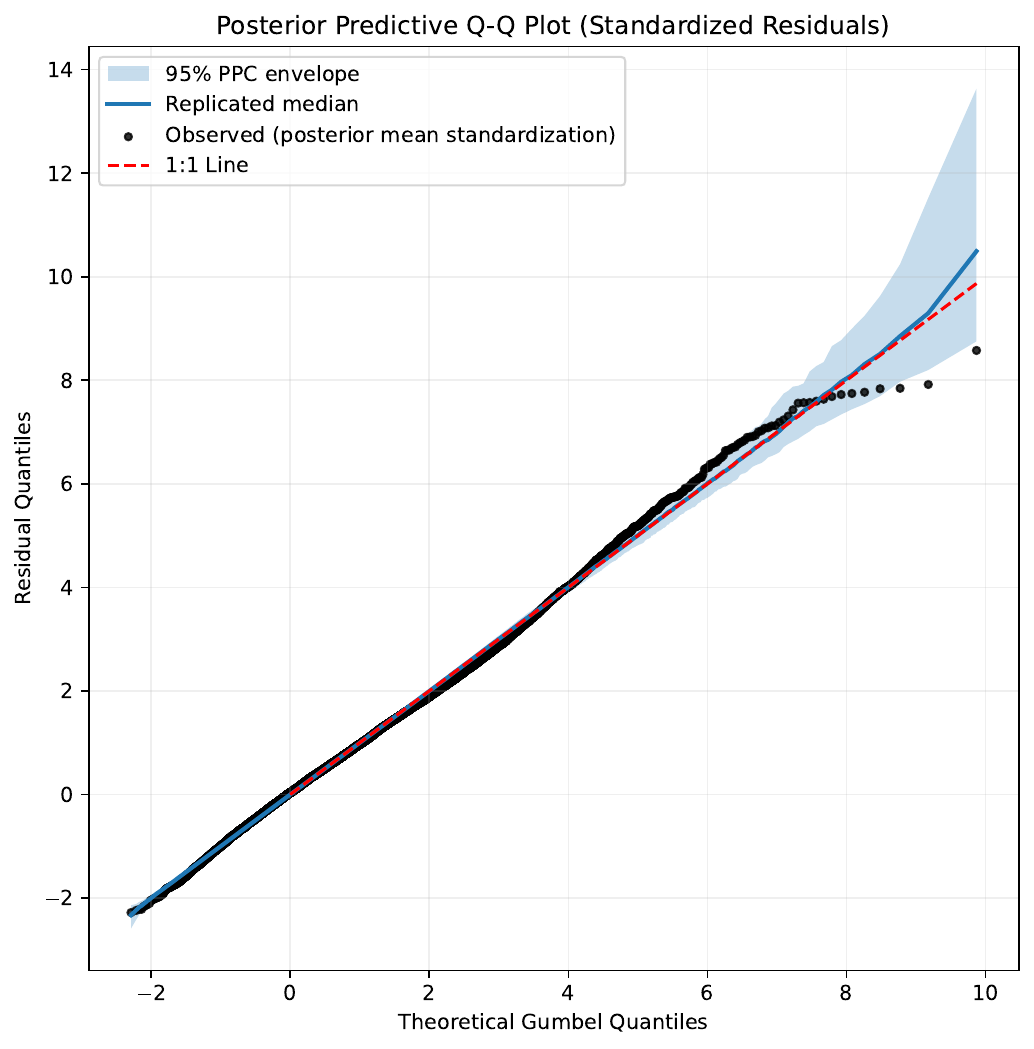}
    \caption{Factual (F) climate.}
    \label{fig:qq_ppc_gev_f}
  \end{subfigure}

  \caption{
    Posterior predictive Q-Q diagnostics for the fitted GEV model based on
    standardized residuals pooled over Cluster~3. The black curve corresponds to
    observed residual quantiles (computed using posterior mean parameters), the
    blue curve is the posterior median replicated Q--Q curve, the shaded region
    is the pointwise posterior predictive envelope, and the red dashed line is
    the $1{:}1$ reference.
  }
  \label{fig:qq_ppc_gev}
\end{figure}

In Figure~\ref{fig:qq_ppc_gev}, the black dots is the Q-Q plot of
the observed residuals computed using posterior mean parameter estimates, the
blue solid line is the posterior median replicated Q-Q curve, the
shaded band is the pointwise posterior predictive envelope, and the
red dashed line is the $1{:}1$ reference.

\newpage

\subsection{Spatial posterior summaries for baseline GEV parameters}\label{sec:supp_gev_maps}
\begin{figure}[!htbp]
  \centering
  \captionsetup[sub]{font=footnotesize, skip=2pt}

  \begin{subfigure}[t]{\textwidth}
    \centering
    \includegraphics[height=0.4\textheight]{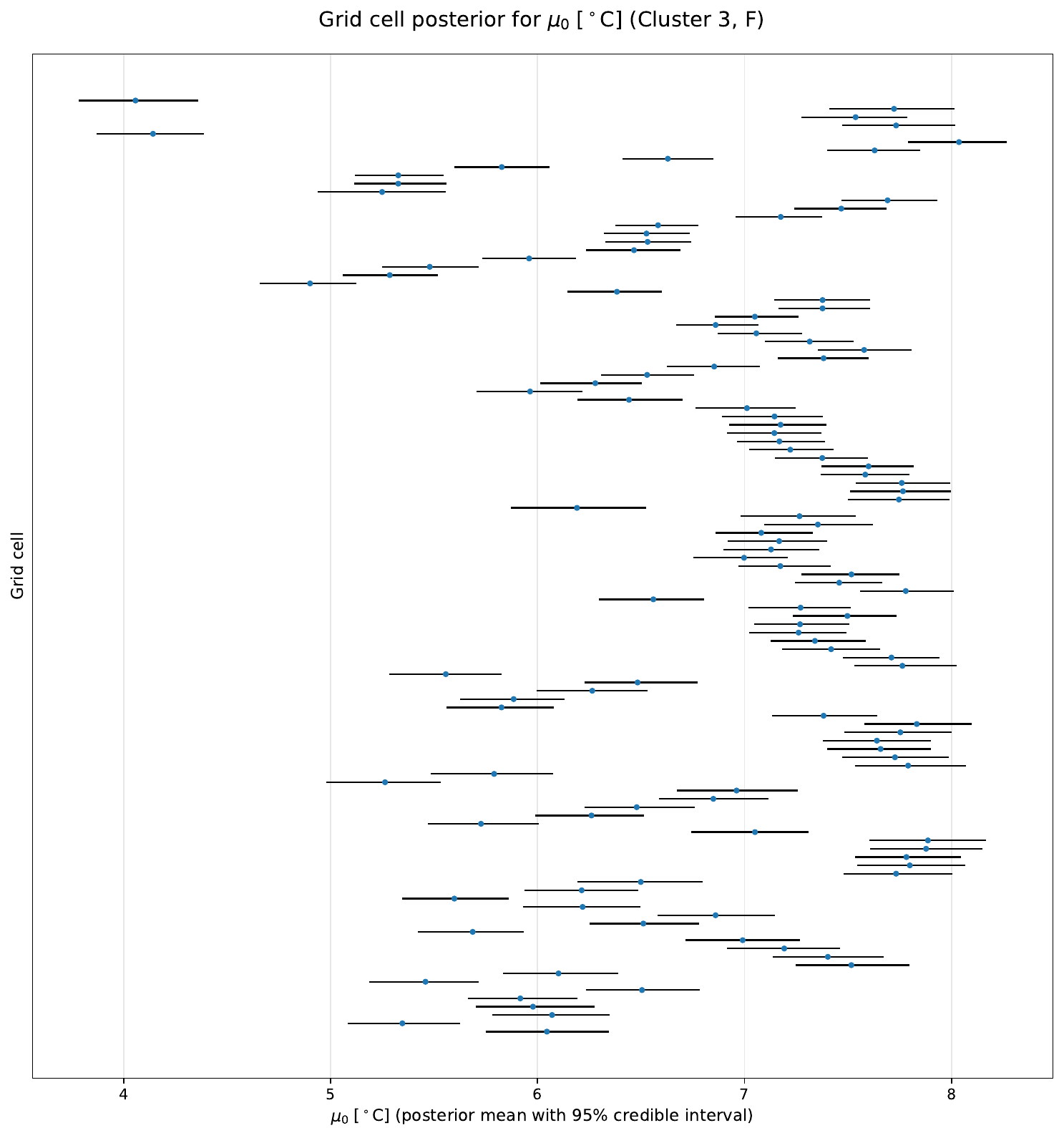}
    \caption{$\mu_0\,[\si{\degreeCelsius}]$}
  \end{subfigure}

  \vspace{0.5em}

  \begin{subfigure}[t]{\textwidth}
    \centering
    \includegraphics[height=0.4\textheight]{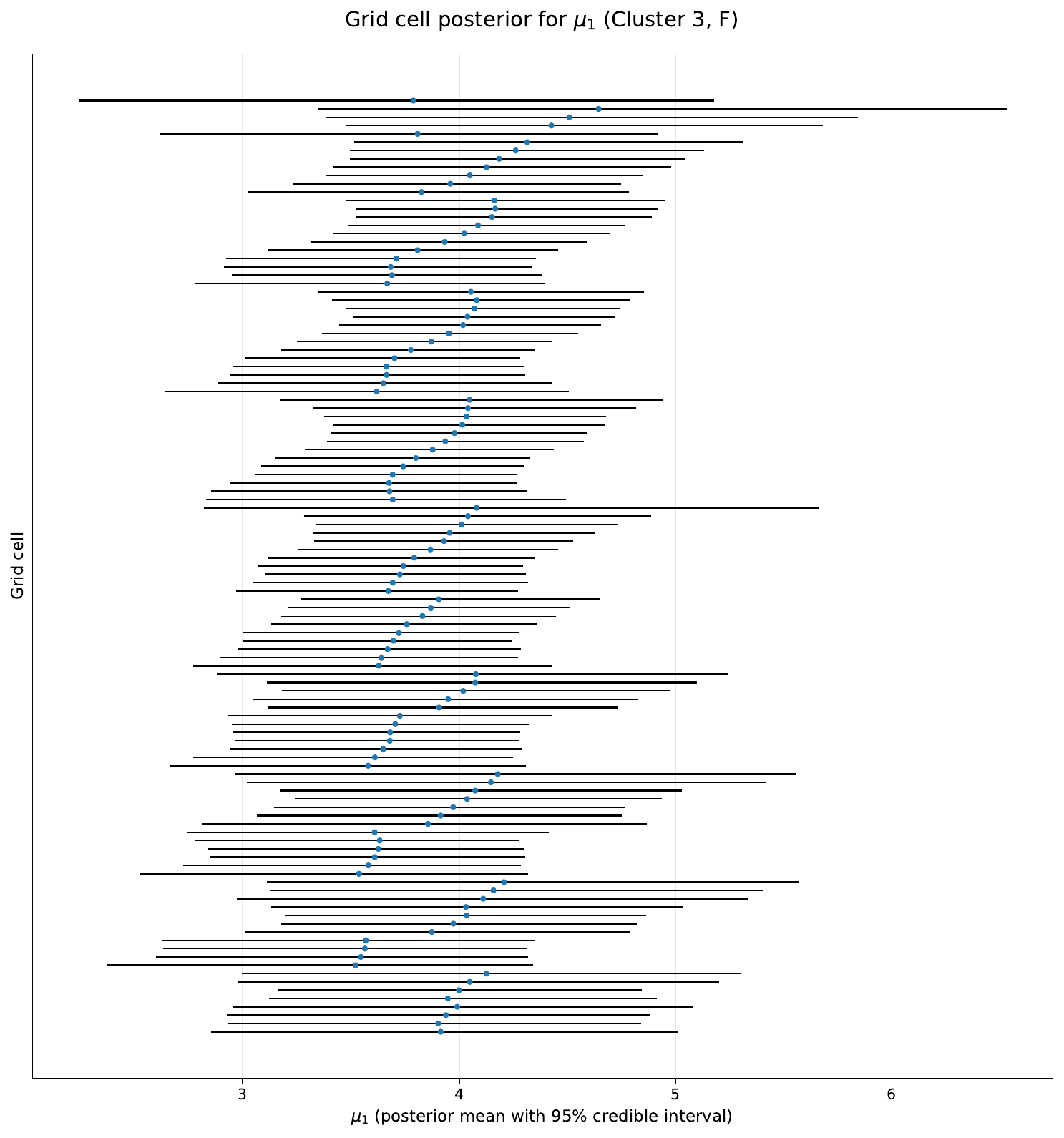}
    \caption{$\mu_1$}
  \end{subfigure}

  \caption{Caterpillar plots of the location parameters for Cluster~3 under the factual climate.}
  \label{fig:gev_caterpillar_mu}
\end{figure}

\begin{figure}[!htbp]
  \centering
  \captionsetup[sub]{font=footnotesize, skip=2pt}

  \begin{subfigure}[t]{\textwidth}
    \centering
    \includegraphics[height=0.4\textheight]{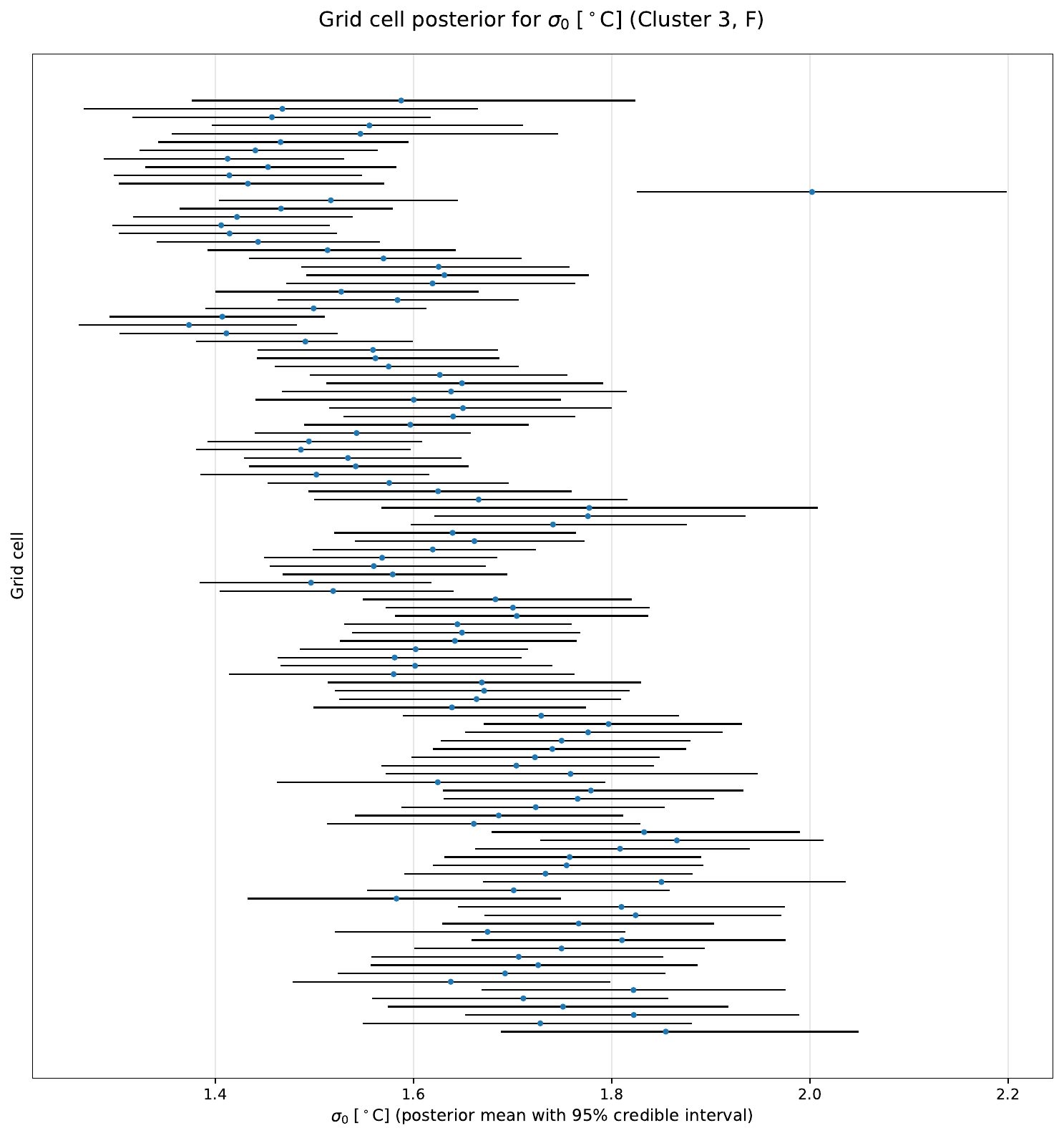}
    \caption{$\sigma_0\,[\si{\degreeCelsius}]$}
  \end{subfigure}

  \vspace{0.8em}

  \begin{subfigure}[t]{\textwidth}
    \centering
    \includegraphics[height=0.4\textheight]{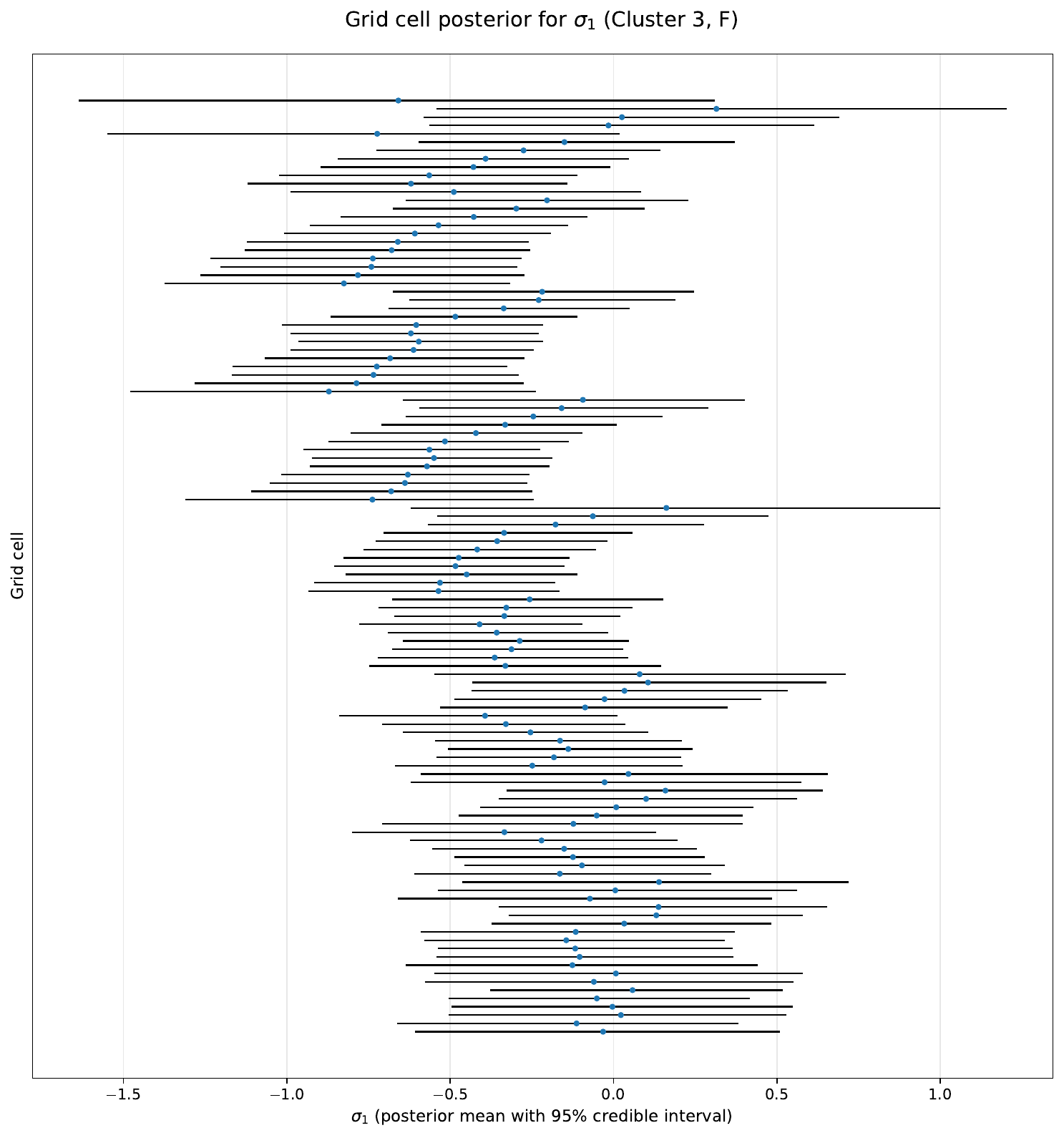}
    \caption{$\sigma_1$}
  \end{subfigure}

  \caption{Caterpillar plots of the scale parameters for Cluster~3 under the factual climate.}
  \label{fig:gev_caterpillar_sigma}
\end{figure}

\begin{figure}[!htbp]
  \centering

  \includegraphics[height=0.4\textheight]{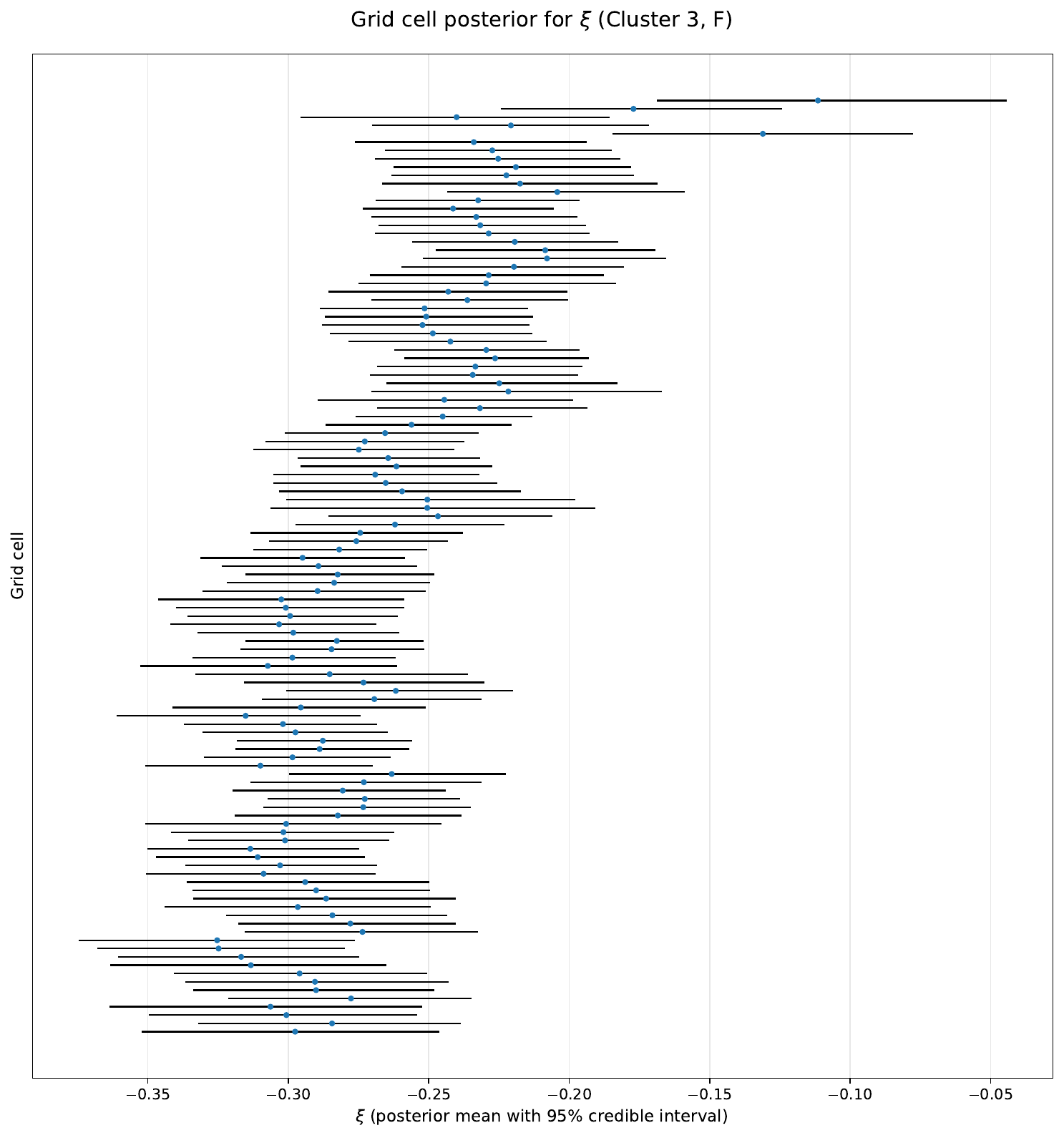}

  \caption{Caterpillar plot of the shape parameter $\xi$ for Cluster~3 under the factual climate.}
  \label{fig:gev_caterpillar_xi}
\end{figure}

This section presents posterior summaries of the spatially varying GEV parameters under the factual climate for Cluster~3. Figures~\ref{fig:gev_caterpillar_mu}--\ref{fig:gev_caterpillar_xi} display caterpillar plots for the location, scale, and shape parameters, respectively. These plots summarize the posterior distributions across grid cells, highlighting spatial variability and associated uncertainty in the estimated GEV coefficients.
%========================================
% Figure : mu0 and mu1
%========================================
%========================================
% Figure 2: sigma0 and sigma1
%========================================

%========================================
% Figure 3: xi
%========================================

\subsection{Diagnostic estimation Neural Bayes estimator} \label{sec:sup_plot_NBE}
\begin{figure}[htbp]
    \centering
    
    \begin{subfigure}[t]{0.45\textwidth}
        \centering
        \includegraphics[width=\textwidth]{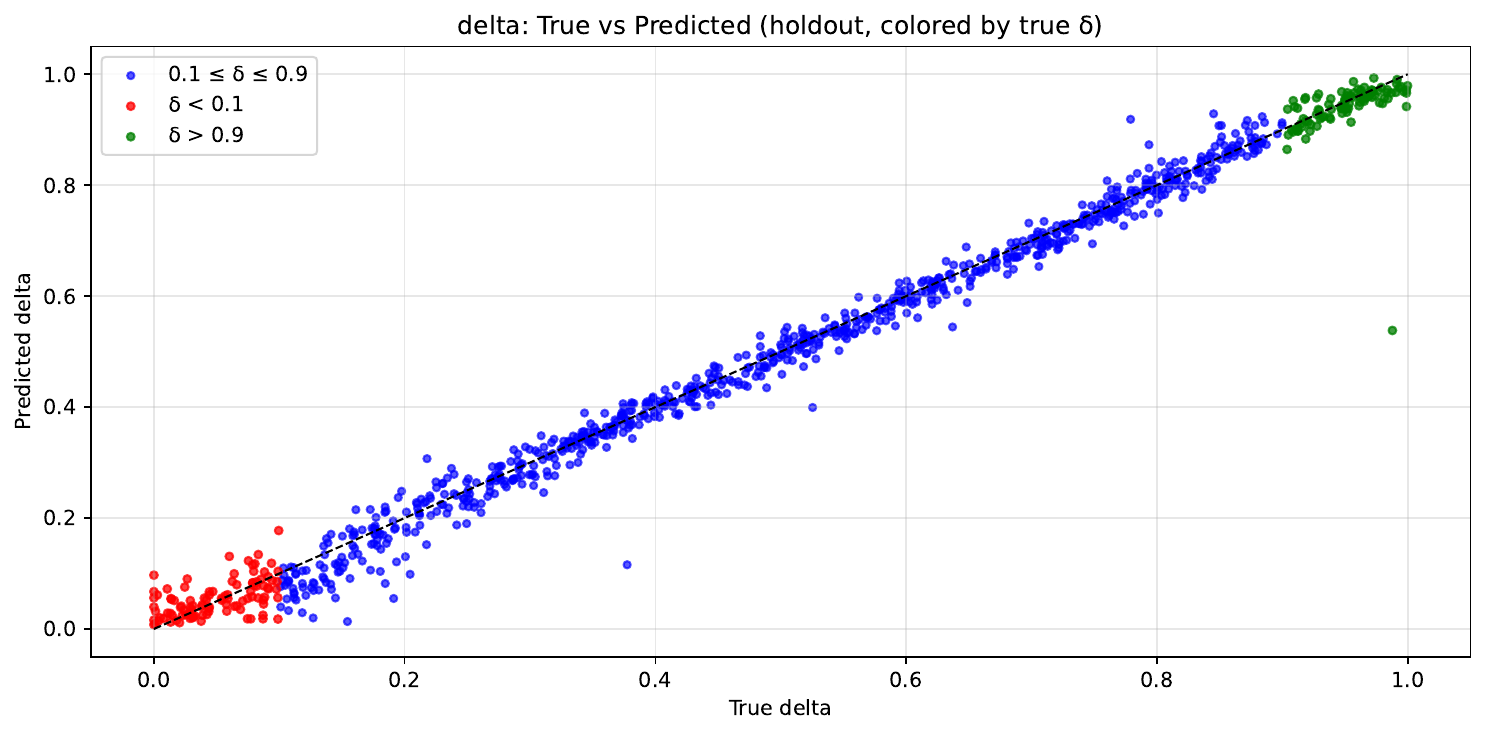}
        \caption{$\delta$}
    \end{subfigure}
    \hfill
    \begin{subfigure}[t]{0.45\textwidth}
        \centering
        \includegraphics[width=\textwidth]{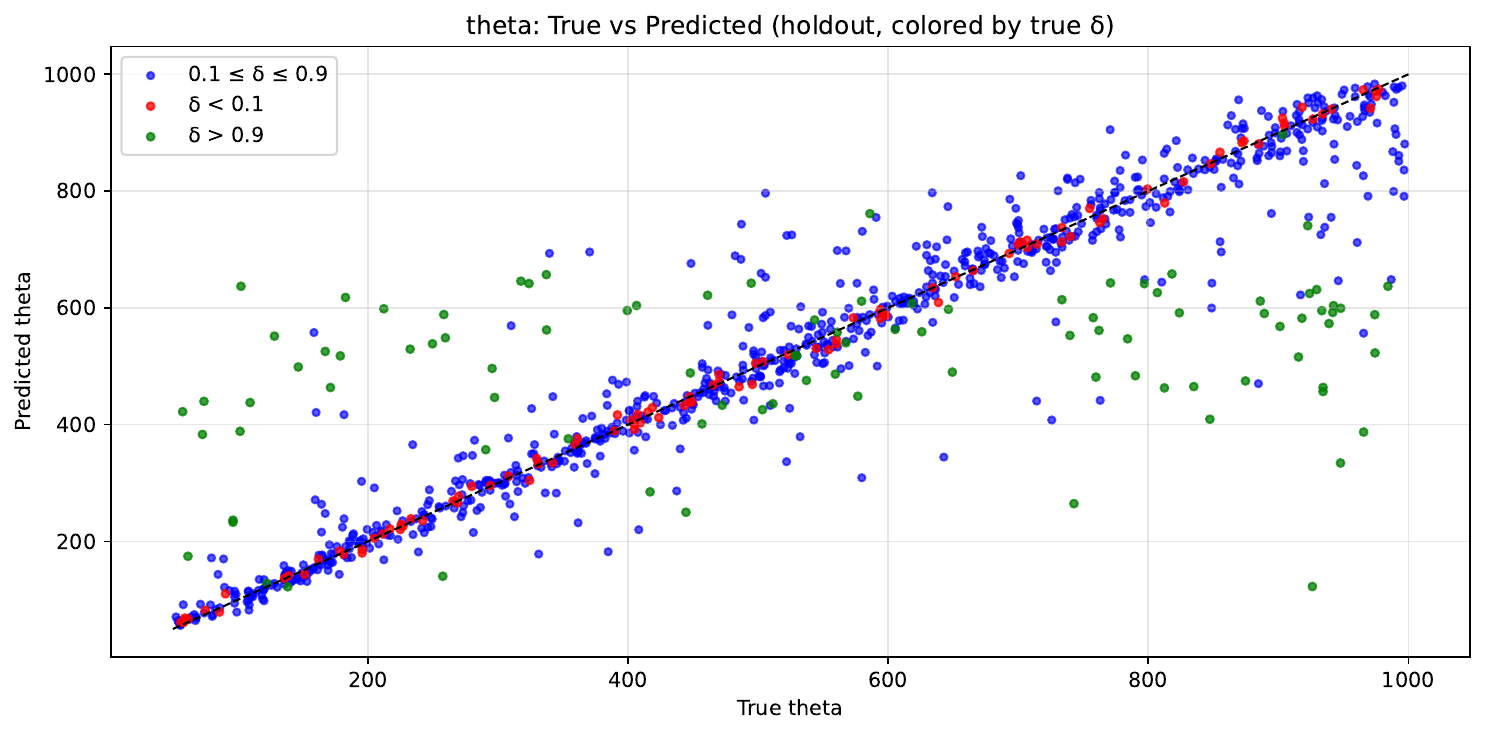}
        \caption{$\theta$}
    \end{subfigure}

    \vspace{0.3cm}

    \begin{subfigure}[t]{0.45\textwidth}
        \centering
        \includegraphics[width=\textwidth]{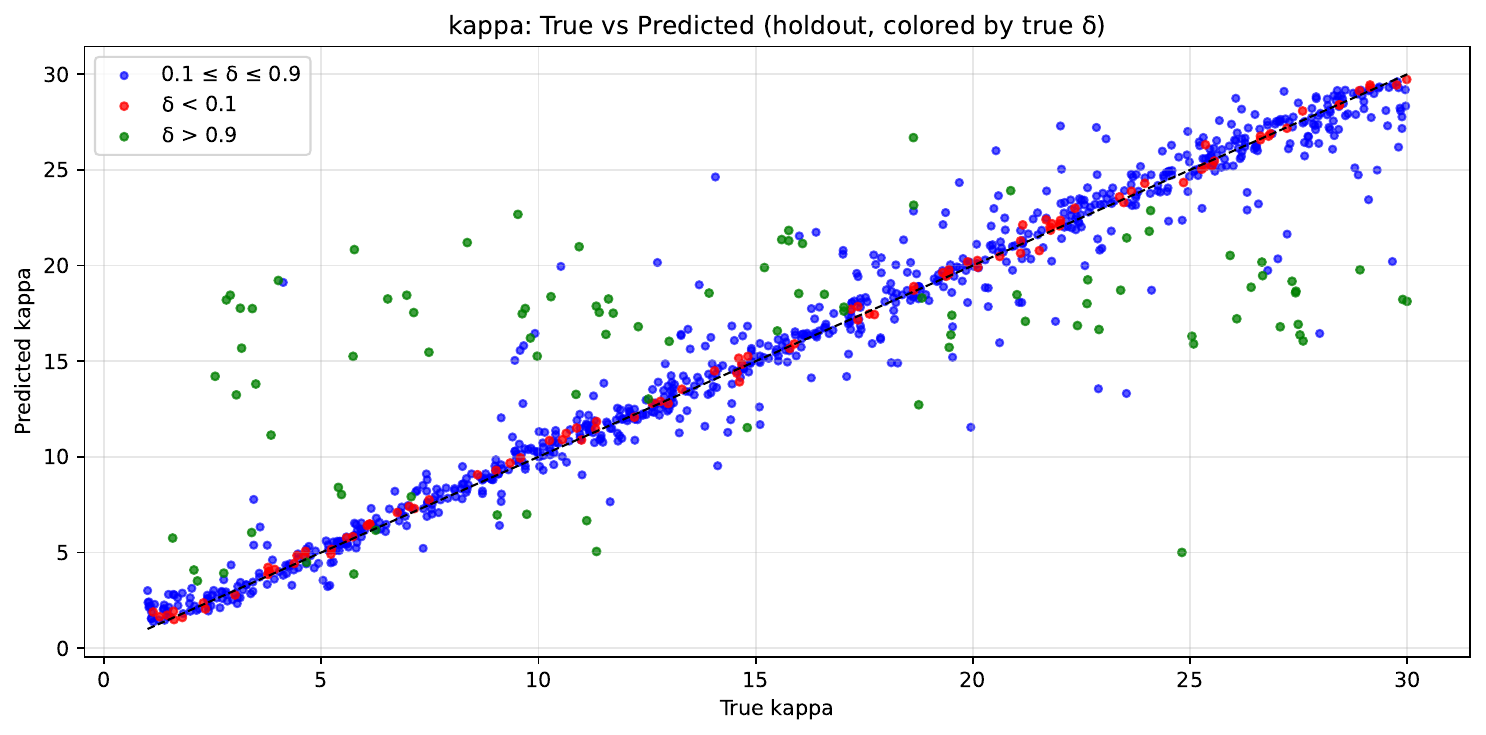}
        \caption{$\kappa$}
    \end{subfigure}
    \hfill
    \begin{subfigure}[t]{0.45\textwidth}
        \centering
        \includegraphics[width=\textwidth]{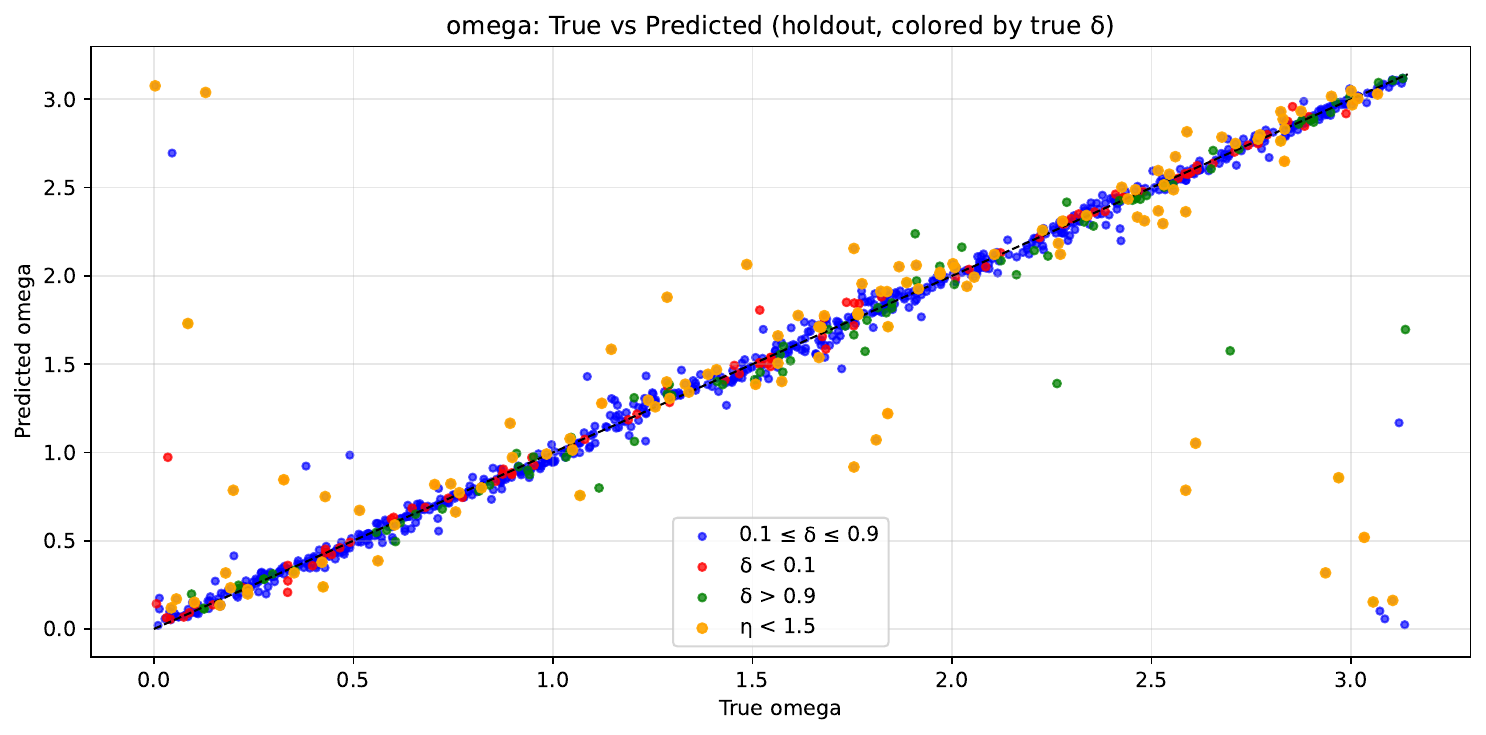}
        \caption{$\omega$}
    \end{subfigure}

    \vspace{0.3cm}

    \begin{subfigure}[t]{0.45\textwidth}
        \centering
        \includegraphics[width=\textwidth]{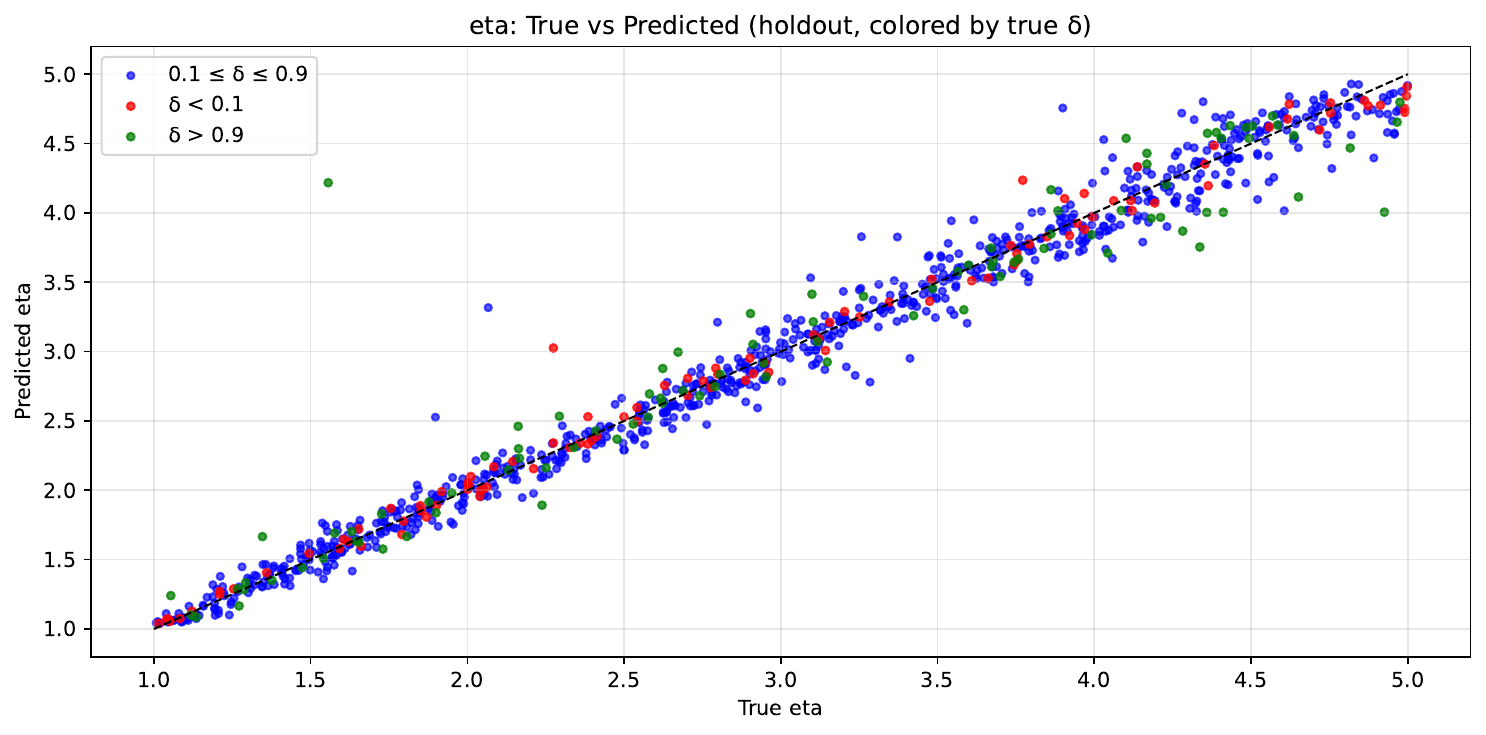}
        \caption{$\eta$}
    \end{subfigure}
    \hfill
    \begin{subfigure}[t]{0.45\textwidth}
        \centering
        \includegraphics[width=\textwidth]{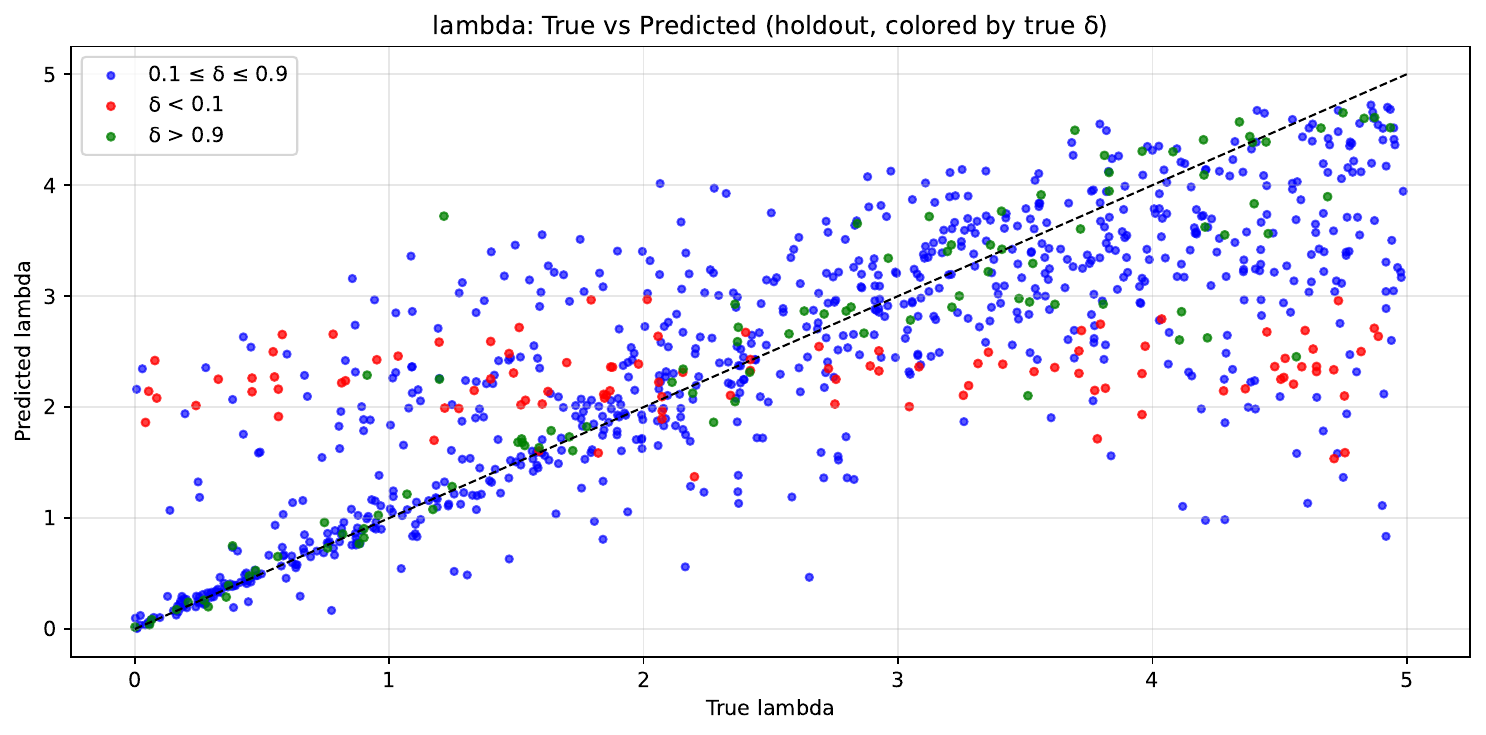}
        \caption{$\lambda$}
    \end{subfigure}

    \vspace{0.3cm}

    \begin{subfigure}[t]{0.45\textwidth}
        \centering
        \includegraphics[width=\textwidth]{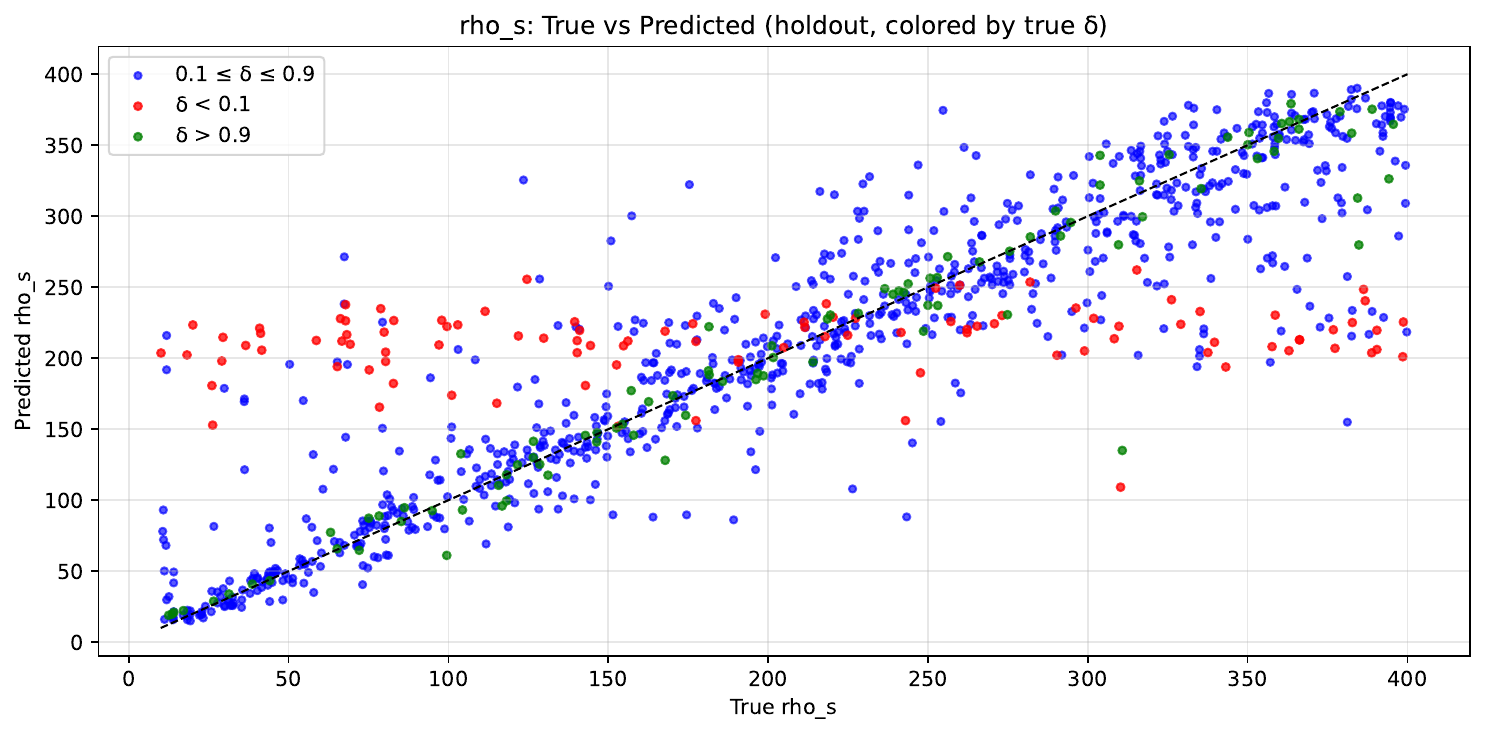}
        \caption{$\rho_s$}
    \end{subfigure}
    \hfill
    \begin{subfigure}[t]{0.45\textwidth}
        \centering
        \includegraphics[width=\textwidth]{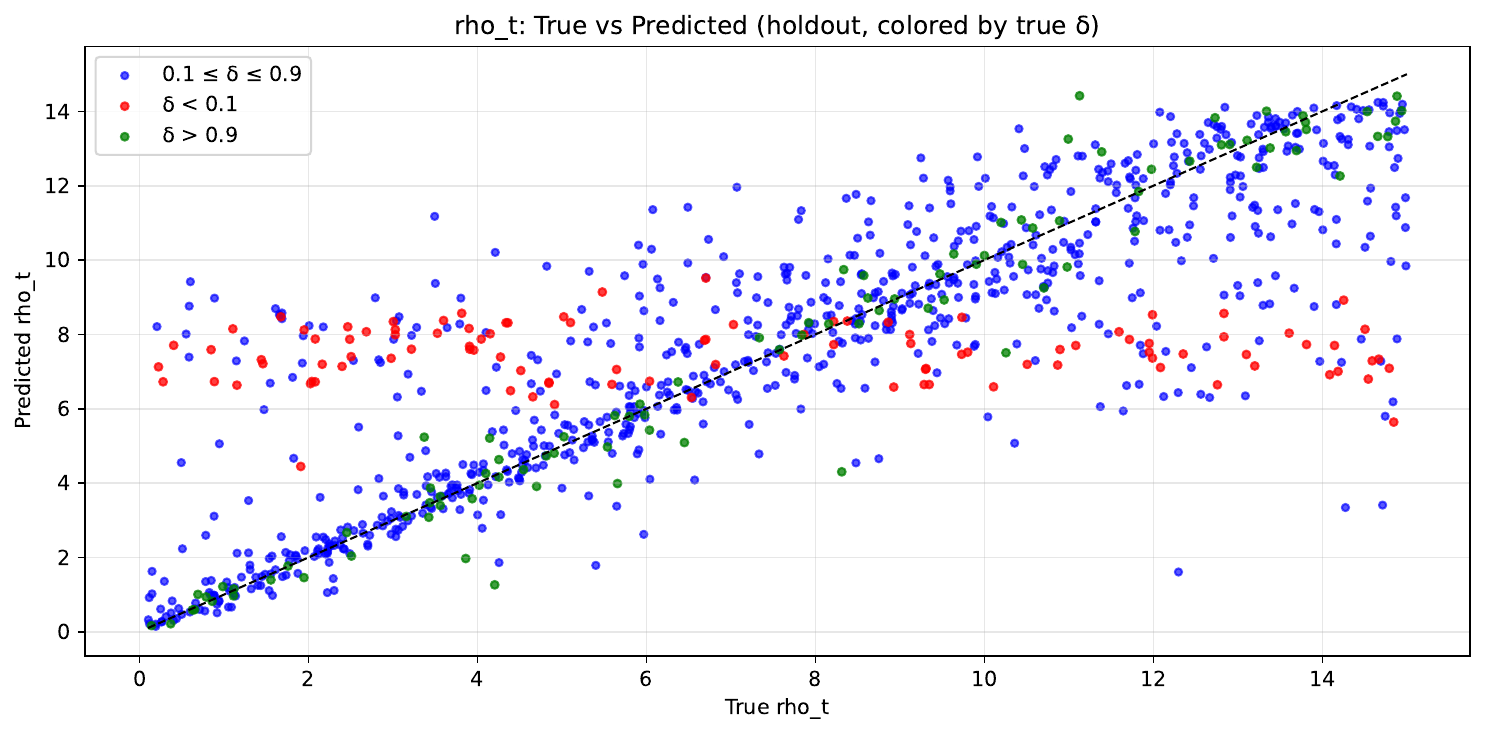}
        \caption{$\rho_t$}
    \end{subfigure}

    \caption{
    Scatterplots of predicted versus true parameter values on the holdout set for Cluster~3.
    The dashed line in each panel indicates perfect agreement between prediction and truth.
    }
    \label{fig:cluster3_scatter_params}
\end{figure}

\begin{figure}[h]
    \centering
    
    \begin{subfigure}[t]{0.24\textwidth}
        \centering
        \includegraphics[width=\textwidth]{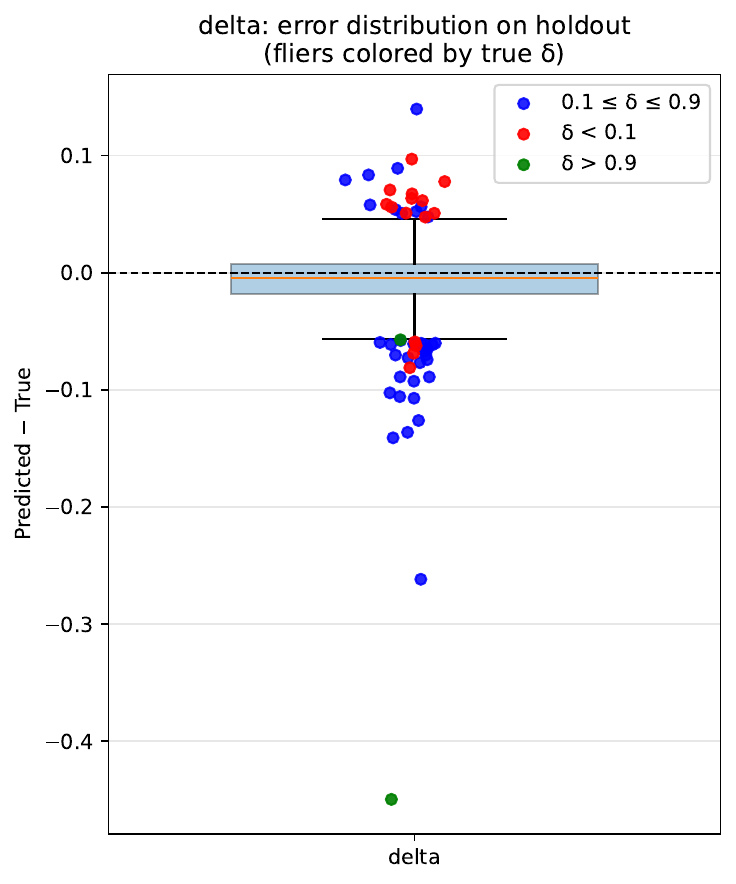}
        \caption{$\delta$}
    \end{subfigure}
    \hfill
    \begin{subfigure}[t]{0.24\textwidth}
        \centering
        \includegraphics[width=\textwidth]{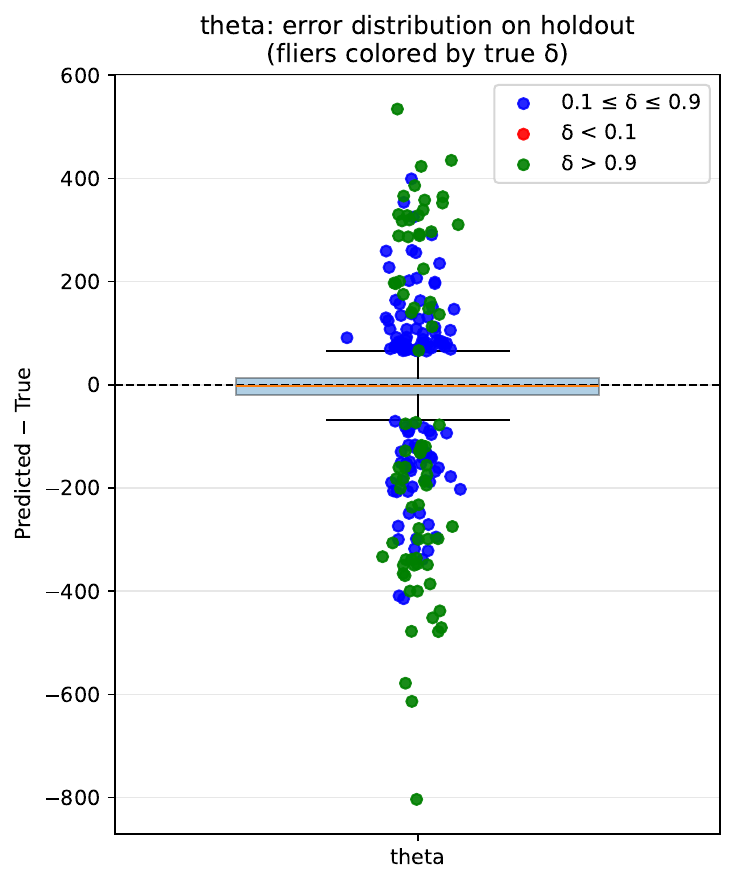}
        \caption{$\theta$}
    \end{subfigure}
    \hfill
    \begin{subfigure}[t]{0.24\textwidth}
        \centering
        \includegraphics[width=\textwidth]{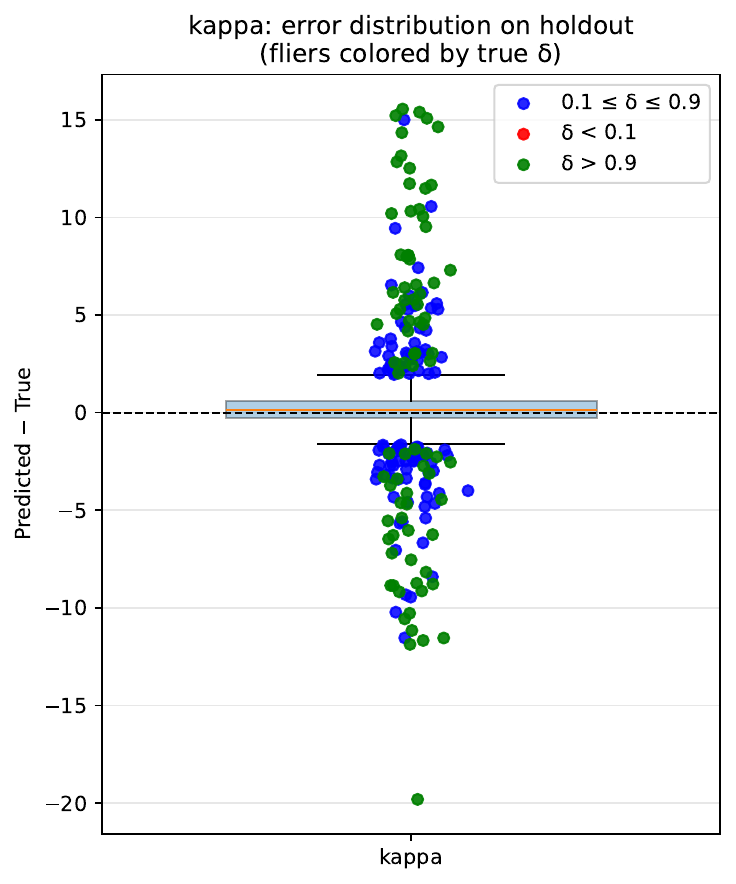}
        \caption{$\kappa$}
    \end{subfigure}
    \hfill
    \begin{subfigure}[t]{0.24\textwidth}
        \centering
        \includegraphics[width=\textwidth]{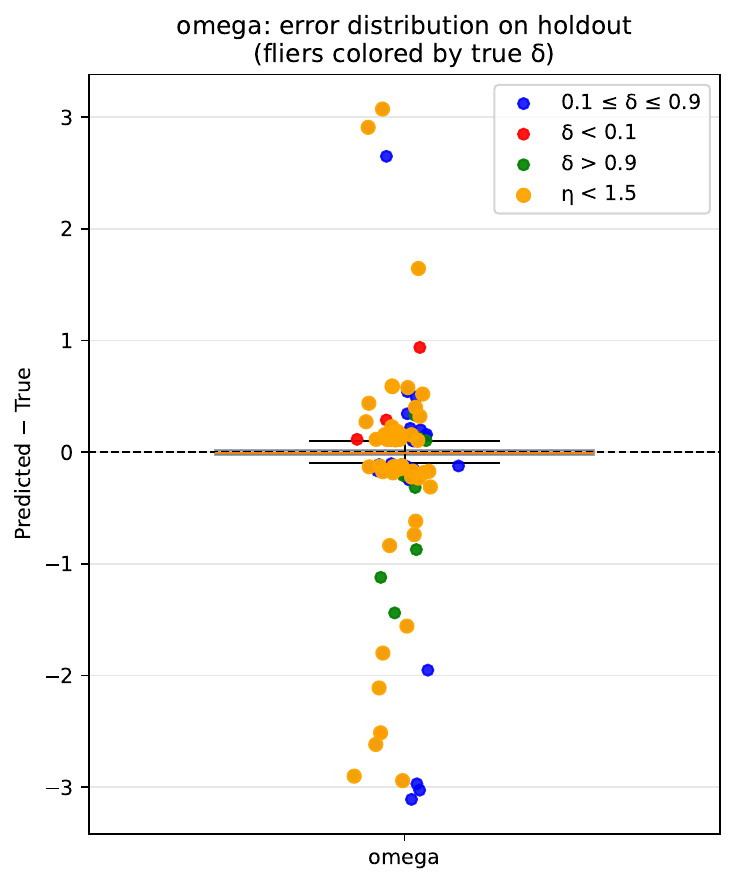}
        \caption{$\omega$}
    \end{subfigure}

    \vspace{0.4cm}

    \begin{subfigure}[t]{0.24\textwidth}
        \centering
        \includegraphics[width=\textwidth]{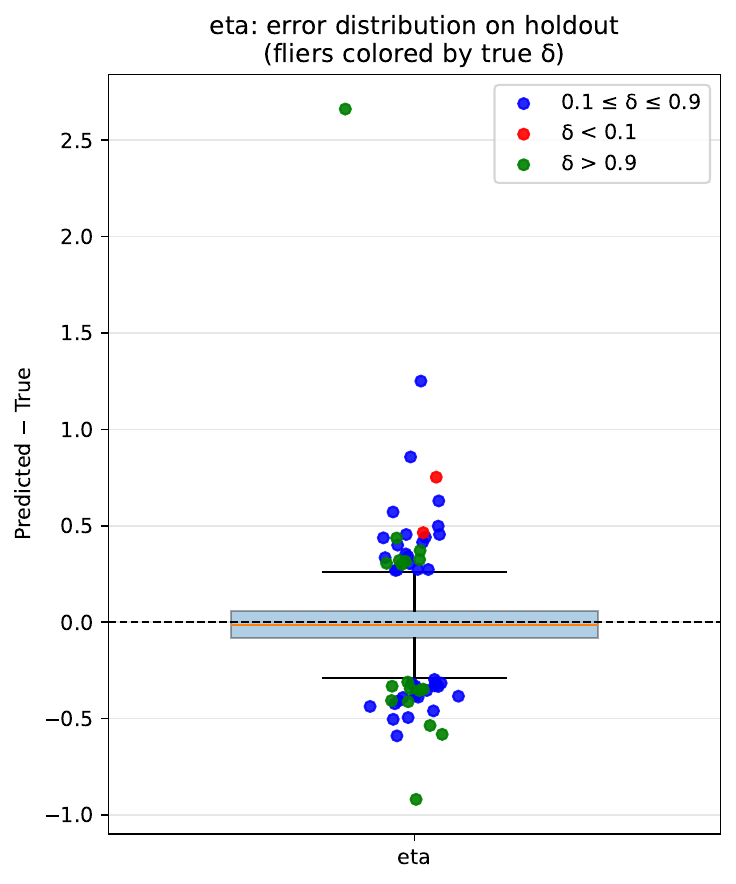}
        \caption{$\eta$}
    \end{subfigure}
    \hfill
    \begin{subfigure}[t]{0.24\textwidth}
        \centering
        \includegraphics[width=\textwidth]{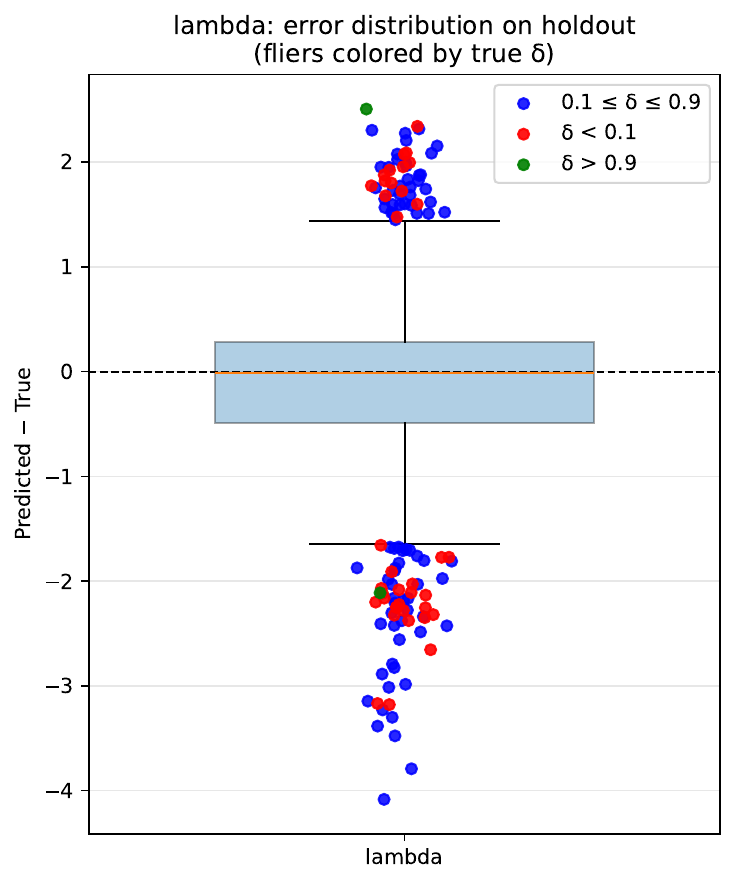}
        \caption{$\lambda$}
    \end{subfigure}
    \hfill
    \begin{subfigure}[t]{0.24\textwidth}
        \centering
        \includegraphics[width=\textwidth]{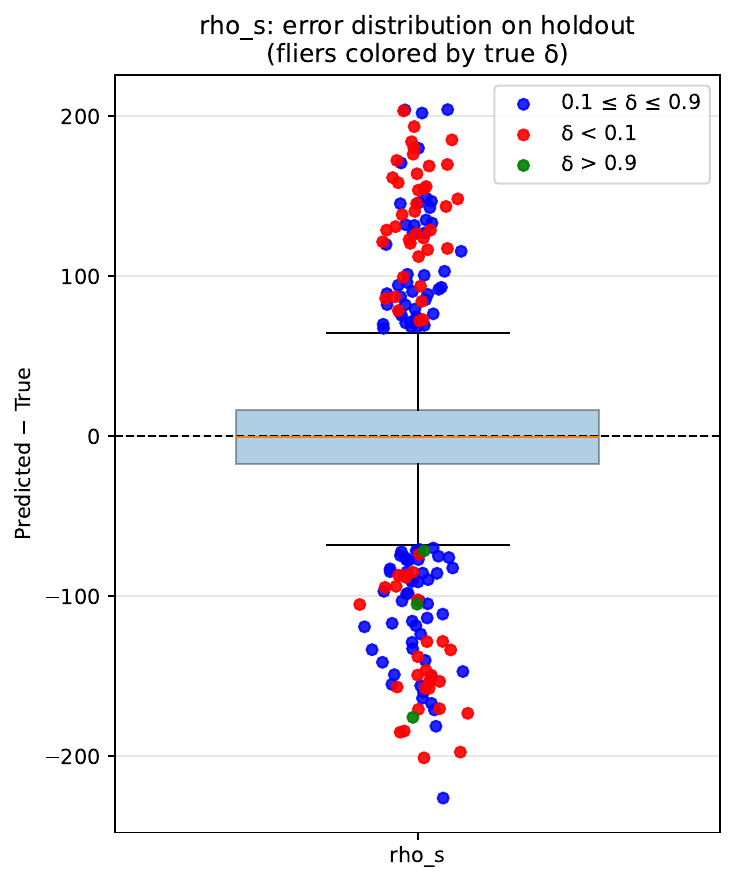}
        \caption{$\rho_s$}
    \end{subfigure}
    \hfill
    \begin{subfigure}[t]{0.24\textwidth}
        \centering
        \includegraphics[width=\textwidth]{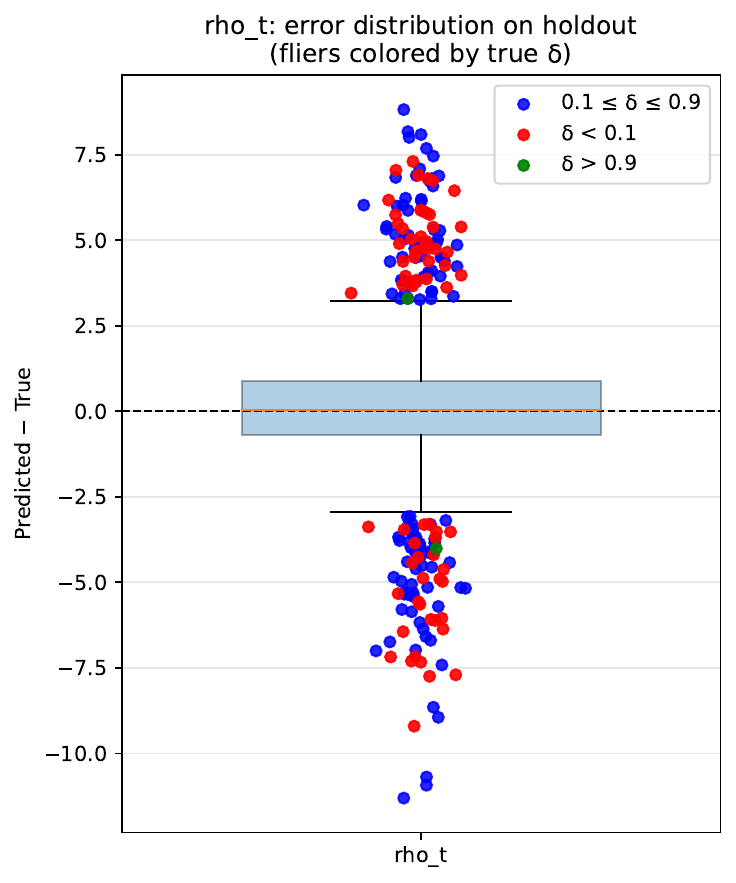}
        \caption{$\rho_t$}
    \end{subfigure}

    \caption{
    Boxplots of the prediction errors (predicted minus true value) on the holdout set for Cluster~3.
    Values centered around zero indicate low bias, while the spread reflects estimation uncertainty.
    }
    \label{fig:cluster3_boxplot_params}
\end{figure}

This section presents validation diagnostics for the neural Bayes estimator on a holdout dataset. Figures~\ref{fig:cluster3_scatter_params} and~\ref{fig:cluster3_boxplot_params} compare predicted and true parameter values for Cluster~3. The scatterplots assess agreement between predictions and ground truth, while the error boxplots summarize bias and variability of the estimates. Together, these diagnostics provide an overall evaluation of the predictive performance of the neural network.

%%%%%%%%%%%%%%%% END OF MAIN TEXT %%%%%%%%%%%%%%%

\end{document}